%% file: main.tex
\documentclass[11pt,a4paper]{article}
\usepackage[T1]{fontenc}
\usepackage{lmodern}
\usepackage[margin=1.05in,includefoot]{geometry}
\usepackage{amsmath,amssymb,amsthm,mathtools}
\usepackage{booktabs,framed}
\usepackage[protrusion=false]{microtype}
\usepackage{xurl}
\usepackage[colorlinks=true,linkcolor=blue,citecolor=blue,urlcolor=blue]{hyperref}
\newtheorem{theorem}{Theorem}[section]
\newtheorem{lemma}[theorem]{Lemma}

\theoremstyle{remark}

\newcommand{\F}{\mathbb F}
\newcommand{\R}{\mathbb R}
\newcommand{\Z}{\mathbb Z}
\newcommand{\softO}{\widetilde O}
\newcommand{\calF}{\mathcal F}
\newcommand{\calG}{\mathcal G}

\newcommand{\DocumentTitle}{Faster Deterministic All-Pairs Shortest Paths with Real Edge Weights}
\title{Faster Deterministic All-Pairs Shortest Paths\\with Real Edge Weights}
\author{
  \normalsize Isaac Barouch Essayag\\[0.5em]
  \small MIGAL--Galilee Research Institute, Kiryat Shmona, Israel\\
  \small Tel-Hai University of Kiryat Shmona and the Galilee, Israel\\[0.25em]
  \small\href{mailto:isaac.es@migal.org.il}{\texttt{isaac.es@migal.org.il}}\\[2em]
  \normalsize Aryeh Lev Zabokritskiy (Yohananov)\\[0.5em]
  \small Department of Computer Science, MIGAL--Galilee Research Institute,\\
  \small Kiryat Shmona, Israel, and Department of Computer Science, Tel-Hai\\
  \small University of Kiryat Shmona and the Galilee, Kiryat Shmona, Israel\\[0.25em]
  \small\href{mailto:yuhanalev@telhai.ac.il}{\texttt{yuhanalev@telhai.ac.il}}\\[0.5em]
  \small ORCID: \href{https://orcid.org/0000-0003-3151-6192}{0000-0003-3151-6192}
}
\date{}
\input{glyphtounicode}
\pdfglyphtounicode{Omega}{03A9}
\pdfglyphtounicode{Delta}{0394}
\pdfglyphtounicode{mu}{03BC}
\hypersetup{pdftitle={\DocumentTitle},pdfauthor={Isaac Barouch Essayag, Aryeh Lev Zabokritskiy (Yohananov)},pdflang={en}}
\begin{document}
{\hypersetup{urlcolor=black}\maketitle}
\input{opening}
\input{framework}
\input{primitives}

\input{selection}

\input{comparison-proofs}

\input{recovery}
\input{finish}

\input{envelope}
\input{source}
\input{walsh-source}
\input{exponents}

\input{conclusion}
\appendix
\input{numerical}

\input{rectangular}
\input{recovery-proof}
\input{original-embedding}
\bibliographystyle{plain}
\let\APSPthebibliography\thebibliography
\renewcommand{\thebibliography}[1]{\APSPthebibliography{#1}\addcontentsline{toc}{section}{\refname}}
\bibliography{references}
\end{document}

%% file: opening.tex
\begin{abstract}
\fontsize{10.2pt}{12pt}\selectfont
We give a deterministic algorithm for exact all-pairs shortest paths (APSP), that is, all distances, in directed $n$-vertex graphs with arbitrary finite real edge weights and no negative cycles. Its worst-case running time is $O(n^{2.9929729})$ with the recent rectangular matrix multiplication bound $\omega(1,\tfrac12,1)=2$, and $O(n^{2.99306273})$ with earlier rectangular bounds alone; both improve the deterministic bound $O(n^{2.9997})$ and the Las Vegas expected-time bound $O(n^{2.993805})$. The algorithm runs on a real RAM whose only real-number operations are exact addition, subtraction, and comparison, each at unit cost.
\end{abstract}

\section{Introduction}

In the all-pairs shortest paths problem (APSP), we seek distances between every ordered pair of vertices in a directed graph. A distance is the minimum walk weight, or $+\infty$ if the pair is unreachable. We allow finite real weights of either sign and assume there are no negative cycles.

The cubic Floyd--Warshall algorithm adds one permitted intermediate vertex at a time. Floyd~\cite{Floyd1962} gave the weighted shortest-path recurrence and Warshall~\cite{Warshall1962} the Boolean recurrence for reachability. Fredman obtained an $o(n^3)$ algorithm~\cite{Fredman1976}, and Chan obtained an $O(n^3/\log n)$ bound~\cite{Chan2008}. Williams~\cite{Williams2018} obtained a running time of $n^3/2^{\Omega(\sqrt{\log n})}$ with high probability in the real-RAM model, and Chan and Williams~\cite[Theorem 1.1]{ChanWilliams2021} obtained this bound deterministically.

Chan, Itakura, Ma and Yu~\cite[Section 4]{CIMY} gave a deterministic $O(n^{2.9997})$ algorithm for real-weight APSP. Alman and Vassilevska Williams~\cite[Theorem 35]{AlmanVassilevskaWilliams2026} gave a Las Vegas algorithm with expected running time $O(n^{2.998})$. A public working note~\cite{WorkingNote} improved this expected-time bound to $O(n^{2.993805})$. We give a deterministic algorithm for the same graphs with worst-case running time $O(n^{2.9929729})$.{\interfootnotelinepenalty=10000\footnote{\fontsize{10.2pt}{12pt}\selectfont Using Uzna\'nski's bounds for all rectangular shapes~\cite[Theorem~1.1]{Uznanski}, we also obtain a deterministic worst-case bound of $O(n^{2.99296306733})$ for the same real-weight problem, and, in our model, deterministic worst-case time $O(n^{2.992676})$ for polynomially bounded integer weights, improving the deterministic $O(n^{2.9926862626})$ bound in~\cite{WorkingNote}. We do not prove these two refinements here: their small exponent gains require substantially longer proofs, which are not yet publicly available.}}

Fischer, Polak and Schmidt~\cite[Theorem~1.1 and Section~1.2]{FischerPolakSchmidt2026} give a randomized reduction that transfers a truly subcubic integer-weight APSP bound to real-weight APSP. The reduction uses comparisons and additions of real weights. Combined with an integer APSP algorithm in the real RAM with multiplication of real numbers, which our model does not permit, it gives a randomized truly subcubic bound for real weights~\cite[Corollary~1.2]{FischerPolakSchmidt2026}. Unlike our algorithm, this route is randomized, and its reduction turns a saving~$\varepsilon$ in the integer-weight exponent into a saving of~$\Omega(\varepsilon^3)$ for real weights~\cite[Theorem~1.1]{FischerPolakSchmidt2026}.

The min-plus product of finite matrices $A,B\in\R^{n\times n}$ is $(A\star B)_{ij}\coloneqq\min_k(A_{ik}+B_{kj})$. Also called the distance product, it chooses the cheapest route through an intermediate vertex, and APSP reduces to it by standard repeated squaring~\cite[Section 4]{CIMY}.

We use a restricted real RAM in the setting of~\cite{Williams2018}. On real numbers it permits only exact addition, subtraction and comparison, each at unit cost, as in~\cite[Section 4]{CIMY}, \cite[Theorem 35]{AlmanVassilevskaWilliams2026} and~\cite{WorkingNote}. Auxiliary integer arithmetic and indexed memory access are charged at bit cost, including the address. The running times below use the rectangular matrix multiplication bounds of Le Gall and Urrutia~\cite{LeGallUrrutia} and Uzna\'nski~\cite[Corollary 1.2]{Uznanski}.

For a shape parameter $\alpha$, $\omega(1,\alpha,1)$ denotes the arithmetic exponent for multiplying an $n\times\lfloor n^\alpha\rfloor$ matrix by a $\lfloor n^\alpha\rfloor\times n$ matrix.

\begin{theorem}\label{thm:main}
In this model, our deterministic algorithm computes the exact min-plus product of two $n\times n$ matrices with arbitrary finite real entries in $O(n^{2.9929729})$ time using Uzna\'nski's theorem $\omega(1,\tfrac12,1)=2$ together with the Le Gall--Urrutia bounds. Using the Le Gall--Urrutia bounds alone gives $O(n^{2.99306273})$ time. Both time bounds hold for all-pairs distances on directed $n$-vertex graphs given by adjacency matrices, with arbitrary finite real edge weights, missing edges and no negative cycles.
\end{theorem}

The algorithm uses its fast routine for computing selected entries of ordinary matrix products only at extremely large input sizes, because of the dimension tests in Section~\ref{sec:point}.

The algorithm saves over cubic time by computing, jointly for many outputs, exact counts of the intermediate vertices in a set that improve on a given intermediate vertex. The joint computation works even when the comparisons behind these counts are encoded by integer ranks and tested modulo different, deterministically chosen integers for different sets. Section~\ref{sec:framework} gives the min-plus algorithm and its use for APSP. Sections~\ref{sec:primitives}--\ref{sec:finish} implement its selection and recovery phases. Section~\ref{sec:envelope} bounds the total cost in terms of ordinary-product computations. Section~\ref{sec:ordinary} supplies those computations, and Section~\ref{sec:point} derives the running time.

%% file: framework.tex
\section{The min-plus algorithm}\label{sec:framework}

To compute a min-plus product, we must find the best intermediate vertex for every pair of endpoints. Testing all $n$ candidates at all $n^2$ outputs takes cubic time. Our algorithm first selects a small set of candidates shared by the outputs. At each output, it then recovers every candidate that improves on the best choice so far. Taking the minimum over these candidates and the best choice so far gives the exact answer. The saving comes from finding the improving candidates through comparison counts computed jointly for many outputs.

This product algorithm gives APSP by repeated min-plus squaring, as Algorithm~\ref{alg:apsp} at the end of this section shows.

Our selection and recovery carry out deterministically the plan of Chan, Vassilevska Williams and Xu~\cite[Claim 8.10 in the full version]{CVWX}. To share work across recovery tasks even though their numerical inputs differ, we adapt in Section~\ref{sec:ordinary}, with proofs, the ordinary-product constructions of Alman and Vassilevska Williams~\cite{AlmanVassilevskaWilliams2026}, Majdoddin~\cite{Majdoddin} and the working note~\cite{WorkingNote}.

\subsection{Block minima and improving candidates}\label{sec:blocks}

Partition the intermediate vertices into blocks of size $d$, where $d\le n$ is a power of two chosen later. The comparison operations of Section~\ref{sec:primitives} require $d^2\le n$ and $d^4(1+\log_2(d^2))\le n^\alpha$ for the shape parameter $\alpha$ of their rectangular products, whose short dimension is at most $n^\alpha$; these conditions bound the block size (line~\ref{alg:mp-guard} of Algorithm~\ref{alg:minplus}). For one block, the operands have dimensions $n\times d$ and $d\times n$; we keep the names $A$ and $B$ for them and index the candidates of the block by $[d]$. Compute its minimum at every output, and take the entrywise minimum over the $\lceil n/d\rceil$ blocks.

Give candidate $k\in[d]$ the score
\[
s_{ij}(k)\coloneqq(A_{ik}+B_{kj},k)
\]
in the lexicographically ordered additive group $\R\times\Z$. Every output then has a strict candidate order, even when real sums coincide. The integer tags $k$ use $O(\log n)$ bits; only the real component is returned. An incumbent table $\pi$ assigns a candidate to each output. In the terminology of Alman and Vassilevska Williams~\cite{AlmanVassilevskaWilliams2026}, the incumbent is the pivot. The table's exact zero-based ranks, the number of outputs and the total rank are
\[
r_{ij}\coloneqq|\{k:s_{ij}(k)<s_{ij}(\pi_{ij})\}|,
\qquad P\coloneqq n^2,\qquad \Phi\coloneqq\sum_{i,j}r_{ij}.
\]
The candidates counted by $r_{ij}$ are the \emph{improving candidates} at $(i,j)$: precisely the intermediate vertices giving a better tagged score than the incumbent. If $r_{ij}=0$, the incumbent is already the answer. Otherwise the minimum lies among these improving candidates. Recovering their indices, then evaluating their scores, therefore suffices to compute the block minimum.

For a short, nonempty last block, copy one genuine column of~$A$ and the matching row of~$B$, assigning the extra candidates distinct tags after the original ones. Every repeated real score already occurs in the block, so the block minimum is unchanged.

\subsection{Selection, recovery and exact completion}\label{sec:phases}

Fix an integer rank scale $1\le R\le d$. The computation of a block has three phases: \emph{selection} of incumbents from candidates shared by all outputs, \emph{recovery} of their improving candidates, and \emph{completion} of the block minimum. The selection routine of Lemma~\ref{lem:selection} leaves total rank $\Phi\le PR$, keeps every rank $O(R\log(2n))$ and uses $h=O(d/R)$ distinct \emph{incumbent labels}, that is, distinct values of $\pi$. The total rank limits how many recovered scores must be evaluated, the pointwise cap on the ranks limits the work of recovering one set of improving candidates, and the number of labels limits the number of different incumbents against which we must count comparisons.

For recovery, regard the set of improving candidates at $(i,j)$ as its indicator vector $z^{ij}\in\{0,1\}^d$, $z^{ij}_k\coloneqq\mathbf1[s_{ij}(k)<s_{ij}(\pi_{ij})]$; a row of a binary matrix measures how many of those improving candidates lie in its subset of candidates. One such matrix serves all outputs, and its construction and decoding costs are paid separately from the computation of the counts. During measurement and decoding the incumbents stay fixed, so every count describes the same set it is meant to recover.

Algorithm~\ref{alg:minplus} lists the complete product computation, with the three phases of each block.

\paragraph{Model and parameters.}
The algorithm runs in the following model: the only operations on real numbers are exact additions, subtractions and comparisons, each at unit cost; integer arithmetic and indexed memory access are charged at bit cost. Its parameters are one column of~\eqref{eq:point} in Section~\ref{sec:point}, for Uzna\'nski's or for the Le Gall--Urrutia bounds: the rationals $c,v,\zeta,r,u,e,\eta$, the target saving $S$, and the two rectangular algorithms of Section~\ref{sec:ordinary-guarantee} that Section~\ref{sec:point} fixes for this column; put $x\coloneqq e/2$ and $b\coloneqq x+\eta$. Selection uses the shape parameter $\alpha=3/10$ (Section~\ref{sec:primitives}) and the loss $\varepsilon=10^{-4}$ (the $\varepsilon$ in its time $\softO(n^{2+\varepsilon}d/R)$, line~\ref{alg:mp-select}). Appendix~\ref{app:numerical} certifies the slack of the column. Line~\ref{alg:mp-dim} fixes the common inner dimension $D=4^m$ of the ordinary products in line~\ref{alg:mp-count} (Sections~\ref{sec:envelope} and~\ref{sec:point}).

\paragraph{Guarantees of the subroutines.}
Lines~\ref{alg:mp-select}, \ref{alg:mp-matrix}, \ref{alg:mp-count} and~\ref{alg:mp-decode} use subroutines of Sections~\ref{sec:selection}--\ref{sec:ordinary}; the correctness of the algorithm relies only on the following guarantees. Given one block and the rank scale $R$, Algorithm~\ref{alg:select} (Section~\ref{sec:selection}) returns incumbents $\pi$ with exact ranks $(r_{ij})$ such that $\Phi\le PR$, every $r_{ij}\le\lambda_{\rm cap}R$, where $\lambda_{\rm cap}=128(1+82\lceil\log_2n\rceil)$, and fewer than $1+5268d/R$ distinct labels (Lemma~\ref{lem:selection}); it requires the hypotheses of Lemma~\ref{lem:weighted} on $\alpha$ and $d$, which hold after line~\ref{alg:mp-guard}.

If $\Phi>0$, line~\ref{alg:mp-kappa} rounds $\min(d-1,\lambda_{\rm cap}R)$ up to the least power of two $\kappa$. This bounds the number of improving candidates at every output.

For integers $1\le\kappa\le d$, Lemma~\ref{lem:recovery} constructs a binary matrix $A_{d,\kappa}$ with $d$ columns, at most $\kappa f(d)$ rows and at most $df(d)$ nonzero entries, together with its row and column lists, where $f(d)=(1+\lceil\log_2(2d)\rceil)^{O(1)}$; from the exact integer measurements $A_{d,\kappa}z$, its decoder recovers every $z\in\{0,1\}^d$ with at most $\kappa$ ones in $O(\kappa f(d))$ bit operations. Line~\ref{alg:mp-chunks} splits the rows of $A_{d,\kappa}$ into \emph{chunks} of candidates, whose counts are computed separately and summed to the measurement of their row.

Given a nonempty chunk $C$, a label $p$ and a set $W$ of requested positions, Algorithm~\ref{alg:count} returns the exact forward and reverse counts at every $w=(i,j)\in W$, the forward count being the number of $k\in C$ with $s_{ij}(k)<s_{ij}(p)$ and the reverse count the number of the other candidates of $C$. It checks the pool test~\eqref{eq:prime-guard}, which its prime selection requires, and counts directly if the test fails (Section~\ref{sec:composite-primes}), so that it is exact on its own. Inside Algorithm~\ref{alg:minplus}, a failed pool test triggers the fallback of line~\ref{alg:mp-fail}. For each incumbent label $p$, line~\ref{alg:mp-count} runs Algorithm~\ref{alg:count} on every chunk with the positions $W_p$ of that label, all calls together in \emph{sweeps}, the slots of the common schedule of Section~\ref{sec:assembly}; a sweep consists of $J'=\max(J,\lceil n^\eta\rceil)$ ordinary-product jobs on $W_p$, one for each of the $J$ chunks and the others all-zero.

Given the $J'$ jobs of a sweep and their common requested positions, the entry procedure (Algorithm~\ref{alg:entry}, Section~\ref{sec:ordinary}) returns the requested entries of all $J'$ products exactly, at a total cost that, divided by $J'$, has the form~\eqref{eq:abstract-source}, provided the tests~\eqref{eq:blocked-width-test} and the other hypotheses of Lemma~\ref{lem:walsh-finite} hold; if the window-length test in line~\ref{alg:en-wtest} of Algorithm~\ref{alg:entry} fails, it reports a failure instead. After every failed test, Algorithm~\ref{alg:minplus} returns $A\star B$ computed from its definition.

Phase~1 produces $(\pi,(r_{ij}))$ with the three bounds of Lemma~\ref{lem:selection}. If $\Phi>0$ and $\kappa\le d/2$, Phase~2 produces, at every output $(i,j)$, the set $Z_{ij}$ of improving candidates of $\pi_{ij}$, decoded from the measurements $A_{d,\kappa}z^{ij}$. Phase~3 produces the block minimum at every output, and line~\ref{alg:mp-merge} keeps the \emph{running minimum}, the entrywise minimum of the real components of the block minima found so far.

\makeatletter
\newcounter{algorithmbox}
\newcounter{algline}[algorithmbox]
\newdimen\AL@indent \AL@indent=1.3em
\newdimen\AL@numwd \AL@numwd=1.5em
\newdimen\AL@numsep \AL@numsep=0.6em
\newbox\AL@box
\newcount\AL@lev
\def\fps@algorithmbox{!htb}
\def\ftype@algorithmbox{4}
\renewcommand{\topfraction}{0.85}
\renewcommand{\bottomfraction}{0.6}
\renewcommand{\textfraction}{0.1}
\renewcommand{\floatpagefraction}{0.8}
\newenvironment{algorithmbox}[1]{%
  \@float{algorithmbox}%
  \fontsize{10.2pt}{12pt}\selectfont\setlength{\parskip}{0pt}\setlength{\parindent}{0pt}%
  \lineskip\z@%
  \refstepcounter{algorithmbox}%
  \hrule height 0.4pt\relax\vskip 3pt
  \noindent\hspace*{0.4em}\textbf{Algorithm~\thealgorithmbox:} #1\par\vskip 2pt
  \hrule height 0.4pt\relax\vskip 1pt
}{%
  \par\vskip 3pt\hrule height 0.4pt\relax
  \end@float
}
\newcommand{\AL@open}[1]{\par\addvspace{\intextsep}\noindent
  \begin{minipage}{\linewidth}%
  \fontsize{10.2pt}{12pt}\selectfont\setlength{\parskip}{0pt}\setlength{\parindent}{0pt}%
  \lineskip\z@
  \hrule height 0.4pt\relax\vskip 3pt
  \noindent\hspace*{0.4em}#1\par\vskip 2pt
  \hrule height 0.4pt\relax\vskip 1pt}
\newcommand{\AL@close}{\par\vskip 3pt\hrule height 0.4pt\relax\end{minipage}\par\addvspace{\intextsep}}
\newenvironment{algorithmhere}[1]{%
  \refstepcounter{algorithmbox}%
  \AL@open{\textbf{Algorithm~\thealgorithmbox:} #1}%
}{\AL@close}
\newcommand{\ALbreak}{\AL@close\newpage\AL@open{\textbf{Algorithm~\thealgorithmbox} (continued)}}
\newcommand{\AL@line}[3]{%
  \setbox\AL@box=\vtop{\hsize=\dimexpr\linewidth-\AL@numwd-\AL@numsep-#1\AL@indent\relax
    \lineskip\normallineskip\raggedright\noindent\strut#3\strut}%
  \makebox[\AL@numwd][r]{#2}\hspace*{\AL@numsep}%
  \AL@lev=0
  \loop\ifnum\AL@lev<#1\relax
    \advance\AL@lev 1
    \hbox to\AL@indent{\hskip 0.25em\vrule width 0.3pt height\ht\AL@box depth\dp\AL@box\hss}%
  \repeat
  \box\AL@box}
\newcommand{\ALstep}[2][0]{\par\noindent\refstepcounter{algline}%
  \AL@line{#1}{\scriptsize\bfseries\thealgline}{#2}}
\newcommand{\ALtext}[2][0]{\par\noindent\AL@line{#1}{}{#2}}
\makeatother
\newcommand{\kw}[1]{\textbf{#1}}
\newcommand{\ALmark}{\ensuremath{\triangleright}\,}
\newcommand{\ALnote}[1]{\unskip\nobreak\hfill\mbox{\ALmark\itshape #1}}
\begin{algorithmhere}{Exact min-plus product}\label{alg:minplus}
\ALtext{\kw{Input:} $A,B\in\R^{n\times n}$ with finite real entries}
\ALtext{\kw{Output:} the min-plus product $A\star B$}
\ALstep{$d\gets$ the largest power of two with $d\le n^b$;\enspace $R\gets\min(d,\lceil n^\eta\rceil)$}\label{alg:mp-param}
\ALstep{$D\gets$ the least grid value $4^m$ satisfying~\eqref{eq:effective-grid}\ALnote{Section~\ref{sec:envelope}}}\label{alg:mp-dim}
\ALstep{\kw{if} $d^2>n$ \kw{or} $d^4(1+\log_2(d^2))>n^\alpha$ \kw{or} $d=1$ \kw{or} $R=d$ \kw{then}}\label{alg:mp-guard}
\ALstep[1]{\kw{return} $A\star B$ computed from its definition}
\ALstep{\kw{for each} block of $d$ consecutive intermediate vertices \kw{do}}\label{alg:mp-block}
\ALstep[1]{\kw{if} the block is short \kw{then} pad it to $d$ candidates as in Section~\ref{sec:blocks}}\label{alg:mp-pad}
\ALtext[1]{\ALmark\textbf{Phase 1: Selection}}
\ALstep[1]{$(\pi,(r_{ij}))\gets$ Algorithm~\ref{alg:select} on this block with rank scale $R$\ALnote{Section~\ref{sec:selection}}}\label{alg:mp-select}
\ALtext[1]{\ALmark\textbf{Phase 2: Recovery}}
\ALstep[1]{\kw{if} $\Phi>0$ \kw{then}}\label{alg:mp-phi}
\ALstep[2]{$\kappa\gets$ the least power of two with $\kappa\ge\min(d-1,\lambda_{\rm cap}R)$}\label{alg:mp-kappa}
\ALstep[2]{\kw{if} $\kappa\le d/2$ \kw{then}}\label{alg:mp-kappatest}
\ALstep[3]{construct $A_{d,\kappa}$ with its row and column lists\ALnote{Lemma~\ref{lem:recovery}}}\label{alg:mp-matrix}
\ALstep[3]{split each nonempty row of $A_{d,\kappa}$ into as few disjoint chunks of at most $\lceil d/\kappa\rceil$ candidates as possible; $J\gets$ the number of chunks}\label{alg:mp-chunks}
\ALstep[3]{$J'\gets\max(J,\lceil n^\eta\rceil)$}\label{alg:mp-jprime}
\ALstep[3]{\kw{for each} incumbent label $p$ \kw{do}\ALnote{$\pi$ and $(r_{ij})$ stay fixed through line~\ref{alg:mp-min}}}\label{alg:mp-label}
\ALstep[4]{$W_p\gets\{(i,j):\pi_{ij}=p\}$, including outputs of rank zero}\label{alg:mp-wp}
\ALstep[4]{run Algorithm~\ref{alg:count} on $(C,p,W_p)$ for every chunk $C$, all calls together in the synchronized sweeps of Section~\ref{sec:assembly}; before the entry procedure allocates its arrays for the $J'$ jobs of a sweep, test~\eqref{eq:blocked-width-test}\ALnote{Algorithm~\ref{alg:entry}; Section~\ref{sec:point}}}\label{alg:mp-count}
\ALstep[4]{\kw{if} a test of~\eqref{eq:blocked-width-test}, another condition of the entry procedure or the prime pool fails in line~\ref{alg:mp-count} \kw{then}\ALnote{Sections~\ref{sec:composite-primes}, \ref{sec:walsh-sharing} and~\ref{sec:point}}}\label{alg:mp-fail}
\ALstep[5]{\kw{return} $A\star B$ computed from its definition}
\ALstep[3]{\kw{for each} output $(i,j)$ \kw{do}}\label{alg:mp-outputs}
\ALstep[4]{$y\gets$ the vector of the sums, over the chunks of each row of $A_{d,\kappa}$, of their forward counts at $(i,j)$}\label{alg:mp-measure}
\ALstep[4]{$Z_{ij}\gets$ the support decoded from $y$ by Lemma~\ref{lem:recovery}}\label{alg:mp-decode}
\ALtext[1]{\ALmark\textbf{Phase 3: Completion}}
\ALstep[1]{\kw{if} $\Phi=0$ \kw{then} evaluate the incumbent scores and store them as the block minima}\label{alg:mp-zero}
\ALstep[1]{\kw{else if} $\kappa>d/2$ \kw{then} take the minimum of all $d$ scores at every output\ALnote{direct completion}}\label{alg:mp-direct}
\ALstep[1]{\kw{else} at every output $(i,j)$, evaluate the scores $s_{ij}(k)$, $k\in Z_{ij}$, and store the least among them and the incumbent score as the block minimum, in a table separate from $\pi$}\label{alg:mp-min}
\ALstep[1]{merge the real components of the block minima into the running minimum}\label{alg:mp-merge}
\ALstep{\kw{return} the running minimum}\label{alg:mp-return}
\end{algorithmhere}

\paragraph{Correctness and charged costs.}
Lines~\ref{alg:mp-param} and~\ref{alg:mp-dim} of Algorithm~\ref{alg:minplus} need only integer arithmetic and comparisons. Since $4b<\alpha$ and $b>\eta>0$, the guard of line~\ref{alg:mp-guard} holds only for bounded $n$ (Section~\ref{sec:envelope}). After line~\ref{alg:mp-pad} every candidate $k\in[d]$ has the tagged score $s_{ij}(k)=(A_{ik}+B_{kj},k)$, and the padding leaves the block minimum unchanged (Section~\ref{sec:blocks}). Given the bounds checked in line~\ref{alg:mp-guard}, line~\ref{alg:mp-select} takes time $\softO(n^{2+\varepsilon}d/R)$ (Lemma~\ref{lem:selection}). If $\Phi=0$, the incumbent scores of line~\ref{alg:mp-zero}, evaluated in time $\softO(P)$, are the block minima. The case $\kappa>d/2$ of line~\ref{alg:mp-direct} happens only for bounded $n$ (Section~\ref{sec:envelope}); there the cap is too large for recovery to help, and direct completion takes time $\softO(Pd)$.

Line~\ref{alg:mp-matrix} costs $O(df(d))$ bit operations, and line~\ref{alg:mp-chunks} gives $J\le2\kappa f(d)$ chunks. The fallback of line~\ref{alg:mp-fail} charges the work already done and happens only for bounded $n$ (Section~\ref{sec:point}). Otherwise the jobs of a sweep in line~\ref{alg:mp-count} cost an amortized amount of the form~\eqref{eq:abstract-source} per job (Lemma~\ref{lem:walsh-finite}); each label of line~\ref{alg:mp-label} needs $n^{o(1)}$ sweeps, and a block at most $hJn^{o(1)}$ jobs with request volume (the total number of entries they request) $O(PJ)n^{o(1)}$ before zero-job padding. Since the forward counts of line~\ref{alg:mp-count} are exact and $\pi$ and its ranks stay fixed from line~\ref{alg:mp-label} through line~\ref{alg:mp-min}, at each output $(i,j)$ the vector $y$ of line~\ref{alg:mp-measure} is the measurement $A_{d,\kappa}z^{ij}$ of the indicator of the improving candidates, which has $r_{ij}\le\kappa$ ones; hence line~\ref{alg:mp-decode} returns the set $Z_{ij}$ of improving candidates, and line~\ref{alg:mp-min} stores the block minimum (Lemma~\ref{lem:finish}). With $r_A\le\kappa f(d)$ rows of $A_{d,\kappa}$, lines~\ref{alg:mp-outputs}--\ref{alg:mp-decode} and~\ref{alg:mp-min} take time $\softO(P(J+r_A)+P\kappa f(d)+\Phi)$ over the $P$ outputs of the block, where $\Phi\le PR$ counts the evaluated scores, and line~\ref{alg:mp-merge} takes time $\softO(P)$. Algorithm~\ref{alg:minplus} takes time $O(n^{3-S})$, the product bound of Theorem~\ref{thm:main} for the chosen column: Section~\ref{sec:envelope} adds these costs over the $n^{1-b+o(1)}$ blocks of line~\ref{alg:mp-block}, and Section~\ref{sec:point} shows that every saving exceeds $S$.

Two different counting tasks occur. During selection, in line~\ref{alg:mp-select}, we need weighted sums of comparisons and the complete rank table of a newly chosen candidate. During recovery, in line~\ref{alg:mp-count}, we need subset counts against the fixed incumbents. Section~\ref{sec:finish} expresses the recovery counts as prescribed entries of ordinary integer matrix products. There each comparison becomes an equality of binary prefixes of ranks in a sorted list of keys, tested modulo small integers; the modular tests can create false matches. Counting the reverse comparisons as well certifies the counts: the true forward and reverse counts have a known sum, so the number of false matches is observable. Deterministic choices of the moduli remove the false matches.

The remaining task is to compute these ordinary-product entries cheaply enough. For each incumbent label, many recovery counts ask for the same output positions. Their numerical matrices can differ, because they belong to different chunks and their moduli are chosen separately, but the common positions let these products share the work that depends only on the requested positions.

\subsection{All-pairs distances}\label{sec:apsp}

Algorithm~\ref{alg:apsp} computes all distances by repeated min-plus squaring with Algorithm~\ref{alg:minplus}. A finite sentinel weight replaces the missing edges before the first product, so Algorithm~\ref{alg:minplus} receives only finite real entries.

\begin{algorithmbox}{All-pairs distances}\label{alg:apsp}
\ALtext{\kw{Input:} a directed graph on $[n]$ with finite real edge weights $w_a$ and no negative cycle; edges may be missing or parallel}
\ALtext{\kw{Output:} all distances, with $+\infty$ for unreachable pairs}
\ALstep{$W\gets\max(\{1\}\cup\{|w_a|:a\text{ an edge}\})$; form $nW$ and the finite sentinel $H\gets(4n+1)W$ by addition and binary doubling}\label{alg:apsp-sentinel}
\ALstep{\kw{for each} ordered pair $(i,j)$ \kw{do}}\label{alg:apsp-init}
\ALstep[1]{\kw{if} $i=j$ \kw{then} $A_{ij}\gets0$ \kw{else} $A_{ij}\gets$ the least weight of an edge from $i$ to $j$, or $H$ if none}
\ALstep{\kw{repeat} $\lceil\log_2(\max\{1,n-1\})\rceil$ times\ALnote{no squaring for $n=1$}}\label{alg:apsp-square}
\ALstep[1]{$A\gets A\star A$, computed by Algorithm~\ref{alg:minplus}}
\ALstep{\kw{return}, for every $(i,j)$, the distance $A_{ij}$ from $i$ to $j$ if $A_{ij}\le nW$, and $+\infty$ otherwise}\label{alg:apsp-detect}
\end{algorithmbox}

\paragraph{Correctness and cost.} Algorithm~\ref{alg:apsp} puts zero on the diagonal and the finite sentinel $H=(4n+1)W$ on missing edges, where $W\ge1$ and $|w_a|\le W$ for every edge $a$. After its $\lceil\log_2(\max\{1,n-1\})\rceil$ min-plus squarings, the entries represent walks of at most $2^{\lceil\log_2(\max\{1,n-1\})\rceil}<2n$ edges, padded by zero-weight self-loops (the zero diagonal entries). A walk containing a sentinel weighs at least $H-(2n-1)W>nW$. Without negative cycles a reachable pair has a shortest simple path of weight at most $(n-1)W$. Exactly the entries exceeding $nW$ therefore represent unreachable pairs. For a list of $m_{\rm in}$ input edges, reading and consolidating them costs an additional $O(m_{\rm in})$ real comparisons and paid indexing. The proof of Theorem~\ref{thm:main} in Section~\ref{sec:point} absorbs the $O(\log(2n))$ calls of Algorithm~\ref{alg:minplus} in the fixed slack.

%% file: primitives.tex
\section{Comparison operations for selection}\label{sec:primitives}

The selection phase (line~\ref{alg:mp-select} of Algorithm~\ref{alg:minplus}) uses two operations: weighted sums of candidate comparisons, which determine a shared choice, and the rank of that choice at every output, which updates the incumbents. Section~\ref{sec:selection} chooses each batch of candidates by conditional expectations on a quadratic upper bound for the ranks that remain. The coefficients of this bound are assembled from sums, over the outputs of one incumbent label, of one or two comparisons weighted by integers that depend on the output. We state the two operations here and give their implementations in Section~\ref{sec:comparison-proofs}. The scores, incumbents and ranks are those of Section~\ref{sec:framework}.

Fredman's trick~\cite{Fredman1976,Williams2018} compares two sums $A_{ik}+B_{kj}$ and $A_{iq}+B_{qj}$ by comparing the row difference $A_{ik}-A_{iq}$ with the column difference $B_{qj}-B_{kj}$. Our candidate-index tags also break ties between scores with equal real parts. Every predicate used below has the form
\begin{equation}\label{eq:comparison}
s_{ij}(k)<s_{ij}(q)
\quad\Longleftrightarrow\quad
(B_{qj}-B_{kj},0)>(A_{ik}-A_{iq},k-q).
\end{equation}
For fixed $k,q$, the right-hand inequality compares a column key depending only on $j$ with a row threshold depending only on $i$, as in Lemma~\ref{lem:weighted} below; both require only real differences and comparisons. Two such predicates, with possibly different candidate pairs, form a \emph{predicate pair} of Lemma~\ref{lem:weighted}; a single predicate is paired with an always-true one. Forming at most $d^2$ predicate pairs takes $O(nd^2)$ real subtractions and indexed operations, at a cost of $\softO(nd^2)$; this cost is charged to each application of Lemma~\ref{lem:weighted}, outside the bound of the lemma.

\subsection{Weighted comparison counts}

The weighted operation sums one or two simultaneous comparisons over columns. Its weights may depend on both output indices, but one weight matrix is shared by all predicate pairs in an invocation. The rectangular-product bound below gives the cost of this operation.

For the products of Lemma~\ref{lem:rectangular} below, with short dimension $H\le n^\alpha$, we call $\alpha$ the \emph{shape parameter}. Le Gall and Urrutia~\cite[Theorem 1.1 and Table 3]{LeGallUrrutia} give $\omega(1,0.31389,1)=2$ and $\omega(1,17/50,1)\le2.001118$. Both algorithms of Theorem~\ref{thm:main} use the shape parameter $3/10$ for selection. The $O(n^{2.99306273})$ algorithm uses only these bounds; the $O(n^{2.9929729})$ algorithm also uses Uzna\'nski's theorem $\omega(1,\tfrac12,1)=2$ over $\mathbb C$~\cite[Corollary 1.2]{Uznanski}. Section~\ref{sec:point} assigns the remaining shape parameters. Appendix~\ref{app:rectangular} derives Lemma~\ref{lem:rectangular} from them, with exact rational bilinear algorithms and integer costs for both product shapes.

\begin{lemma}\label{lem:rectangular}
Take $\alpha\in\{3/10,31389/100000,17/50\}$ under the Le Gall--Urrutia bounds, or $\alpha=1/2$ under Uzna\'nski's theorem. Put $\omega_*\coloneqq2.001118$ at $\alpha=17/50$ and $\omega_*\coloneqq2$ otherwise. For fixed $\varepsilon,K>0$, the products $n\times H$ by
$H\times n$ and $n\times n$ by $n\times H$ can be computed exactly
and deterministically in $O_{\varepsilon,K}(n^{\omega_*+\varepsilon})$
bit operations for integer entries of magnitude at most $n^K$ and
$H\le n^\alpha$, including input formation and all output entries.
\end{lemma}

Appendix~\ref{app:rectangular} also bounds the bit cost when these products are used in Section~\ref{sec:ordinary} with longer entries and addresses.

\begin{lemma}[Weighted comparison counts]\label{lem:weighted}
Suppose $\alpha$ is one of the shape parameters with $\omega_*=2$ in Lemma~\ref{lem:rectangular}, $d^2\le n$ and $d^4(1+\log_2(d^2))\le n^\alpha$. For each of at most $d^2$ predicate pairs $f$, let $x_f(j),y_f(j)$ be column keys and $a_{if},b_{if}$ row thresholds in $\R\times\Z$, with $O(\log n)$-bit integer components. Given integer weights $W(i,j)$ of magnitude $n^{O(1)}$, all sums
\begin{equation}\label{eq:orthant}
Q_{if}\coloneqq\sum_j W(i,j)
\mathbf1[x_f(j)>a_{if}]\mathbf1[y_f(j)>b_{if}]
\end{equation}
can be computed in $\softO(n^{2+\varepsilon})$ time for every fixed $\varepsilon>0$. Keys and thresholds may be tagged real differences as in~\eqref{eq:comparison}. Their formation is charged separately when used.
\end{lemma}

Lemma~\ref{lem:weighted} adapts the classical algorithms for the dominance product described by Vassilevska and Williams~\cite[Sections 3 and 6]{VW2006}; Section~\ref{sec:comparison-proofs} proves it after its use in choosing a candidate batch.

\subsection{The rank table of one candidate}

When the selection procedure tries a new candidate, it can improve many outputs at once. To update them, we compute the candidate's rank at every output in one table.

\begin{lemma}\label{lem:rank}
Under the hypotheses of Lemma~\ref{lem:weighted} on $\alpha$ and $d$, the complete table
\[
\operatorname{rank}_{ij}(q)\coloneqq|\{k:s_{ij}(k)<s_{ij}(q)\}|
\]
for one specified candidate $q$ can be computed in $\softO(n^{2+\varepsilon})$ time for every fixed $\varepsilon>0$.
\end{lemma}

The construction of this table is given in Section~\ref{sec:comparison-proofs}.

After choosing $q$, compare it with every current incumbent and replace both the incumbent and its stored rank wherever $q$ wins. This adds $O(P)$ comparisons and indexed updates, costing $\softO(P)$ with address bits. The table $(\operatorname{rank}_{ij}(q))_{i,j}$ is absolute: it depends only on the block and on $q$. Once $q$ has been applied globally, applying it again cannot improve any output. Hence a used-candidate flag permits discarding its table after the update.

%% file: selection.tex
\section{A common candidate set with capped ranks}\label{sec:selection}

This section proves Lemma~\ref{lem:selection}, which supplies the selection phase of Algorithm~\ref{alg:minplus} with the three bounds whose roles Section~\ref{sec:phases} describes. Throughout this section, $\alpha$ is a shape parameter with $\omega_*=2$, as in Lemma~\ref{lem:weighted}.

\begin{lemma}\label{lem:selection}
For an integer $1\le R\le d$, one can deterministically obtain incumbents and their exact ranks satisfying
\[
\Phi\le PR,\qquad
\max_{i,j}r_{ij}\le\lambda_{\rm cap} R,\qquad
h<1+5268d/R,
\]
where the returned common set is $\{\pi_{ij}:i,j\in[n]\}$, of size $h\le d$, and
\[
\lambda_{\rm cap}\coloneqq128(1+82\lceil\log_2n\rceil).
\]
Under the hypotheses of Lemma~\ref{lem:weighted} on $\alpha$ and $d$, the complete cost is $\softO(n^{2+\varepsilon}d/R)$ for every fixed $\varepsilon>0$.
\end{lemma}

Algorithm~\ref{alg:select} obtains these incumbents with batch rounds (Algorithm~\ref{alg:batch}, Section~\ref{sec:batch}), each of which lowers the total rank on a designated set of outputs of mean rank above~$4$. It applies them first to all outputs until their total rank is at most $PR$, and then to those outputs whose ranks still exceed the cap. A batch uses candidates common to the designated outputs, and applying it globally cannot increase any rank.

\begin{algorithmbox}{Selection with capped ranks}\label{alg:select}
\ALtext{\kw{Input:} one block, with tagged scores $s_{ij}(k)$ for $k\in[d]$, and an integer $1\le R\le d$; $\alpha$ and $d$ satisfy the hypotheses of Lemma~\ref{lem:weighted}}
\ALtext{\kw{Output:} incumbents $\pi$ with exact ranks $(r_{ij})$ satisfying the bounds of Lemma~\ref{lem:selection}}
\ALstep{\kw{if} $R<4$ \kw{then}}\label{alg:sel-small}
\ALstep[1]{at every output $(i,j)$, compare all $d$ candidates; $\pi_{ij}\gets$ the candidate of least tagged score and $r_{ij}\gets0$}
\ALstep[1]{\kw{return} $(\pi,(r_{ij}))$}
\ALstep{choose one candidate $q$ and compute its rank table (Lemma~\ref{lem:rank}); at every output, $\pi_{ij}\gets q$ and $r_{ij}\gets\operatorname{rank}_{ij}(q)$}\label{alg:sel-init}
\ALstep{\kw{while} $\Phi>PR$ \kw{do}}\label{alg:sel-total}
\ALstep[1]{apply Algorithm~\ref{alg:batch} to the set $\Omega$ of all outputs\ALnote{Section~\ref{sec:batch}}}
\ALstep{\kw{while} the high set $\Omega\gets\{(i,j):r_{ij}>\lambda_{\rm cap}R\}$ is nonempty \kw{do}}\label{alg:sel-cap}
\ALstep[1]{apply Algorithm~\ref{alg:batch} to $\Omega$}
\ALstep{\kw{return} $(\pi,(r_{ij}))$}\label{alg:sel-return}
\end{algorithmbox}

For real 3SUM, Chan et al.~\cite[Section 5.1]{CIMY} select pivots using low averages of squared deviations from a midpoint rank. Our quadratic estimator below bounds the remaining ranks under a common candidate set and, analogously to those averages~\cite[Lemma~5.7]{CIMY}, is evaluated from comparison counts. Chan, Vassilevska Williams and Xu~\cite[Claim 8.10 in the full version]{CVWX} group outputs by sampled incumbents and recover sparse improving witnesses; Sections~\ref{sec:selection}--\ref{sec:finish} carry out this plan deterministically. Alon and Naor~\cite[Sections 2.1--2.3]{AlonNaor} already use shared choices, contraction of aggregate witness counts and a changing set of unresolved outputs for explicit Boolean products. Here we implement the required common-candidate choices for implicit real comparisons: the weighted counts of Lemma~\ref{lem:weighted} evaluate every coefficient of the estimator, and each global update uses one absolute rank table.

The deterministic reduction of Chan et al.~\cite[Section 4]{CIMY} halves ranks separately for each output; the rounds below instead reduce the total rank with candidates common to all outputs. Saha and Ye~\cite[Section 3.1]{SahaYe} use a global potential to count false modular witnesses on polynomially bounded integer inputs; Section~\ref{sec:finish} likewise counts false modular matches over all active positions when it chooses its moduli, with the prime selection credited in Section~\ref{sec:composite-count}.

\subsection{One deterministic batch}\label{sec:batch}

A round must reduce the total rank by a fixed factor while adding few candidates.

Fix a nonempty set $\Omega$ of outputs $o=(i,j)$ whose incumbents and ranks are held fixed during the round; we write $r_o,\pi_o,s_o$ for $r_{ij},\pi_{ij},s_{ij}$. Write $\Psi\coloneqq\sum_{o\in\Omega}r_o$ and $\phi\coloneqq\Psi/|\Omega|$. For a candidate set $G$, let $r'_o\coloneqq\min(\{r_o\}\cup\{\operatorname{rank}_o(g):g\in G\})$ be the rank at $o$ after applying $G$, so $r'_o=r_o$ when $G$ is empty.

\begin{lemma}[Batch round]\label{lem:round}
Suppose $\phi>4$. We can deterministically construct a batch $G$ satisfying
\[
\sum_{o\in\Omega}r'_o\le\rho\Psi,
\qquad 1\le|G|<128d/\phi,
\qquad \rho\coloneqq281/288.
\]
Under the hypotheses of Lemma~\ref{lem:weighted} on $\alpha$ and $d$, for every fixed $\varepsilon>0$, if $h$ distinct incumbent labels occur in $\Omega$, its construction costs
$\softO(hn^{2+\varepsilon}+P+d^2)$, followed by
$\softO(|G|n^{2+\varepsilon})$ to apply its candidates globally and update exact ranks.
\end{lemma}

Algorithm~\ref{alg:batch} constructs the batch. Write $g\prec q$ for $s_o(g)<s_o(q)$. The algorithm fixes the bits $z_g\coloneqq\mathbf1[g\in G]$ one candidate at a time by the sign of a slope $D_g$ computed from weighted comparison counts (Lemma~\ref{lem:weighted}). The slope $D_g$ is the scaled change in the conditional expected cost from excluding $g$ to including it. The cost combines a quadratic upper bound on the total remaining rank on $\Omega$ with a penalty for the size of the batch. Line~\ref{alg:b-scale} fixes the sampling probability $\gamma$ of that estimator, the size penalty $\mu$ added to it and the integer scale $C$ of the weights, and lines~\ref{alg:b-c}--\ref{alg:b-round} fix its constants $c_o$. Lines~\ref{alg:b-apply}--\ref{alg:b-global} apply every chosen candidate at every output of the block, not only at those of $\Omega$.

\begin{algorithmhere}{Batch round}\label{alg:batch}
\ALtext{\kw{Input:} a nonempty set $\Omega$ of outputs with mean rank $\phi=\Psi/|\Omega|>4$; on $\Omega$, the incumbents and ranks stay fixed until line~\ref{alg:b-apply}}
\ALtext{\kw{Output:} a batch $G$ satisfying the bounds of Lemma~\ref{lem:round}, applied at every output of the block}
\ALstep{$\theta\gets$ the largest integer $\theta\ge1$ with $2^{\theta+1}|\Omega|\le\Psi$;\enspace $\gamma\gets2^{-\theta}$;\enspace $\mu\gets2^\theta\Psi/(32d)$;\enspace $C\gets2^{\lceil\log_2d\rceil+1}$}\label{alg:b-scale}
\ALstep{\kw{for each} $o\in\Omega$ \kw{do}}\label{alg:b-c}
\ALstep[1]{\kw{if} $r_o=0$ \kw{then} $c_o\gets1$}\label{alg:b-zero}
\ALstep[1]{\kw{else} $c_*\gets2r_o(r_o-1+2^\theta)/(2^\theta(r_o+1))$, $e\gets\lfloor\log_2c_*\rfloor$, and $c_o\gets2^e$ if $c_*/2^e\le4/3$, $c_o\gets2^{e+1}$ otherwise}\label{alg:b-round}
\ALstep{\kw{for each} incumbent label $q$ occurring in $\Omega$ and each coefficient family $u\in\{1,2,3\}$ \kw{do}}\label{alg:b-counts}
\ALstep[1]{$Q^{(u,q)}\gets$ the sums of Lemma~\ref{lem:weighted} with the weights $W^{(u)}_q(i,j)\coloneqq\mathbf1[o\in\Omega,\ \pi_o=q]\,w^{(u)}_o$, where $(w^{(1)}_o,w^{(2)}_o,w^{(3)}_o)\coloneqq(2C^2r_o/c_o,\ C^2r_o/c_o^2,\ 2C^2/c_o)$, and the predicate pairs $g\prec q$ for $u=1$, $(g\prec q,\ g'\prec q)$ for $u=2$ and $(g\prec q,\ k\prec g)$ for $u=3$}\label{alg:b-sums}
\ALstep{$\Lambda_2(g,g')\gets\sum_{q,i}Q^{(2,q)}_{i,(g,g')}$ and $\kappa_g\gets\sum_{q,i,k}Q^{(3,q)}_{i,(g,k)}-\sum_{q,i}Q^{(1,q)}_{i,g}+\Lambda_2(g,g)$ for all candidates $g,g'$}\label{alg:b-coef}
\ALstep{$D_g\gets\kappa_g+2\gamma\sum_{g'\ne g}\Lambda_2(g,g')+C^2\mu$ for every candidate $g$}\label{alg:b-slope}
\ALstep{\kw{for each} candidate $g$ in turn \kw{do}}\label{alg:b-fix}
\ALstep[1]{$z_g\gets1$ if $D_g<0$, and $z_g\gets0$ otherwise}\label{alg:b-bit}
\ALstep[1]{$D_{g'}\gets D_{g'}+2(z_g-\gamma)\Lambda_2(g',g)$ for every candidate $g'$ not yet fixed}\label{alg:b-update}
\ALstep{$G\gets\{g:z_g=1\}$}
\ALstep{\kw{for each} $g\in G$ \kw{do}}\label{alg:b-apply}
\ALstep[1]{compute $\operatorname{rank}_{ij}(g)$ at every output of the block\ALnote{Lemma~\ref{lem:rank}}}
\ALstep[1]{at every output $(i,j)$ of the block with $s_{ij}(g)<s_{ij}(\pi_{ij})$, $\pi_{ij}\gets g$ and $r_{ij}\gets\operatorname{rank}_{ij}(g)$}\label{alg:b-global}
\ALstep{\kw{return} $G$}
\end{algorithmhere}

\begin{proof}
\emph{A bound on the remaining rank.} Let $Y_o$ be the number of candidates in $G$ better than the incumbent at output $o$, and let $Z_o$ be the sum of their absolute ranks, the values $\operatorname{rank}_o(g)$ of Lemma~\ref{lem:rank}. For every $c_o>0$,
\begin{equation}\label{eq:majorant}
r'_o\le v_o(G)\coloneqq r_o(1-Y_o/c_o)^2+2Z_o/c_o.
\end{equation}
Since $0\le r'_o\le r_o$ and $Z_o\ge Y_or'_o$, setting $y\coloneqq Y_o/c_o$ gives $v_o-r'_o\ge(r_o-r'_o)(1-y)^2+r'_oy^2\ge0$. For a random batch $G$, the expectation of the nonnegative quadratic $v_o(G)$ depends only on the first two moments of $Y_o$ and the mean of $Z_o$, quantities we can compute from the comparison counts.

\emph{Choosing a small batch.} With $\theta$ and $\gamma$ from line~\ref{alg:b-scale} of Algorithm~\ref{alg:batch},
$2/\phi\le\gamma<4/\phi$ and $\gamma\le1/2$.
To define the estimator, let $G$ be random, containing each candidate independently with probability $\gamma$, and write $z_g\coloneqq\mathbf1[g\in G]$. For one output of rank $r>0$, its improving candidates have ranks $0,\ldots,r-1$, so
\[
\mathbb E v=r-a/c+b/c^2,
\quad a\coloneqq\gamma r(r+1),\quad
b\coloneqq\gamma r^2[1+\gamma(r-1)].
\]
The largest possible decrease is $a^2/(4b)$, attained at
\[
c_*\coloneqq\frac{2b}{a}
=\frac{2r(r-1+2^\theta)}{2^\theta(r+1)}.
\]
Line~\ref{alg:b-round} rounds $c_*$ to $c=c_o$ using integer arithmetic and comparisons; this gives $|c_*/c-1|\le1/3$, hence
\[
r-\mathbb E v\ge\frac89\frac{a^2}{4b}
\ge\frac29 r\frac{\gamma r}{1+\gamma r}.
\]
For rank zero, line~\ref{alg:b-zero} sets $c\coloneqq1$; its contribution is zero. Outputs with $r\ge\phi/2$ carry at least $\Psi/2$ rank mass and have $\gamma r\ge1$. Therefore, with $V(G)\coloneqq\sum_{o\in\Omega}v_o(G)$,
\[
\mathbb E V\le17\Psi/18.
\]
With $\mu$ from line~\ref{alg:b-scale}, put $F(G)\coloneqq V(G)+\mu|G|$. Then $\mathbb EF\le(17/18+1/32)\Psi=\rho\Psi$. Lines~\ref{alg:b-fix}--\ref{alg:b-update} apply the method of conditional expectations: they fix candidate bits successively, each time choosing a branch of no larger conditional expectation of $F$, as shown below. The final value is at most $\rho\Psi$. By~\eqref{eq:majorant} and $V\ge0$, this gives both the rank bound and $|G|\le32\rho\gamma d<128d/\phi$. The empty set has $F=\Psi$, so the batch is nonempty.

\emph{Computing the estimator.} It remains to form the quadratic estimator $V$ and evaluate it together with the linear size penalty at the claimed cost. Write $I_o(g)\coloneqq\mathbf1[s_o(g)<s_o(\pi_o)]$. The three coefficient families of $V$ are
\begin{align*}
L_1(g)&\coloneqq\sum_{o\in\Omega}\frac{2r_o}{c_o}I_o(g),\\
L_2(g,g')&\coloneqq\sum_{o\in\Omega}\frac{r_o}{c_o^2}I_o(g)I_o(g'),\\
L_3(g)&\coloneqq\sum_{o\in\Omega}\frac2{c_o}I_o(g)\operatorname{rank}_o(g).
\end{align*}
The polynomial is $V=\Psi-\sum_gL_1(g)z_g+\sum_{g,g'}L_2(g,g')z_gz_{g'}+\sum_gL_3(g)z_g$. The pair sum is ordered and includes its diagonal.

Each $c_o$ divides the integer $C$ of line~\ref{alg:b-scale}, since $1\le c_*<2r<2d$ for positive rank. Line~\ref{alg:b-sums} forms, for each incumbent label $q$ and each coefficient family, one weight matrix shared by all predicate pairs of that family; the weights $w^{(1)}_o,w^{(2)}_o,w^{(3)}_o$ of that line belong to $L_1,L_2,L_3$, respectively. All entries are integers of magnitude $O(d^3)$. For fixed $q$, the tagged comparison~\eqref{eq:comparison} separates into a row threshold and a column key. The incumbent label depends on the output, so we process its possible values separately. With the sums $Q^{(u,q)}_{if}$ of line~\ref{alg:b-sums},
\[
\begin{aligned}
 C^2L_1(g)&=\sum_{q,i}Q^{(1,q)}_{i,g},&
 C^2L_2(g,g')&=\sum_{q,i}Q^{(2,q)}_{i,(g,g')},\\
 C^2L_3(g)&=\sum_{q,i,k}Q^{(3,q)}_{i,(g,k)}.
\end{aligned}
\]
The last equality uses $\operatorname{rank}_o(g)=\sum_k\mathbf1[k\prec g]$. Each family has at most $d^2$ predicate pairs. The masks, the work of Lemma~\ref{lem:weighted} and the row sums are charged separately for each label.

Line~\ref{alg:b-coef} therefore computes $\Lambda_2=C^2L_2$ and
$\kappa_g=C^2[L_3(g)-L_1(g)+L_2(g,g)]$.
Folding $z_g^2=z_g$ is essential: an unfixed diagonal has expectation $\gamma$, not $\gamma^2$. With unassigned bits still of mean $\gamma$, lines~\ref{alg:b-slope}--\ref{alg:b-update} maintain the scaled conditional slope
\[
D_g\coloneqq\kappa_g+2\sum_{g'\ne g}p_{g'}\Lambda_2(g,g')+C^2\mu,
\]
where $p_{g'}$ is its assigned bit or $\gamma$. Conditional on the assignments already made, $D_g=C^2\bigl(\mathbb E[F\mid z_g=1]-\mathbb E[F\mid z_g=0]\bigr)$. Line~\ref{alg:b-bit} therefore takes a branch of no larger conditional expectation, and line~\ref{alg:b-update} is the corresponding update. Symmetry of $\Lambda_2$ gives exactly this slope and update. Initialization and all decisions cost $O(d^2)$ arithmetic operations.

\emph{Exact arithmetic and updates.} For the unscaled objective $F$, put $N\coloneqq32dC^2 2^{2\theta}<128d^5$. The $C^2$ clears the dyadic coefficient denominators, $32d$ clears the penalty denominator, and a quadratic term contains at most two free bits, contributing at most $2^{2\theta}$ to the denominator. Thus $N\mathbb E[F\mid\text{assignments}]$ is integral, and its branch difference is $(N/C^2)D_g$, with the same sign as $D_g$. Coefficients and their sums have magnitude polynomial in $P,d$, so all integer numerators have $O(\log n)$ bits. The same holds for the slope updates. Scales are recomputed each round; denominators are never multiplied across rounds. Finally, lines~\ref{alg:b-apply}--\ref{alg:b-global} apply every selected candidate using Lemma~\ref{lem:rank} and the global update after it. This proves the complete cost. Previously applied candidates $g$ have $I_o(g)=0$ at every output and positive penalty, so they cannot be selected again.
\end{proof}

\subsection{Mean rank and pointwise cap}

The first loop of Algorithm~\ref{alg:select} bounds the total number of improving candidates across the outputs, and the second bounds how many the recovery procedure will have to find at any one output.

\begin{proof}[Proof of Lemma~\ref{lem:selection}]
For $R<4$, the branch of line~\ref{alg:sel-small} compares all candidates directly at every output. The scan uses $O(Pd)$ comparisons and costs $\softO(Pd/R)$ with indexed updates, since $R<4$. Every returned rank is zero. Otherwise $d\ge R\ge4$.

\emph{Reducing the total rank.} Line~\ref{alg:sel-init} starts from one common candidate and its rank table, and the loop of line~\ref{alg:sel-total} applies Lemma~\ref{lem:round} with all outputs while $\Phi>PR$; its hypothesis holds because $\phi=\Phi/P>R\ge4$. Number the $t_*$ rounds by $0\le t<t_*$, with batch $G_t$ and initial mean $\phi_t$. Contraction gives $t_*=O(1+\log(d/R))$ and, backwards from the last active round,
\[
\frac1{\phi_t}<\frac{\rho^{t_*-1-t}}R.
\]
Consequently
\[
\sum_t|G_t|<\frac{128d}{R(1-\rho)}<5267d/R.
\]
Sum the number of active incumbent labels over all rounds to count the label-round occurrences. Counting each new label from its own round gives the upper bound
\[
t_*+\sum_t(t_*-t)|G_t|
<t_*+\frac{128d}{R(1-\rho)^2}<t_*+216700d/R.
\]
Each label-round occurrence pays for its coefficient computations. Each applied candidate pays for one rank table and a dense update. With the exact conditional-expectation work, these costs total $\softO(n^{2+\varepsilon}d/R)$. This loop gives $h_0<1+5267d/R$ and $\Phi\le PR$.

\emph{Capping every rank.} Put $\tau\coloneqq\lambda_{\rm cap} R$. At the start of each round of the loop of line~\ref{alg:sel-cap}, the high set is $\Omega=\{(i,j):r_{ij}>\tau\}$, and the loop stops if it is empty. Otherwise its mean exceeds $\tau>4$, so the same batch round applies to this fixed set and updates all outputs globally. Ranks never increase, so the next high set is a subset of the old one. By Lemma~\ref{lem:round}, the new ranks on the old high set total at most $\rho$ times the old high-set total; since ranks are nonnegative, the same bound holds for the next high set.

Bernoulli's inequality gives $(288/281)^{41}>2$, so $\rho^{82\lceil\log_2n\rceil}P<1$. Starting from high-set total at most $PR$, after at most $82\lceil\log_2n\rceil$ rounds it would be less than $R$, whereas a nonempty high set has total exceeding $\tau$. Hence the loop terminates within that many rounds. Each round selects fewer than $128d/(\lambda_{\rm cap} R)$ candidates, giving fewer than $d/R$ additional applications. Total rank stays at most $PR$ and the final label bound follows.

There are $O(\log n)$ cap rounds, each with at most $h_0+d/R$ labels. Their weighted masks merely acquire the known high-set indicator. All coefficient computations, new rank products, global updates, exact scales, $O(d^2)$ conditional-expectation work and the final stopping test are included. The resulting cost remains $\softO(n^{2+\varepsilon}d/R)$. If every rank is already at most $\tau$, this loop consists only of its first stopping test, whose cost is included.
\end{proof}

%% file: comparison-proofs.tex
\section{Implementing the comparison operations}\label{sec:comparison-proofs}

The quadratic estimator in Section~\ref{sec:selection} uses one comparison with the incumbent for $L_1$, two for $L_2$, and a comparison with the incumbent together with an absolute-rank comparison for $L_3$. Lemma~\ref{lem:weighted} computes these sums with one output-dependent weight matrix for each incumbent label and coefficient family. Lemma~\ref{lem:rank} then supplies the exact ranks needed to apply a chosen candidate globally. We prove Lemmas~\ref{lem:weighted} and~\ref{lem:rank} by separating comparisons into stored prefixes and directly scanned remainders.

Classical algorithms for the dominance product combine ordinary products for separated rank blocks with direct scans for residual comparisons; Vassilevska and Williams describe this scheme and its weighted extension~\cite[Sections 3 and 6]{VW2006}. We adapt this scheme to two simultaneous inequalities and arbitrary weights depending on both output indices. The proof supplies the range decomposition and uses the $n\times n$ by $n\times H$ product of Lemma~\ref{lem:rectangular}, including dense formation and all integer costs.

\begin{proof}[Proof of Lemma~\ref{lem:weighted}]
Set $m_0\coloneqq\lceil n/d^2\rceil$. For each predicate pair, sort the columns by $x_f$ and divide them into $d^2$ consecutive blocks of size at most $m_0$. Build the complete binary tree on the blocks. At each node, sort its columns by decreasing $y_f$ and store indicator columns for prefixes of lengths $m_0,2m_0,\ldots$. At a fixed tree depth the node lists are disjoint and have total length $n$. The number of stored prefixes for one predicate pair is therefore at most
\[
\sum_{\text{depths}}\sum_{u\text{ at that depth}}
\left\lfloor\frac{|S_u|}{m_0}\right\rfloor
\le d^2(1+\log_2(d^2)),
\]
where $S_u$ is the node's column set. Concatenate all predicate pairs' indicator columns into an $n\times H$ Boolean matrix $T$. Then $H\le d^4(1+\log_2(d^2))$.

Compute the full ordinary product $WT$. To answer one query, that is, to compute one sum $Q_{if}$, binary search the first-coordinate list for its qualifying suffix. This suffix is a disjoint union of at most one partial block and $O(\log(d^2))$ whole tree nodes. Inspect the partial block directly. In a whole node, binary search the qualifying second-coordinate prefix, of length $v$. Read the product entry for its first $m_0\lfloor v/m_0\rfloor$ positions, if nonempty, and add the weights of the fewer than $m_0$ remaining positions. These disjoint pieces count each qualifying column once. Binary searches use the coordinate inequalities, including their strictness. Index tie-breaking only orders equal keys inside a list; it does not decide inclusion.

There are at most $nd^2$ queries. Their direct probes and product reads cost $\softO(n^2)$. Sorting costs $\softO(nd^2)$, while dense formation, all product outputs and their addresses cost $\softO(nH+n^2)$. Lemma~\ref{lem:rectangular} bounds the cost of the product with the stated integer weights. The weights, accumulated counts, product entries and addresses use $O(\log n)$ bits. Comparisons of the real key components are unit-cost comparisons of the real RAM. These charges prove the bound. For $d=1$, use the direct scan of its single block.
\end{proof}

For a fixed chosen candidate $q$, the rank table sums the indicators of the candidates better than $q$ at every output. The same scheme now uses the rectangular product in the other orientation.

\begin{proof}[Proof of Lemma~\ref{lem:rank}]
For each candidate $k$, sort the column keys $B_{qj}-B_{kj}$ of~\eqref{eq:comparison} in decreasing order for this fixed $q$. Put $m_1\coloneqq\lceil n/d\rceil$ and store indicator rows for prefixes of lengths $m_1,2m_1,\ldots$. There are at most $d$ such rows per candidate and at most $d^2$ rows in total; let $T_q$ be their matrix. For row $i$ and candidate $k$, binary search the qualifying prefix length $v_{ik}$. An $n\times H$ Boolean matrix $U_q$ selects, in row $i$ and for each candidate $k$, the single stored prefix of $k$ of length $m_1\lfloor v_{ik}/m_1\rfloor$, if nonempty. Different candidates have distinct column addresses.

The full product $U_qT_q$ counts all selected prefixes at every output. For each $(i,k)$, increment the entries belonging to the remaining qualifying positions. The residual and stored prefix are disjoint and exhaustive. There are fewer than $ndm_1\le2n^2$ residual increments. Sorting, binary searches, dense $O(nd^2)$ formation, all $n^2$ outputs and addresses are paid. Lemma~\ref{lem:rectangular}, with $H\le d^2$, proves the bound. Strict searches treat equal keys exactly as~\eqref{eq:comparison} requires, and $k=q$ contributes zero.
\end{proof}

%% file: recovery.tex
\section{Binary measurements and exact support recovery}\label{sec:recovery}

Selection leaves only a bounded number of candidates better than each incumbent. To identify them in the recovery phase (Phase~2 of Algorithm~\ref{alg:minplus}), we will count their intersections with a fixed family of subsets of $[d]$. Write $z\in\{0,1\}^d$ for the indicator of one set of improving candidates. A binary row applied to $z$ counts the improving candidates in that row's subset. The following result supplies one family that works for every output; its construction is independent of the input matrices.

\begin{lemma}[Binary support recovery]\label{lem:recovery}
There is a fixed function $f(d)=(1+\lceil\log_2(2d)\rceil)^{O(1)}$, with a sufficiently large fixed constant factor, for which the following holds. For integers $1\le\kappa\le d$, one can deterministically construct a binary matrix $A_{d,\kappa}$ with $d$ columns, at most $\kappa f(d)$ rows and at most $df(d)$ nonzero entries, together with its row lists and column lists. Construction takes $O(df(d))$ bit operations. From $A_{d,\kappa}z$ one can recover every $z\in\{0,1\}^d$ of support size at most $\kappa$ in $O(\kappa f(d))$ bit operations, including reading all supplied measurements and writing the recovered indices. Lemma~\ref{lem:finish} charges the formation of these measurements in our application. Matrix entries are binary and all measurements are exact integer sums.
\end{lemma}

The decoder repeatedly finds a subset containing exactly one remaining index, determines that index from binary tests, and removes it from all affected counts. It therefore suffices to construct a small family of subsets that always isolates an element of every nonempty support of size at most $\kappa$, with each index belonging to only polylogarithmically many subsets. Appendix~\ref{app:recovery-proof} constructs that family and turns its subsets into the rows of $A_{d,\kappa}$, including the row and column lists and the construction and decoding bit costs.

Indyk~\cite[Theorem 3.1 and Section 3.1]{Indyk2002Selectors} constructs selectors from a disperser and a low-collision hash family, and uses index-bit tests and subtractive recovery. Aumann et al.~\cite[Section 2, Lemmas 1--2 and Theorem~3]{AumannLewensteinLewensteinTsur2011} formulate peeling families with bounded element occurrence and give the corresponding peeling procedure. Their Theorem~3, which they credit to Indyk, gives a deterministic peeling family for all subsets of size at most $k$ of an $m$-element universe, with $O(k\operatorname{polylog}m)$ sets, polylogarithmic element occurrence and $O(m\operatorname{polylog}m)$ construction time. With their Lemma~2 and Indyk's index-bit tests, this gives the statement of Lemma~\ref{lem:recovery}. We rely on the self-contained proof that Appendix~\ref{app:recovery-proof} gives, based on the explicit disperser of Ta-Shma, Umans and Zuckerman~\cite{TaShmaUmansZuckerman2007}, because our application needs a binary measurement matrix with exact construction and decoding bit costs.

Applied to the measurements of the indicator of the improving candidates at each output, the decoder of Lemma~\ref{lem:recovery} returns exactly the candidates whose scores still have to be compared with the incumbent.

%% file: finish.tex
\section{Recovering improving candidates with composite moduli}\label{sec:finish}

We now complete a min-plus block from its selected incumbents, as in Phases~2 and~3 of Algorithm~\ref{alg:minplus}. At each output, the incumbent together with its improving candidates contains the minimum. Lemma~\ref{lem:recovery} recovers those improving candidates once their binary measurements are known. The task here is to compute those measurements and to bound the number of ordinary-product entries that this requires.

A row of the measurement matrix specifies a subset of candidates. Split that subset into disjoint nonempty chunks; summing the counts of improving candidates of its chunks gives the measurement for the row. The length of a chunk bounds the inner dimension of the ordinary products that count it: a chunk of $s_C$ candidates needs at most $s_C^2n^{o(1)}$ inner indices (Section~\ref{sec:composite-primes}), so all chunks fit one common inner dimension (Section~\ref{sec:envelope}). Write $r_A$ for the number of measurement rows and $J$ for the total number of chunks after choosing the maximum chunk size below.

A \emph{job} computes prescribed entries of one ordinary product of bounded-integer matrices. We call the procedure performing such jobs the \emph{entry procedure}; Section~\ref{sec:ordinary} constructs it for all jobs of one sweep (Section~\ref{sec:assembly}) together. The shared computation of Section~\ref{sec:ordinary} relies on at least $n^\eta$ jobs of one label in each sweep; Section~\ref{sec:assembly} shows that the chunks always supply them, and it adds all-zero jobs only as a safeguard. The counts in Lemma~\ref{lem:finish} are taken before this zero-job padding. The following lemma separates the number of jobs and requested entries from the work of decoding their answers.

\begin{lemma}[Recovering the block minima]\label{lem:finish}
Suppose the selection of Lemma~\ref{lem:selection} returns $h$ incumbent labels with total rank $\Phi\le PR$ and every rank at most the pointwise cap $\rho_{\max}\coloneqq\min(d-1,\lambda_{\rm cap} R)$, where $P=n^2$. If $\Phi=0$, the incumbent scores give the answer. Otherwise let $\kappa$ be the smallest power of two at least $\rho_{\max}$. If $\kappa>d/2$, use \emph{direct completion}, which takes the minimum of all $d$ scores at every output. In the recovery branch $\kappa\le d/2$, take $s_C^{\max}\coloneqq\lceil d/\kappa\rceil$. The block minima can be recovered with a common inner dimension, at most $hJ n^{o(1)}$ jobs, and total request volume $O(PJ)n^{o(1)}$ before zero-job padding. Here $J\le2\kappa f(d)$ is the number of nonempty chunks. In addition to the entry procedure and the work of Algorithm~\ref{alg:count} outside it (dense initialization, lists and residues, progress scans and the sieve), whose cost per call is given in Section~\ref{sec:composite-primes} and summed over all calls in Section~\ref{sec:envelope}, the block uses
\[
 \softO\big(n^{2+\varepsilon}d/R+d^2 f(d)+df(d)
             +P(J+r_A)+P\kappa f(d)+P+\Phi\big)
\]
time, where $r_A\le\kappa f(d)$ is the number of measurement rows.
\end{lemma}

These jobs arise from one smaller task: for a fixed incumbent label, count the improving candidates in one chunk at a prescribed set of outputs. As in the working note~\cite{WorkingNote}, we represent a comparison by a match of binary prefixes. Reducing prefixes modulo an integer keeps every true match and may introduce false ones. Computing the two directions gives a test for exactness, because their true counts sum to the known chunk size. We first construct and certify these counts, then assemble the measurements in the proof of Lemma~\ref{lem:finish}.

\subsection{Two directional counts for a chunk}\label{sec:composite-count}

Fix a nonempty chunk $C$ of $s_C$ candidates and a label $p$ to be tested as incumbent. For each candidate $k\in C$ put the row and column keys $A_{ik}-A_{ip}$ and $B_{pj}-B_{kj}$ on the two corresponding lists. By~\eqref{eq:comparison} an improving candidate has row key smaller than its column key, or equal keys and $k<p$. Sort the $2n$ occurrences of each candidate separately and assign them distinct integer ranks from zero. On equal keys put rows before columns when $k<p$, and columns before rows otherwise, breaking remaining ties by index. Thus a row rank below a column rank is exactly the desired comparison. In particular $k=p$ is never an improving candidate. These occurrence ranks encode comparisons within a chunk; the incumbent ranks $r_{ij}$ count improving candidates in the whole block.

Use $\ell_C\coloneqq\lceil\log_2(2ns_C+1)\rceil$ bits for each occurrence rank; every rank is below $2n<2^{\ell_C}$. Write $H_b(z)$ for the prefix preceding bit $b$ of a rank $z$, numbered from the most significant end. Distinct ranks have exactly one position where their prefixes agree and their current bits differ. The ordered bits $0,1$ there give the forward comparison, and $1,0$ the reverse one. Matching exact prefixes would need an inner index of the ordinary product for every occurring prefix, up to $2ns_C\ell_C$ indices, far more than the common inner dimension chosen in Section~\ref{sec:envelope}. For an integer modulus $m'\ge1$, replace equality of the two prefixes by equality modulo $m'$. This leaves $s_C\ell_Cm'$ inner indices but can create false matches. Two jobs of the entry procedure compute the resulting counts $U_{m'}^+$ and $U_{m'}^-$ at an arbitrary requested set $W$. Their inner indices are $(k,b,\sigma)$, with $k\in C$, $1\le b\le\ell_C$ and $0\le\sigma<m'$. On each side enter the number of occurrences with the required current bit and prefix residue $\sigma$. Swap the two required bits for the other direction. The entries are nonnegative bounded integers.

Every actual row and column contains exactly one occurrence of every candidate in the chunk. Thus the true counts sum to $s_C$ at every genuine matrix position $(i,j)\in[n]^2$, whatever its incumbent label. The observable integer
\[
 F_{m'}(w)\coloneqq U_{m'}^+(w)+U_{m'}^-(w)-s_C
\]
counts all false matches, with their multiplicities over candidates and bit positions. Each has a nonzero prefix difference $\delta$ divisible by $m'$ and $|\delta|<2^{\ell_C}$. In particular $0\le F_1(w)\le s_C\ell_C$, and $F_{m'}(w)=0$ proves both directional counts exact, since both are upper bounds.

We use the rank-prefix representation in the working note's lemma \emph{Certified comparison counts}~\cite{WorkingNote}, which also certifies the two directional counts against their exact total, with random primes. Majdoddin's certification with composite moduli~\cite[Section 15.4]{Majdoddin} and its presentation in the working note's \emph{Composite incidence certification} lemma~\cite{WorkingNote} select small prime factors by an aggregate count on active incidences. The working note also purifies integer gap windows by comparing a modular count to a known genuine multiplicity~\cite[lemma \emph{Purifying a window with known genuine multiplicity}]{WorkingNote}. Here the bounded integers are ranks obtained by sorting real comparison keys. The two directional counts have a known sum at each output, which lets us use this prime selection for the recovery measurements of the common incumbents.

\subsection{Choosing prime factors and certifying outputs}\label{sec:composite-primes}

We choose the same next factor for every position still active in a stage of Algorithm~\ref{alg:count}. A bound on the number of prime divisors of a nonzero difference guarantees that one choice decreases the total number of false matches. After a prescribed number of factors, at least half the active positions have no false matches.

Put $J_0\coloneqq\lceil\sqrt{\lceil\log_2(2n)\rceil}\rceil$. Let $P_0$ be the least power of two at least $\max\{4,\ell_C^4\lceil s_C^{1/J_0}\rceil\}$, and let the \emph{pool} consist of the $m_P$ primes in $[P_0,2P_0)$. Let $\lambda_P$ be the largest nonnegative integer with $P_0^{\lambda_P}\le 2^{\ell_C}$. Before using the pool, Algorithm~\ref{alg:count} tests
\begin{equation}\label{eq:prime-guard}
 m_P>0,\qquad 2s_C\ell_C\,\lambda_P^{J_0}\le m_P^{J_0}.
\end{equation}
The test fails only for bounded $n$, uniformly in $1\le s_C\le n$. Indeed the prime number theorem gives $m_P\ge P_0/(4\ln P_0)$ for large $P_0$ and $\lambda_P\le\ell_C\ln2/\ln P_0$. Hence $s_C\ell_C(\lambda_P/m_P)^{J_0}\le\ell_C(4/\ell_C^3)^{J_0}<1/2$ eventually. Integer roots, powers and dyadic ceilings in these tests are computed by exact integer arithmetic and comparison.

Algorithm~\ref{alg:count} uses this pool to certify the two directional counts of one chunk $C$ against a label $p$.

\begin{algorithmbox}{Certified directional counts of one chunk}\label{alg:count}
\ALtext{\kw{Input:} a nonempty chunk $C$ of $s_C$ candidates, a label $p$ to be tested as incumbent, and a set $W$ of requested positions}
\ALtext{\kw{Output:} the exact forward and reverse counts at every $w=(i,j)\in W$; the forward count is the number of $k\in C$ with $s_{ij}(k)<s_{ij}(p)$}
\ALstep{\kw{if} $W=\emptyset$ \kw{then return}}\label{alg:ct-empty}
\ALstep{\kw{for each} $k\in C$ \kw{do}}\label{alg:ct-rank}
\ALstep[1]{sort the row keys $A_{ik}-A_{ip}$ and the column keys $B_{pj}-B_{kj}$ and number these $2n$ occurrences from zero, breaking ties as in Section~\ref{sec:composite-count}}
\ALstep{sieve the pool and test~\eqref{eq:prime-guard}}\label{alg:ct-pool}
\ALstep{\kw{if} the test fails \kw{then}\ALnote{see line~\ref{alg:mp-fail} of Algorithm~\ref{alg:minplus}}}\label{alg:ct-guard}
\ALstep[1]{count directly at the positions of $W$ and \kw{return}}
\ALstep{$S\gets W$}
\ALstep{\kw{while} $S\ne\emptyset$ \kw{do}\ALnote{a \emph{stage}; the active set $S$ stays fixed until line~\ref{alg:ct-certify}}}\label{alg:ct-stage}
\ALstep[1]{$m'\gets1$}\label{alg:ct-reset}
\ALstep[1]{\kw{repeat} $J_0$ times\ALnote{a \emph{prime round}}}\label{alg:ct-round}
\ALstep[2]{\kw{for each} prime $q$ of the pool \kw{do}}
\ALstep[3]{compute $U^+_{m'q}$ and $U^-_{m'q}$ at the full integer modulus $m'q$ and at all positions of $W$, by two jobs of the entry procedure, one per direction}
\ALstep[3]{keep the two count vectors of the prime minimizing $\sum_{w\in S}F_{m'q}(w)$ among the primes tested so far, breaking ties by numerical order}
\ALstep[2]{$q\gets$ the prime whose two count vectors are kept;\enspace $m'\gets m'q$}
\ALstep[1]{\kw{for each} $w\in S$ with $F_{m'}(w)=0$ \kw{do}}\label{alg:ct-certify}
\ALstep[2]{output the two counts of $w$ at modulus $m'$ and remove $w$ from $S$}
\ALstep{\kw{return} the output counts}
\end{algorithmbox}

For any false match still present at modulus $m'$ in a prime round of Algorithm~\ref{alg:count}, extension by $q$ retains it precisely when $q\mid\delta/m'$. At most $\lambda_P$ distinct pool primes divide this nonzero integer. This remains valid for $q$ already used: testing modulo $m'q$ tests one further power of that prime. A match excluded at $m'$ cannot reappear. Averaging the integral multiplicities gives
\[
 \min_q\sum_{w\in S} F_{m'q}(w)
 \le \frac{\lambda_P}{m_P}\sum_{w\in S}F_{m'}(w).
\]
By~\eqref{eq:prime-guard}, after $J_0$ rounds the total is at most $|S|/2$. Each uncertified position contributes at least one. After at most $\lceil\log_2(|W|+1)\rceil$ stages every position has its exact count. Instead of keeping the best prime's two count vectors during the scan, a stage may recompute the two count vectors at its final modulus $m'$ once at its end.

Every trial modulus $m'q$, including one whose prime $q$ is not kept, is at most $(2P_0)^{J_0}$. Uniformly $P_0\le4\ell_C^4 s_C^{1/J_0}=n^{o(1)}$ and $(2P_0)^{J_0}\le s_C(8\ell_C^4)^{J_0}=s_C n^{o(1)}$. Indeed $J_0\log\ell_C=o(\log n)$. The number of inner indices is at most $s_C\ell_C(2P_0)^{J_0}\le s_C^2 n^{o(1)}$. The number of directional products is at most $2(J_0m_P+1)\lceil\log_2(|W|+1)\rceil=n^{o(1)}$. We build new arrays for each trial prime, in each direction, at a common padded inner dimension $D$ fixed in Section~\ref{sec:envelope}. Beyond the work of the entry procedure, dense initialization costs $O(nD)$ writes, constructing lists and residues costs $O(ns_C\ell_C)$ writes and the progress scan costs $O(|S|)$ operations, each with its logarithmic bit cost. The sieve costs $P_0\operatorname{polylog} P_0$. Moduli, counts, progress sums, guard powers and addresses have $O(\log(2n))$ bits. The two best-so-far vectors need $O(|W|\log(2n))$ bits besides the working space of the entry procedure on one sweep. Every step is deterministic.

\subsection{Assembling the recovery measurements}\label{sec:assembly}

\paragraph{Synchronized sweeps.}
For each incumbent label $p$, every chunk requests all of $W_p\coloneqq\{(i,j):\pi_{ij}=p\}$, including the positions of rank zero. The calls of Algorithm~\ref{alg:count} for the chunks of label $p$ follow one common schedule indexed by the loops of Algorithm~\ref{alg:count}: the stage, the prime round, the index of the trial prime in the chunk's own sorted pool, and the direction. Each slot of this schedule, a \emph{sweep}, holds one ordinary-product job per chunk. A chunk is \emph{finished} once its active set is empty, and its trial prime is \emph{missing} in a slot whose index exceeds the number of primes in its own pool; a finished chunk or a missing trial prime supplies zero arrays in its slot. All-zero ordinary-product jobs extend every sweep to $J'\coloneqq\max(J,\lceil n^\eta\rceil)$ jobs on the positions $W_p$. Here $\eta>0$ is the exponent of the rank scale $R=\min(d,\lceil n^\eta\rceil)$ fixed in Section~\ref{sec:envelope}. In the recovery branch no all-zero job is added: every candidate lies in a row of $A_{d,\kappa}$, whose decoder recovers each unit vector, and a chunk has at most $d/\kappa$ candidates; moreover $\kappa\le d/2$ forces $\kappa\ge\lambda_{\rm cap}R$, since line~\ref{alg:mp-guard} of Algorithm~\ref{alg:minplus} excludes $d=1$ directly and $d=2$ through its test $R=d$; hence $J\ge\kappa>\lceil n^\eta\rceil$ and $J'=J$.

\begin{proof}[Proof of Lemma~\ref{lem:finish}]
The term $n^{2+\varepsilon}d/R$ in the lemma is the selection cost of Lemma~\ref{lem:selection}. If $\Phi=0$, evaluating the $P$ incumbent scores costs $\softO(P)$. Direct completion evaluates and compares all $Pd$ scores, costing $\softO(Pd)$, and uses no chunk, measurement row or job; since $d<2\kappa$ in its branch, this lies within the term $P\kappa f(d)$. In the recovery branch, $\kappa\le2\lambda_{\rm cap} R$, since $\rho_{\max}\le\lambda_{\rm cap}R$; moreover $\kappa\ge R$ whenever $\lambda_{\rm cap}R\le d-1$, a bound used in Section~\ref{sec:envelope}. Apply Lemma~\ref{lem:recovery} once and retain its row lists. Their total length is $N_A\le df(d)$. Splitting each nonempty row into as few disjoint chunks of size at most $s_C^{\max}$ as possible gives $J\le r_A+N_A/s_C^{\max}\le2\kappa f(d)$. The constructor costs $O(df(d))$ by Lemma~\ref{lem:recovery}, which the conservative charge $d^2f(d)$ covers, and traversing the lists costs $df(d)$.

\emph{Counting on each incumbent class.} The sets $W_p$ of the synchronized sweeps partition all $P$ positions. Each chunk maintains its own active subset of $W_p$ and minimizes its number of false matches there. A value recomputed at a previously certified position does not enter that objective. Thus all $J$ chunks of label $p$ may request $W_p$ throughout the prime search, giving $PJ$ original requests in total.

\emph{Synchronizing the product jobs.} The bound $J'\ge n^\eta$ gives the group products of Section~\ref{sec:walsh-sharing} enough data columns, as checked in Section~\ref{sec:point}. An all-zero job has no recovered score or measurement. Once both directions of every trial in a round have been processed, each genuine chunk picks its own best prime; its full count vectors and its active set belong to that chunk alone. The uniform bounds above permit $n^{o(1)}$ synchronized sweeps for a label, including any final recomputation.

\emph{Returning the block minima.} Initialize the $P r_A$ integer counters, route the $PJ$ genuine chunk answers to their parent rows and sum. These are the exact binary measurements of the indicator vector $z^{ij}$ of the improving candidates (Section~\ref{sec:phases}) for the frozen incumbent. Its support size is $r_{ij}\le\kappa$. Decode each vector in $O(\kappa f(d))$ time, evaluate each recovered score and retain the minimum of these scores and the incumbent score. Exactly $\Phi$ scores of improving candidates are evaluated. Store improvements separately. Initializing the $P$ best scores, merging min-plus blocks, and returning exact answers are all paid. A short, nonempty last min-plus block is padded as in Section~\ref{sec:blocks}. Zero entries pad only ordinary products and all-zero jobs.
\end{proof}

%% file: envelope.tex
\section{The cost of deterministic comparison counts}\label{sec:envelope}

Lemma~\ref{lem:finish} reduces each block minimum to ordinary-product jobs and bounds their requested entries. We now derive a cost bound for those jobs that suffices to make the complete min-plus product faster than cubic. This separates the time analysis from the construction of the entry procedure.

We write the cost for $w$ prescribed entries of one ordinary-product job as follows. For the entry procedure it is an amortized bound: the total for a sweep of $J^\prime$ jobs on the same $w$ positions, divided by $J^\prime$. For $n\times D$ and $D\times n$ bounded-integer inputs, it is
\begin{equation}\label{eq:abstract-source}
 nD^{E+o(1)}+n^2D^{-V+o(1)}
       +w(D^{q_1+o(1)}+D^{q_2+o(1)})+
       O(nD\operatorname{polylog}(2n)).
\end{equation}
Here $E$ measures operand preparation, $V$ the saving in the work that the requested positions share, and $q_1,q_2$ the exponents of the two costs per requested entry. The final term is the cost of forming the dense inputs. The construction in Section~\ref{sec:ordinary} has three preparation costs, for its root, residual and high encoders, with exponents $A_i$, and $E=\max_i A_i$; Section~\ref{sec:ordinary-guarantee} describes these encoders, the joint tables and the two reads. Each preparation uses only one operand; the cost table below charges all three on each side. The construction also requires an outer dimension of at least $D^{T+o(1)}$ to fit its arrays. Lemma~\ref{lem:walsh-finite} supplies these exponents, and Section~\ref{sec:point} checks the finite dimensions before allocation.

\subsection{From product costs to the total running time}

Fix rational exponents $e>0$, $x\coloneqq e/2$, $\eta>0$, $b\coloneqq x+\eta$ and a selection loss $\varepsilon>0$. Choose the largest power of two $d\le n^b$ and $R\coloneqq\min(d,\lceil n^\eta\rceil)$. Test the actual selection conditions $d^2\le n$ and $d^4(1+\log_2(d^2))\le n^\alpha$; failure of a test, $d=1$ or $R=d$ permits direct multiplication. If $4b<\alpha\le1$, these exceptions have bounded $n$. In each nontrivial block select and cap the incumbents, checking the direct-completion condition $\kappa>d/2$ of Lemma~\ref{lem:finish}. There $\kappa=O(R\log(2n))$ and $d/R=n^{x+o(1)}$, so $\kappa\le d/2$ eventually. The total number of chunks per block is $J\le n^{\eta+o(1)}$, with at most $h=O(d/R)\le n^{x+o(1)}$ labels. Hence $n^\eta\le J^\prime=\max(J,\lceil n^\eta\rceil)\le n^{\eta+o(1)}$. The individual lengths of genuine chunks are at most $\lceil d/\kappa\rceil\le\lceil n^x\rceil$, since eventually $\kappa\ge R$ by the proof of Lemma~\ref{lem:finish}. Since a chunk of length $s_C$ needs at most $s_C^2n^{o(1)}$ inner indices (Section~\ref{sec:composite-primes}), the choice $x=e/2$ lets one inner dimension $D=n^{e+o(1)}$ hold the products of every chunk.

We choose a common padded inner dimension satisfying $n^e\le D\le n^{e+o(1)}$. The explicit choice below accommodates every trial modulus, including those of short chunks.

In the table, $P=n^2$ is the number of outputs, $\Phi=\sum_{i,j}r_{ij}$ is the total incumbent rank, $r_A$ is the number of measurement rows, and $f(d)$ is the polylogarithmic recovery factor of Lemma~\ref{lem:recovery}. For each kind of work, the table gives its cost per job or per block and its total saving~$S'$, which stands for a total cost of $n^{3-S'+o(1)}$ over the whole computation.
\begin{center}\begin{tabular}{lcc}
\toprule Work & Per job or per block & Total saving\\
\midrule
Root, residual and high encoders & $nD^{A_i}$ per job and side & $1-eA_i$ each\\
Joint tables and ordinary products & $n^2D^{-V}$ per job & $eV$\\
Low and high reads & $D^{q_i}$ per request & $x-eq_i$, $i=1,2$\\
Dense inputs and rank-list construction & $O(nD)$ per job, up to logs & $1-e$\\
Selection and rank products & $n^{2+\varepsilon}d/R$ per block & $\eta-\varepsilon$\\
Recovery construction and row lists & $d^2f(d),\,df(d)$ per block & $2-b,\ 2$\\
Measurement assembly and decoding & $P(J+r_A+\kappa f(d))$ per block & $x$\\
Recovered scores and label updates & $\Phi,\,O(Pd/R)$ per block & $x,\ \eta$\\
Block initialization and merging & $P$ per block & $b$\\
Cubic and quadratic selection work & $d^3,\,d^2$ per block & $2-2b,\ 2-b$\\
\bottomrule\end{tabular}\end{center}

\begin{lemma}[Sufficient cost criterion]\label{lem:criterion}
Suppose that the cost for $w$ prescribed entries of one ordinary-product job has the amortized form~\eqref{eq:abstract-source}. If every saving displayed in the table above exceeds $S>0$, $eT<1$, and the dimension tests of the entry procedure and the selection tests eventually succeed, Algorithm~\ref{alg:minplus} with these exponents computes the exact min-plus product in $O(n^{3-S})$ time.
\end{lemma}

\begin{proof}
There are $n^{1-b+o(1)}$ min-plus blocks. Counting both directions and all trial slots gives at most $n^{1-b+o(1)}hJ^\prime n^{o(1)}\le n^{1+o(1)}$ jobs. Because $\sum_p|W_p|=P$, their request volume is at most $n^{1-b+o(1)}P J^\prime n^{o(1)}\le n^{3-x+o(1)}$. Thus the savings $S'$ of the table give total costs $n^{3-S'+o(1)}$.
The per-job and per-block costs in the table suppress uniform subpolynomial overhead. The conditional-expectation decisions use $O(d^2)$ arithmetic operations in each of $O(\log n)$ rounds. The cubic row adds a conservative $d^3$ charge to this quadratic cost. Measurement assembly uses only the answers of genuine chunks, and recovered scores cost $\Phi\le PR$ per block while label updates cost $O(Pd/R)$.
The finitely many savings of the table exceed $S$ by a positive minimum slack. The rounding of $D$ to its grid, the prime-trial overhead, the uniform recovery factor and indexed bit costs together use $n^{o(1)}$, which eventually fits within any fixed smaller slack.
\end{proof}

\subsection{A common inner dimension for all trial moduli}

The cost calculation used the same $D$ for every job. Its upper bound limits the cost of operand formation and reads, while its lower bound retains the saving in $D^{-V}$. We now choose this inner dimension so that even the largest trial modulus fits.

For the entire computation set
\[
 s_0\coloneqq\max(1,\lceil2n^x\rceil),\quad
 L_0\coloneqq\lceil\log_2(2ns_0+1)\rceil,\quad
 J_0\coloneqq\lceil\sqrt{\lceil\log_2(2n)\rceil}\rceil .
\]
Take the least power of two $P_*$ at least $\max\{4,L_0^4\lceil s_0^{1/J_0}\rceil\}$, and take the least inner dimension $D\coloneqq4^m$ on the grid satisfying the following bound. The grid consists of the values $4^m$ with $m$ on the integral grid of Section~\ref{sec:ordinary-guarantee}, that is, a positive multiple of the fixed denominator in Section~\ref{sec:point}, so successive values of $D$ have a fixed ratio. We require
\begin{equation}\label{eq:effective-grid}
 D\ \ge\ \max\{\lceil n^e\rceil,\ s_0L_0(2P_*)^{J_0}\}.
\end{equation}
This choice of $D$ bounds all $s_C\ell_C m'$ inner indices of every trial, including the first, in both directions and in all padded sweeps. Since $J_0\log L_0=o(\log n)$, $(2P_*)^{J_0}\le s_0(8L_0^4)^{J_0}=s_0 n^{o(1)}$. Thus the right-hand side is at most $n^{2x+o(1)}$. A fixed grid ratio then gives $n^e\le D\le n^{e+o(1)}$ uniformly. The lower bound retains the saving in the joint term $D^{-V}$ even for a short chunk. Dyadic roots, the ceiling of each rational power and the grid index use integer arithmetic and comparisons before any large array is allocated.

The prime sieve, progress scans and the formation of the two directional products are covered by the job, input and request charges. In particular $ns_0\le O(nD)$ eventually. The common inner dimension therefore satisfies the assumptions used in the cost table.

%% file: source.tex
\section{Computing prescribed entries of ordinary products}\label{sec:ordinary}

The recovery algorithm of Section~\ref{sec:finish} needs prescribed entries of many ordinary integer matrix products. In each sweep the jobs request the same positions. We use these common positions to compute the entries jointly within the cost bound~\eqref{eq:abstract-source}.

\subsection{The jobs of one sweep and what they share}\label{sec:ordinary-plan}

\paragraph{The counts that remain.}
Fix an incumbent label $p$ and one sweep, that is, one slot of the common schedule of Section~\ref{sec:assembly}: a stage, a prime round, the index of a trial prime and a direction. In this sweep every genuine chunk of label $p$ supplies one job, the ordinary product that gives its directional count at its current trial modulus (Section~\ref{sec:composite-count}); a finished chunk or a missing trial prime supplies zero arrays in this slot. All-zero jobs extend the list to $J'$ jobs. A job thus consists of an $N\times D$ integer matrix and a $D\times N$ integer matrix, where $N=n$ and $D$ is the common padded inner dimension of Section~\ref{sec:envelope}. Their entries count occurrences with a prescribed current bit and prefix residue, and the entries of the product at the requested positions are the directional counts of the chunk. All $J'$ jobs of the sweep request the same set $W_p$ of $w\coloneqq|W_p|$ positions, but their matrices differ: they belong to different chunks, and each chunk chooses its own trial moduli.

\paragraph{The cost target and two straightforward methods.}
Section~\ref{sec:envelope} needs these entries at the cost~\eqref{eq:abstract-source} per job (Lemma~\ref{lem:criterion}), averaged over the $J'$ jobs of the sweep, and its accounting rules out the two straightforward methods. In the table there, a cost of $D^{q}$ per requested entry saves $x-eq$; since $x=e/2$, a request may cost only $D^{q}$ with $q<1/2$. An inner product of length $D$ for each request has $q=1$, so its saving $x-e=-x$ is negative. The work that a job shares among its requests must also stay below $n^2$ by a factor $D^{V}$ with $V>0$. The complete product of a job has $n^2$ entries, so computing it for every job rules out $V>0$.

\paragraph{Low and high parts.}
For a single product, Alman and Vassilevska Williams~\cite{AlmanVassilevskaWilliams2026} and Majdoddin~\cite{Majdoddin} avoid both costs. After a re-indexing of the inner dimension, they write a requested entry as a sum of products indexed by the leaves of a recursive bilinear identity. Leaves of high order are grouped into boxes shared by many requested positions, and the values of the boxes are tabulated once; the other leaves are evaluated at each request. We use the same split. The high part of a requested entry is read from tables at a cost $D^{q_2}$ per request, and the products that build the tables cost the joint term $n^2D^{-V}$; the low part costs $D^{q_1}$ per request. The complete sum of the two parts reconstructs the entry (Lemma~\ref{lem:low-high}). The restrictions of the operands in Section~\ref{sec:walsh-cover} preserve this complete sum (Lemma~\ref{lem:restricted-sum}), but they may change each part separately.

\paragraph{Data shared by the jobs of a sweep.}
In our construction the inner dimension of every job is split into $K$ consecutive middle components. The low part of a request is an integer combination of products of a left and a right operand value, and each operand value is a linear combination of stored operand values whose coefficients depend only on the requested position and on fixed parameters. These coefficients form the selector rows of Section~\ref{sec:walsh-sharing}. The operand values for many requests are therefore computed by rectangular matrix products: a matrix of selector rows times a matrix of data columns, one column for each job and middle component. The rectangular algorithm used for these products has a fixed shape. It distributes its cost over its columns only when there are enough of them, as the third condition of Lemma~\ref{lem:walsh-finite} requires; with fewer columns its cost per column grows by the missing factor. One job supplies only one column for each middle component. At both parameter choices of Section~\ref{sec:point} these columns fall short by a factor of about $D^{0.19}$ and $D^{0.33}$, respectively, and padding the products of a single job would raise the cost of its low part far beyond the slack certified in Appendix~\ref{app:numerical}. The $J'$ jobs of a sweep, however, request the same positions and therefore use the same selector rows. Together they supply $J'$ times as many columns, and $J'\ge n^\eta$ makes this number sufficient (Section~\ref{sec:point}).

Each job supplies its own data columns, determined by its chunk and trial modulus. The jobs share the re-indexing maps, the position encodings, the translations and masks, and the selector rows. A group product applies those rows to all jobs' columns over~$\Z$ and returns separate values for each job. The high tables are computed separately for each job.

We use the five-term identity and squared construction of Majdoddin~\cite[Lemmas 3.1 and 4.1, Theorem 5.1]{Majdoddin}, following the construction of Alman and Vassilevska Williams~\cite{AlmanVassilevskaWilliams2026} for wanted entries of thin matrix products, and the working note's prefix query identities~\cite[lemmas \emph{Prefix contraction} and \emph{Fixed-profile inner-basis prefix queries}]{WorkingNote}. The working note's grouped completion shares the selector rows across middle components~\cite[lemma \emph{Grouped completion with asymmetric row chunks}]{WorkingNote}; here the requested positions also determine the same selector rows for all comparison jobs, even when their numerical inputs use different moduli.

\subsection{The grouped entry procedure}\label{sec:ordinary-guarantee}

\paragraph{Parameters.}
The procedure depends on fixed rationals $r,c,v,u,\zeta$ and on the exponent $m$ of the inner dimension $D=4^m$. Put
\[
 s\coloneqq1-r,\qquad M\coloneqq sm,\qquad \ell\coloneqq rm,\qquad L\coloneqq cM,\qquad k\coloneqq L-M,\qquad t\coloneqq2vM,
\]
and $D_s\coloneqq4^M$, $K\coloneqq4^\ell$ and $B\coloneqq4^{um}$, so that $D=KD_s$. We say that $m$ lies on the \emph{integral grid} if $\ell$, $M$, $L$, $t$ and $um$ are integers; these $m$ are the positive multiples of a fixed integer, given in Section~\ref{sec:point}. The parameters have the following roles. The inner dimension splits into $K$ consecutive \emph{middle components} of inner dimension $D_s$. A component entry is encoded by words with two halves of $L$ levels, of which $M$ in each half are inner, and $t$ separates the low part of a requested entry from its high part. An outer block of rows or columns consists of $B$ component bands (Section~\ref{sec:blocked-source}), and the high tables multiply $B\times K$ by $K\times B$ matrices. Finally, $\zeta$ fixes the common length $C_*=\lceil D_s^{\zeta}\rceil$ of the stored vectors that represent the low part (Section~\ref{sec:walsh-low}).

\paragraph{Two rectangular algorithms.}
The high tables use a fixed exact rectangular multiplication algorithm of shape parameter $\lambda_h$ with positive loss $\xi_h$: if $K\le B^{\lambda_h}$, it multiplies a $B\times K$ matrix by a $K\times B$ matrix in $O(B^{2+\xi_h})$ operations. The low part uses a fixed exact algorithm that multiplies a $C_*^{a_*}\times C_*$ matrix by a $C_*\times C_*^{\kappa_*}$ matrix in $O(C_*^{w_*})$ operations, with a fixed positive loss included in $w_*$. Write
\[
 \alpha_*\coloneqq w_*-\kappa_*-1,\qquad
 \beta_*\coloneqq w_*-\kappa_*-a_* .
\]
The two choices made in Section~\ref{sec:point} have $a_*=\kappa_*\ge2$, $\alpha_*\ge1$, $\beta_*>0$.

\begin{lemma}[Grouped entry procedure for comparison jobs]\label{lem:walsh-finite}
Let $c,v,\zeta,r,u$ be fixed rationals with $c>5$, $0<2v<1/2$, $\zeta>0$ and $0<r<1$ that satisfy the guards~\eqref{eq:group-guards}, and let $m$ lie on the integral grid. Suppose
\[
 N\ge B\binom LM4^{L-M},\qquad
 K\le B^{\lambda_h},\qquad J'K\ge\lceil C_*^{\kappa_*}\rceil .
\]
Assume for the rate bound that the bit length $b_{\rm in}$ of the input entries satisfies $b_{\rm in}+\log(2N)+\log(2J')=O(m)$, with a fixed implicit constant. For $J'$ bounded-integer jobs on the same $w$ prescribed positions, there is an exact deterministic procedure whose total cost, divided by $J'$, has the form~\eqref{eq:abstract-source}, with $E=\max(A_{\rm root},A_{\rm res},A_{\rm box})$, where the three operand exponents $A_{\rm root},A_{\rm res},A_{\rm box}$ and the remaining rates $T,V,q_1,q_2$ are given in~\eqref{eq:walsh-rates}. All vanishing errors are uniform over the requested positions and the supplied inputs.
\end{lemma}

The exponents in~\eqref{eq:walsh-rates} are explicit functions of the parameters, and Section~\ref{sec:walsh-rates} derives each of them from a count of the objects of the construction. In~\eqref{eq:abstract-source}, the term $nD^{E}$ bounds the cost of three encoders of each operand: its \emph{root encoder} (the contracted arrays of Section~\ref{sec:walsh-sharing}), its \emph{residual encoder} (the stored vectors of the low part, Section~\ref{sec:walsh-sharing}) and its \emph{high encoder} (the dictionary of suffix bases of Section~\ref{sec:walsh-high}), with exponents $A_{\rm root}$, $A_{\rm res}$ and $A_{\rm box}$. The joint term $n^2D^{-V}$ bounds the cost of the joint tables, that is, the high tables and their ordinary products, which all requests of a pair of outer blocks share. The terms $D^{q_1}$ and $D^{q_2}$ bound the costs of the low and the high read of one request.

The three conditions of the lemma are the outer-block fit (outer blocks have width $D^{T+o(1)}$), the inner-dimension condition for the high products, and the number of data columns needed for the group products (Section~\ref{sec:ordinary-plan}). The guards~\eqref{eq:group-guards} restrict the window length of Section~\ref{sec:walsh-low} and are used to count the groups in Section~\ref{sec:walsh-rates}.

\paragraph{The procedure.}
The procedure writes each requested entry of a job as $4^{-L}$ times the sum of a low and a high part, taken over the middle components, the runs and the translation pairs. A run $\sigma\in\{-M,\ldots,M\}$ pairs two arrays $\mathcal A_\sigma,\mathcal B_\sigma$, and the complete contractions summed over the $2M+1$ runs give $4^L$ times a component entry (Section~\ref{sec:walsh}). Each summand of a component entry belongs to exactly one translation pair; the translations and cropping of Section~\ref{sec:walsh-cover} make the shared evaluation of the low part cheaper and preserve the complete sum (Lemma~\ref{lem:restricted-sum}). A group has one stored vector for each job and middle component, and its group product multiplies these vectors by the selector rows of the requests assigned to it (Section~\ref{sec:walsh-sharing}).

In the low part, a channel set is a subset of the $2M$ inner levels of a request. It selects the levels that use the second term in the inner-level expansion of Section~\ref{sec:suffix-queries}. The two channel counts are the numbers of selected levels in the two halves. Each channel set indexes truncated operand polynomials whose products give the low numerator~\eqref{eq:cover-low}. A formal coefficient is one coefficient of such a polynomial. The window length $h_w$ is chosen from the two channel counts by the rule of Section~\ref{sec:walsh-low}. It controls the length of the stored vectors and the number of groups.

The window components, one for each cyclic window of that length, partition the operand arrays. The forward map puts the opposite forms into the pattern of the normal ones (Section~\ref{sec:walsh-sharing}). In the high part, the boxes group the high terms of a request, and the last-star recurrence~\eqref{eq:last-star} evaluates them from the high-table products on the suffix bases, which all requests on a pair of outer blocks share (Sections~\ref{sec:suffix-queries} and~\ref{sec:walsh-high}). Algorithm~\ref{alg:entry} lists the steps of the procedure as an overview to return to: its terms are glossed above and defined in Sections~\ref{sec:blocked-source}--\ref{sec:walsh-high}, and the text after it says where each block of lines is justified.

\begin{algorithmhere}{Grouped entry procedure for one sweep}\label{alg:entry}
\ALtext{\kw{Input:} the $J'$ jobs of one sweep, each an $N\times D$ and a $D\times N$ integer matrix, and their common set $W_p$ of $w$ requested positions; the parameters and the two rectangular algorithms above, satisfying the conditions of Lemma~\ref{lem:walsh-finite}}
\ALtext{\kw{Output:} the entries at $W_p$ of all $J'$ products, or a reported failure (line~\ref{alg:en-wtest})}
\ALtext{\ALmark\textbf{Shared data:} maps, translations and window lengths; the position encoding and the selector rows of every requested position}
\ALstep{construct the re-indexing maps $\Gamma,\Xi$ with their list of good inner sets (Section~\ref{sec:restricted-supports}) and the translations (Section~\ref{sec:walsh-cover})}\label{alg:en-maps}
\ALstep{\kw{for each} pair of channel counts in the two halves with sum less than $t$ \kw{do}\ALnote{Section~\ref{sec:walsh-sharing}}}\label{alg:en-window}
\ALstep[1]{choose the window length $h_w$;\enspace \kw{if} $h_w\ge k$ \kw{then} report a failure and \kw{return}\ALnote{Section~\ref{sec:walsh-low}}}\label{alg:en-wtest}
\ALstep{\kw{for each} requested position \kw{do}}\label{alg:en-select}
\ALstep[1]{compute its position encoding, the suffix locator of Section~\ref{sec:walsh}}
\ALstep[1]{\kw{for each} operand, translation pair, run, formal coefficient, channel set of fewer than $t$ inner levels and window component \kw{do}}
\ALstep[2]{determine the group of the request and its selector row\ALnote{Section~\ref{sec:walsh-sharing}}}\label{alg:en-selrow}
\ALbreak
\ALtext{\ALmark\textbf{Encoders, separately for each job:} the stored vectors and the dictionaries on suffix bases}
\ALstep{\kw{for each} job \kw{do}}\label{alg:en-encode}
\ALstep[1]{pad both outer axes of its inputs to outer blocks of $B$ component bands and split its inner dimension into the $K$ middle components~\eqref{eq:blocked-dot}\ALnote{Section~\ref{sec:blocked-source}}}
\ALstep[1]{\kw{for each} component band, middle component and translation pair \kw{do}}
\ALstep[2]{mask the component inputs, apply the Walsh transform, and form and crop the arrays $\mathcal A_\sigma,\mathcal B_\sigma$ of every run $\sigma$\ALnote{Sections~\ref{sec:walsh} and~\ref{sec:walsh-cover}}}\label{alg:en-crop}
\ALstep[2]{\kw{for each} run \kw{do}}
\ALstep[3]{contract the inner levels, apply the forward map, and fill the stored vector of every group, for each formal coefficient\ALnote{root and residual encoders, Section~\ref{sec:walsh-sharing}}}\label{alg:en-roots}
\ALstep[3]{build the dictionary on the suffix bases\ALnote{high encoder, Section~\ref{sec:walsh-high}}}\label{alg:en-dict}
\ALtext{\ALmark\textbf{Low parts:} the group products, shared by the jobs, and the low numerators of each job}
\ALstep{\kw{for each} group \kw{do}}\label{alg:en-group}
\ALstep[1]{multiply the matrix of its requested selector rows by the matrix of its $J'K$ stored vectors, one for each job and middle component\ALnote{group products~\eqref{eq:group-products}, Section~\ref{sec:walsh-sharing}}}\label{alg:en-gprod}
\ALstep{\kw{for each} job, middle component, requested position, translation pair and run \kw{do}}\label{alg:en-low}
\ALstep[1]{sum the readings of line~\ref{alg:en-gprod} over the window components, multiply the matched operand polynomials, and form the low numerator~\eqref{eq:cover-low}\ALnote{Section~\ref{sec:suffix-queries}}}\label{alg:en-lownum}
\ALtext{\ALmark\textbf{High parts, separately for each job:} the box values}
\ALstep{\kw{for each} job, suffix base, pair of outer blocks, translation pair and run \kw{do}}\label{alg:en-high}
\ALstep[1]{multiply its $B\times K$ and $K\times B$ matrices of constant coefficients with the algorithm of shape parameter $\lambda_h$\ALnote{high tables; this sums the $K$ middle components}}\label{alg:en-hprod}
\ALstep{evaluate the boxes by the last-star recurrence~\eqref{eq:last-star}, and sum the $4^t\binom{2M}t$ box values of each request into its high part in each run and translation pair\ALnote{Section~\ref{sec:suffix-queries}}}\label{alg:en-box}
\ALtext{\ALmark\textbf{Assembly:} the requested entries}
\ALstep{\kw{return}, for every job and requested position, the sum, over the translation pairs and runs, of the low parts of all middle components plus the high part, divided by $4^L$~\eqref{eq:final-assembly}\ALnote{Section~\ref{sec:walsh-high}}}\label{alg:en-return}
\end{algorithmhere}

Lines~\ref{alg:en-maps}--\ref{alg:en-selrow} of Algorithm~\ref{alg:entry} use only the parameters and the requested positions, so all $J'$ jobs and all $K$ middle components share the groups and selector rows (Lemma~\ref{lem:shared-low}), and line~\ref{alg:en-gprod} multiplies the selector rows of a group once by the stored vectors of all jobs.

The rest of this section proves Lemma~\ref{lem:walsh-finite}. A requested entry is the sum of $K$ component entries with common suffix addresses~\eqref{eq:blocked-dot} (Section~\ref{sec:blocked-source}). With the maps of Lemma~\ref{lem:profile-embedding}, the aligned Walsh embedding writes $4^L$ times a component entry as the complete contraction of the arrays $\mathcal A_\sigma,\mathcal B_\sigma$ at the suffix locator, summed over the $2M+1$ runs (Lemma~\ref{lem:aligned-embedding}). This contraction is the sum of a low and a high part (Lemma~\ref{lem:low-high}), and the translations and cropping of line~\ref{alg:en-crop} preserve the complete sum, although each part may change (Lemma~\ref{lem:restricted-sum}). Lemma~\ref{lem:shared-low} shows that lines~\ref{alg:en-group}--\ref{alg:en-lownum} compute the low parts, Section~\ref{sec:walsh-high} that lines~\ref{alg:en-high}--\ref{alg:en-box} compute the high parts, and~\eqref{eq:final-assembly} that line~\ref{alg:en-return} returns every requested entry. Section~\ref{sec:walsh-rates} counts inner sets, channel sets, groups and suffix bases, which gives the exponents~\eqref{eq:walsh-rates} and completes the proof.

\input{blocked-source}

\subsection{Restricted inner sets and the exact band width}\label{sec:restricted-supports}

The component products in~\eqref{eq:blocked-dot} use one common system of row and column addresses. In the aligned embedding of Section~\ref{sec:walsh}, the restriction of one linear map to an inner set enumerates the inner indices of a component, and the restriction of a second map to the complementary levels is used to invert the Walsh transform on the outer indices (proof of Lemma~\ref{lem:aligned-embedding}); both restrictions must be invertible. This subsection constructs the two maps and a large list of inner sets on which both restrictions are invertible. The length of that list determines how many component blocks fit in a band, independently of the bilinear forms used to evaluate them.

Lemma~\ref{lem:profile-embedding} below is stated for a general number of levels per half, size of an inner set and list of inner sets, because Appendix~\ref{app:original-embedding} uses it in this generality; until the end of its proof, $L$, $M$ and $k$ have the following general meanings. For the common re-indexing maps let $L$ be the number of levels in each half of a word, let $M$ be the size of an inner set and put $k\coloneqq L-M\ge M\ge1$. Take a nonempty list $\calF$ of distinct $M$-subsets of $[L]$, called \emph{inner sets}, put $K'\coloneqq|\calF|$, and write $D_s\coloneqq4^M$, $N_0\coloneqq4^k$. In the procedure, $L$, $M$ and $k$ are the parameters of Section~\ref{sec:ordinary-guarantee}, and $\calF$ consists of all $M$-subsets of $[L]$. In Majdoddin's terminology~\cite{Majdoddin}, our words are strings. The working note's \emph{Fixed-profile inner-basis prefix queries} lemma~\cite{WorkingNote} restricts the conditional expectations in the proof of Majdoddin's Lemma~4.1~\cite{Majdoddin} to a list with prescribed regional occupancies; the proof of Lemma~\ref{lem:profile-embedding} restricts them in the same way to an arbitrary list $\calF$.

\begin{lemma}[Restricted re-indexing maps; after Majdoddin~{\cite[Lemma~4.1]{Majdoddin}}]\label{lem:profile-embedding}
One can deterministically construct linear maps $\Gamma:\F_2^L\to\F_2^M$, $\Xi:\F_2^L\to\F_2^k$, represented by $M\times L$ and $k\times L$ binary matrices, and a list $\calG\subseteq\calF$ of good sets, of size
\[
\frac{81}{1024}K'\le K_g\coloneqq|\calG|\le K',
\]
in $K'(2^M+2^k)\operatorname{poly}(L)$ bit operations. A set $Q\in\calF$ is \emph{good} if the columns of $\Gamma$ indexed by $Q$ and the columns of $\Xi$ indexed by $[L]\setminus Q$ form invertible matrices $\Gamma_Q$ and $\Xi_{[L]\setminus Q}$. A band consisting of $K_g$ blocks, each of outer width $N_0$, has width
\[
\Lambda\coloneqq K_g4^k.
\]
The displayed construction cost covers the maps and lists of inner sets. For outer dimension $N$, the condition $N\ge K'4^k$ is sufficient for fitting a band and implies $L\le\log_2 N$.
\end{lemma}

\begin{proof}
We adapt the argument per inner set in the proof of Majdoddin's Lemma 4.1~\cite{Majdoddin} to the list $\calF$. Process the listed inner sets in order. For the family of all $M$-subsets, listing them in lexicographic order takes $\binom{L}M\operatorname{poly}(L)$ bit operations. For a fixed inner set, independent uniform binary columns make the two restrictions invertible with probability $\mu_M\mu_k$, where
\[
\mu_j\coloneqq\prod_{i=1}^j(1-2^{-i})\ge\frac9{32}.
\]
The lower bound follows from the first two factors and $\prod_{i\ge3}(1-2^{-i})\ge1-\sum_{i\ge3}2^{-i}=3/4$; $j=1$ is immediate. Choose $\Gamma$ one whole column at a time by conditional expectations, then choose $\Xi$ over the retained $\Gamma$-good list. A target restriction whose assigned relevant columns are dependent has completion probability zero; if $a$ of its $u$ columns are assigned and independent, the probability is $\mu_{u-a}$, with $\mu_0\coloneqq1$. Rank computations therefore evaluate every conditional expectation exactly. Its average over the next column equals the previous expectation. Choosing a maximizing column proves the list bound. The possible column values contribute $2^M+2^k$, and the dyadic probabilities have $O(L^2)$ bits. Listing the good sets, restricted inverses and a dictionary of inner sets is included. Only the listed inner sets enter those expectations.

Since $K_g\le K'$, the sufficient fit gives $N\ge\Lambda$. Also $L=M+k\le2k\le\log_2N$ because $k\ge M$ and $N\ge4^k$. Under this fit the map construction takes $O(N\operatorname{poly}(L))$ bit operations. The actual width $\Lambda$ and sufficient threshold $K'4^k$ differ by at most the fixed factor $1024/81$.
\end{proof}

For the component decomposition, take the parameters $L=cM$ and $k=L-M$ of Section~\ref{sec:ordinary-guarantee}, for which $k\ge M\ge1$, use all $M$-subsets of $[L]$, and put $C_s\coloneqq\binom{L}{M}$. The lemma supplies the common good list $\calG_s$ and the two restricted maps. Choose the component band width $\Lambda_s\coloneqq|\calG_s|4^{k}$; then $81C_s/1024\le|\calG_s|\le C_s$. Every component and every job uses this ordered list and its restricted inverses. For a binary matrix $U$, write $U[J]$ for the sum over $\F_2$ of its columns indexed by $J$.

It remains to represent each component entry with these maps. Section~\ref{sec:walsh} gives the aligned embedding used by the grouped computation, including its locator and normalization, and Lemma~\ref{lem:aligned-embedding} states its identity. Appendix~\ref{app:original-embedding} gives the complete original squared construction and its coefficient-one identity; Lemma~\ref{lem:original-embedding} records the corresponding statement for those original forms. The procedure, and hence the proof of Theorem~\ref{thm:main}, uses only the aligned forms.

%% file: blocked-source.tex
\subsection{A common decomposition of the requested products}\label{sec:blocked-source}

Recall from Section~\ref{sec:ordinary-guarantee} that $D=KD_s=4^m$ with $K=4^\ell$ and $D_s=4^M$, where $m=\ell+M$, and that the $K$ consecutive middle components each have inner dimension $D_s$. Let the component band width $\Lambda_s$ be a positive integer, supplied by Section~\ref{sec:restricted-supports}, and recall that an outer block consists of $B$ component bands. Pad both outer axes of every $N\times D$ by $D\times N$ input pair to outer blocks of width $B\Lambda_s$.

Within a pair of outer blocks write a row as $(b_L,\varphi_L)\in[B]\times[\Lambda_s]$, a column as $(b_R,\varphi_R)\in[B]\times[\Lambda_s]$, and a middle index as $(\nu,h)\in[K]\times[D_s]$. Denote the corresponding component inputs by $X_{b_L,\nu}$ and $Y_{\nu,b_R}$. The entry to return is
\begin{equation}\label{eq:blocked-dot}
 \sum_{\nu=1}^{K}\sum_{h=1}^{D_s} X_{b_L,\nu}[\varphi_L,h]\,Y_{\nu,b_R}[h,\varphi_R].
\end{equation}
For fixed $b_L,b_R,\varphi_L,\varphi_R$, the \emph{suffix addresses} $(\varphi_L,\varphi_R)$, which locate the entry inside its row and column component bands, are the same in every component. We will encode these addresses once and use the encoding for all $K$ component products and all $J'$ jobs. The numerical entries remain specific to their own job and component.

This decomposition permits two distinct shared computations. The high part, which is read from tables, can be summed over $\nu$ when the tables are formed, so each request reads the sum once (Section~\ref{sec:walsh-high}). For the low part, the same requested addresses select operand values for all $J'K$ pairs of a job and a component by one family of rectangular products (Section~\ref{sec:walsh-sharing}).

%% file: walsh-source.tex
\subsection{The aligned Walsh embedding}\label{sec:walsh}

For each requested component entry we need a sum of bilinear contractions with an address common to every job and component. We choose an encoding whose operand values can later be selected by common selector rows. It is the aligned five-term identity and two-sided polynomial cache of the working note~\cite[lemmas \emph{Walsh embedding with a shared private factor} and \emph{Two-sided polynomial cache}, equations (A1)--(A11)]{WorkingNote}, with the private factor shifted as in its lemma \emph{Shifted private factor and light source decomposition}.

\paragraph{The component coordinates.}
We use the parameters of Section~\ref{sec:ordinary-guarantee}; in particular $L=cM$ and $k=L-M$. For each $\sigma\in\{-M,\ldots,M\}$, the \emph{run} $\sigma$ pairs the left and right arrays $\mathcal A_\sigma,\mathcal B_\sigma$ defined below. The complete contraction, defined before Lemma~\ref{lem:aligned-embedding} and summed over these runs, equals $4^L$ times the original component entry, as Lemma~\ref{lem:aligned-embedding} below shows.

The good sets and re-indexing maps $\Gamma,\Xi$ are those that Lemma~\ref{lem:profile-embedding} supplies for all $M$-subsets of $[L]$, with good list $\calG_s$. Thus $\Gamma_Q$ and $\Xi_{\overline Q}$, where $\overline Q\coloneqq[L]\setminus Q$, are invertible for each good $Q$, and the component band width is $\Lambda_s=|\calG_s|4^k$. For each good set $Q$, write $X_Q$ for its $N_0\times D_s$ component block and $Y_Q$ for its $D_s\times N_0$ component block. The inner indices lie in $\F_2^M\times\F_2^M$. A component row within a good-set block has two $k$-bit indices $(i_L,a_L)$ and a component column has indices $(a_R,i_R)$. On each component independently, apply the unnormalized Walsh transform along its \emph{opposite outer index}, that is, along $a_L$ for the rows and along $a_R$ for the columns. In a good-set block $Q$, with inner index $h$ and $q\in\F_2^k$, the transformed arrays are
\[
 \widehat X_Q[(i,q),h]\coloneqq\sum_{a\in\F_2^k}(-1)^{q\cdot a}X_Q[(i,a),h],
 \quad
 \widehat Y_Q[h,(q,i)]\coloneqq\sum_{a\in\F_2^k}(-1)^{q\cdot a}Y_Q[h,(a,i)] .
\]
Successively adding and subtracting values along each of the $k$ bit coordinates takes $O(k)$ additions per input entry.

\paragraph{Bilinear forms and runs.}
For each component the normal raw alphabet is $x_1,x_2,p_1,p_2$, and the opposite alphabet is $s_o,v_o,w_o$. A \emph{raw word} of an operand assigns a raw letter to each of the $2L$ levels, $L$ in each half: the left operand uses normal letters in the first half and opposite letters in the second, and the right operand the reverse; we call its two half-words its \emph{normal word} and its \emph{opposite word} accordingly. The local forms are
\[
\begin{array}{c|c|c}
 \text{label}&\text{normal form}&\text{opposite form}\\ \hline
 11&x_1+p_2&s_o+v_o\\
 12&x_1&s_o-v_o\\
 21&x_2-p_2&s_o+v_o-w_o\\
 22&x_2&s_o-v_o+w_o\\
 0&-x_1-x_2+p_1&2s_o
\end{array}
\]
The labels $11,12,21,22$ are \emph{ordinary}, and $0$ is \emph{private}; each label indexes one of the five terms of the identity. The private forms are $\Pi_N\coloneqq-x_1-x_2+p_1$, $\Pi_O\coloneqq2s_o$ and their product is $\Pi\coloneqq\Pi_N\Pi_O$. Adding the five paired products gives the inner kernel $I\coloneqq2p_1s_o+p_2w_o$. These aligned forms are used in the rest of Section~\ref{sec:ordinary}.

For a normal word $\mathbf u$, let $Q(\mathbf u)$ be its $p$ levels, $T(\mathbf u)$ its $p_2$ levels and $i(\mathbf u)$ its $x$-subscript bits on the complement. For an opposite word $\mathbf v$, write $V(\mathbf v),W(\mathbf v)$ for its $v_o,w_o$ levels. Use the subset-sum notation $\Gamma[J],\Xi[J]$ defined above. In run $\sigma\in\{-M,\ldots,M\}$ the left raw word $(\mathbf u,\mathbf v)$ and the right raw word $(\mathbf v,\mathbf u)$ index the two scalar arrays
\[
 \begin{aligned}
 \mathcal A_\sigma[\mathbf u,\mathbf v]&\coloneqq2^{|T(\mathbf u)|}\,
  \widehat X_{Q(\mathbf u)}[(i(\mathbf u),\Xi[V(\mathbf v)]),(\Gamma[T(\mathbf u)],\Gamma[W(\mathbf v)])],\\
 \mathcal B_\sigma[\mathbf v,\mathbf u]&\coloneqq2^{|T(\mathbf u)|}\,
  \widehat Y_{Q(\mathbf u)}[(\Gamma[W(\mathbf v)],\Gamma[T(\mathbf u)]),(\Xi[V(\mathbf v)],i(\mathbf u))].
 \end{aligned}
\]
Set an entry to zero unless $Q(\mathbf u)$ is good and $|W(\mathbf v)|-|T(\mathbf u)|=\sigma$ on the left, or $|T(\mathbf u)|-|W(\mathbf v)|=\sigma$ on the right.

To locate an output with inner sets $Q_L,Q_R$, call the levels of $Q_L$ in the first half and of $Q_R$ in the second its \emph{inner levels} and all other levels its \emph{outer levels}. At the outer levels, the ordinary labels have normal bits $i_L,i_R$ and opposite bits
\[
 j_L\coloneqq\Xi_{\overline Q_L}^{\mathsf T}a_R,\qquad
 j_R\coloneqq\Xi_{\overline Q_R}^{\mathsf T}a_L ,
\]
respectively. At each outer level its normal and opposite bits select the first and second subscripts of the ordinary label (bit zero selects subscript one). Each requested entry determines its component bands and good-set blocks by division with remainder by $\Lambda_s$ and $N_0$. Computing the opposite bits with the transposed maps uses both coordinates of the request and takes polynomial time in $L+\log(2N)$; the row and column inner sets of the request may differ. The resulting word is the \emph{suffix locator} of the requested suffix addresses in~\eqref{eq:blocked-dot}. The component encodings themselves depend on only their own operand.

\paragraph{Reconstructing the component entry.}
At an outer level use the paired product of the two forms in the table row of its ordinary label, and at an inner level use $I$. With these level kernels $\mathcal K_1,\ldots,\mathcal K_{2L}$, the \emph{complete contraction} in run $\sigma$ is
\[
 \sum_{(\mathbf u,\mathbf v),(\mathbf v',\mathbf u')}\mathcal A_\sigma[\mathbf u,\mathbf v]\,\mathcal B_\sigma[\mathbf v',\mathbf u']\prod_{j=1}^{L}\mathcal K_j(\mathbf u_j,\mathbf v'_j)\prod_{j=1}^{L}\mathcal K_{L+j}(\mathbf u'_j,\mathbf v_j),
\]
over left raw words $(\mathbf u,\mathbf v)$ and right raw words $(\mathbf v',\mathbf u')$, where $\mathcal K(a,b)$ is the coefficient of the product of the normal letter $a$ and the opposite letter $b$ in the kernel $\mathcal K$; in the second half the right operand supplies the normal letter.

\begin{lemma}[Aligned embedding; after the working note~\cite{WorkingNote}]\label{lem:aligned-embedding}
Let the blocks $X_Q,Y_Q$, $Q\in\calG_s$, of one component have arbitrary signed entries, and form $\mathcal A_\sigma,\mathcal B_\sigma$ as above. Take a requested entry with row $(i_L,a_L)$ of $X_{Q_L}$ and column $(a_R,i_R)$ of $Y_{Q_R}$, where $Q_L,Q_R$ are good. In a run, a pair of raw words with a nonzero kernel coefficient at the suffix locator of this entry and with both array entries not set to zero has no $w_o$ letter at an outer level; the $w_o$ letters of each of the two words therefore lie exactly at the inner levels where the other has $p_2$. Summed over the $2M+1$ runs, the complete contraction at this locator is $4^L\sum_hX_{Q_L}[(i_L,a_L),h]\,Y_{Q_R}[h,(a_R,i_R)]$, that is, $4^L$ times the requested component entry. The locator depends only on the request, so all $K$ components and all jobs use it.
\end{lemma}

\begin{proof}
\emph{Surviving pairs.} Write $(\mathbf u_L,\mathbf v_L)$ and $(\mathbf v_R,\mathbf u_R)$ for the left and right raw words of such a pair. Every inner factor forces a normal $p$ level. There are exactly $M$ such levels per half, which fixes its good set. In particular, a normal word has no $p$ letter at an outer level, so its outer letters are the $x$ letters that the normal bits $i_L,i_R$ of the locator select. Let $\tau_L\coloneqq|T(\mathbf u_L)|$ and $\tau_R\coloneqq|T(\mathbf u_R)|$ be the two normal $p_2$ counts, and let $e_L,e_R$ count the $w_o$ levels of $\mathbf v_L$ and $\mathbf v_R$ outside the inner levels. Since $I$ pairs $p_2$ only with $w_o$, $|W(\mathbf v_L)|=\tau_R+e_L$ and $|W(\mathbf v_R)|=\tau_L+e_R$, so equality of the two run indices reads $\tau_R+e_L-\tau_L=\tau_R-\tau_L-e_R$. Hence $e_L=e_R=0$, which proves the first claim. Thus $W(\mathbf v_L)=T(\mathbf u_R)$ and $W(\mathbf v_R)=T(\mathbf u_L)$, so both array entries are read at the inner index $(\Gamma[T(\mathbf u_L)],\Gamma[T(\mathbf u_R)])$. The outside opposite factors then have the form $s_o+(-1)^jv_o$.

\emph{Weights and inner indices.} The weights $2^{|T(\mathbf u_L)|}$ and $2^{|T(\mathbf u_R)|}$ from the two raw arrays and the inner $2p_1s_o$ factors together give $2^{\tau_L+\tau_R}2^{2M-\tau_L-\tau_R}=2^{2M}$. The invertible $\Gamma$ restrictions enumerate each of the $4^M$ inner indices once.

\emph{Walsh inversion.} In the second half the left opposite word carries $v_o$ on subsets $J$ outside $Q_R$, signed by the opposite bits $j_R=\Xi_{\overline Q_R}^{\mathsf T}a_L$; these signs are the characters $(-1)^{a_L\cdot\Xi[J]}$. The map $J\mapsto\Xi[J]$ is bijective there, so Walsh inversion gives $2^k$ times the specified left value. In the first half the bits $j_L=\Xi_{\overline Q_L}^{\mathsf T}a_R$ give in the same way $2^k$ times the right value. As $2^{2M}2^k2^k=4^L$, the complete contraction summed over runs is $4^L$ times the desired component product entry. The locator above is computed from the request alone, which gives the last claim.
\end{proof}

\subsection{The low and high parts of a suffix query}\label{sec:suffix-queries}

We now apply the embedding to one middle component, whose words have $M$ inner levels in each of their two halves of length $L$. A \emph{suffix query} computes one requested entry of this component from its suffix addresses: we expand the inner kernels of the entry and group the resulting terms by their ordinary labels. The low terms form the residual, which Section~\ref{sec:walsh-sharing} evaluates for all jobs together; the high terms are grouped into boxes, which Section~\ref{sec:walsh-high} tabulates once per pair of outer blocks from a common set of bases. Use the aligned kernels $I=2p_1s_o+p_2w_o$ and $\Pi=(-x_1-x_2+p_1)(2s_o)$ from Section~\ref{sec:walsh}.

Fix the suffix locator of a request and denote its $2M$ inner levels by $\mathcal H$. All outer labels are prescribed. A \emph{leaf} is a choice of one of the four ordinary labels or the private label at every inner level.

For a fixed component band, component and run $\sigma$, a dictionary key specifies a local linear form $\ell_j$ at each of the $2L$ levels of the aligned arrays. Its left value is the scalar
\[
 {\mathcal D}_{\mathcal A,\sigma}(\ell_1,\ldots,\ell_{2L})
 \coloneqq\sum_{(\mathbf u,\mathbf v)}\mathcal A_\sigma[\mathbf u,\mathbf v]\prod_{j=1}^{L}[\mathbf u_j]\ell_j\prod_{j=1}^{L}[\mathbf v_j]\ell_{L+j} ,
\]
where $[\mathbf u_j]\ell_j$ is the coefficient of the raw letter $\mathbf u_j$ in that form, and $[\mathbf v_j]\ell_{L+j}$ that of $\mathbf v_j$ in $\ell_{L+j}$. The right value is defined on $\mathcal B_\sigma$ with its own forms. Majdoddin's leaf-product formula pairs these forms at the same leaf and run~\cite[Section 6, \emph{Products at leaves and the counts}]{Majdoddin}. Each paired label contributes one left form times one right form. The coefficient of a raw pair thus factors over the operands, so one term of the contraction is the product of these two scalars in the same run. Summing these leaf products over all label choices gives the complete contraction in that run.

\paragraph{The low part.}
For an integer $1\le t\le2M$ (the procedure uses $t=2vM$), the leaves with fewer than $t$ ordinary labels in $\mathcal H$ form the low part of the query, called the \emph{residual}. Write $I^\Theta$ for the tensor product over levels in $\Theta$, omitting the fixed outer factors. The working note's \emph{Prefix contraction} and \emph{Compact five-term prefix data structure} lemmas~\cite{WorkingNote} give the residual kernel
\begin{equation}\label{eq:residual-kernel}
 \sum_{\substack{U\subseteq\mathcal H\\|U|<t}}(I-\Pi)^U\Pi^{\mathcal H\setminus U}
 =\sum_{\substack{\Theta\subseteq\mathcal H\\|\Theta|<t}}
 (-1)^{t-|\Theta|-1}\binom{2M-|\Theta|-1}{t-|\Theta|-1}I^\Theta\Pi^{\mathcal H\setminus \Theta}.
\end{equation}
Indeed, summing the ordinary factors at a level gives $I-\Pi$. For fixed $\Theta$, the coefficient on the right is the alternating partial binomial sum over $U\supseteq\Theta$. In one run, a \emph{transformed key} fixes $\Theta$, a term of $I$ at each of its levels and the private label at the other inner levels. Its term is the product of the matching operand values times the displayed coefficient. The procedure does not enumerate these keys; it evaluates the sum of their terms through~\eqref{eq:cover-low} below.

We collect the contractions involving $I$ and $\Pi$ in~\eqref{eq:residual-kernel} as coefficients of truncated polynomials. Work in $\Z[z]/(z^t)$ and set
\[
 H_0(z)\coloneqq-x_1-x_2+(1+z)p_1,\qquad
 \Pi+zI=H_0(z)(2s_o)+zp_2w_o.
\]
For the $2M$ inner levels $\mathcal H$ of a request, a channel set $C\subseteq\mathcal H$ uses $p_2$ or $w_o$ on its levels, and $H_0(z)$ or $2s_o$ on the other inner levels, with the orientation of the operand and half. Outside $\mathcal H$ use the ordinary forms of the table in Section~\ref{sec:walsh}. For each fixed component band, component and run, the operand value is the tensor contraction defined above, now a truncated polynomial. Denote the two contractions by $l_C(z),r_C(z)$. The low numerator in that run is
\begin{equation}\label{eq:cover-low}
 \sum_{j=0}^{t-1}(-1)^{t-j-1}\binom{2M-j-1}{t-j-1}
 [z^j]\sum_{\substack{C\subseteq\mathcal H\\ |C|<t}}
 z^{|C|}l_C(z)r_C(z).
\end{equation}
This follows by tensoring the displayed identity for $\Pi+zI$ and applying~\eqref{eq:residual-kernel}. It holds for any choice of the operand arrays, and Section~\ref{sec:walsh-cover} applies it separately to each translated and cropped pair. Each product pairs the left and right polynomials of the same channel set, component and run, and is multiplied once by $z^{|C|}$. We reconstruct each operand polynomial completely before multiplying.

\paragraph{The high part.}
For the complementary high part, Alman and Vassilevska Williams~\cite[Sections 4.2--4.3, Lemmas 27--29]{AlmanVassilevskaWilliams2026} group leaves by their last ordinary levels and expand the last star to evaluate each box. Majdoddin~\cite[Lemma 7.1]{Majdoddin} gives the five-term version. Order the first half before the second, in increasing level order within each. Retain the last $t$ ordinary inner levels of a high leaf and their labels as anchors. Star all inner levels before the first anchor and put the private label on the remaining nonanchor inner levels. Each star stands for the full kernel $I$. Expanding all the stars recovers exactly the leaves with those anchors. This starred word is the \emph{box} of these anchors and labels, and its value is the sum of those leaves.

For table construction keep the \emph{suffix words}, which carry an ordinary label, the private label or a star at each of the $2L$ levels, with at most $M$ private-or-star labels in each half and at most $2M-t$ in total, with all stars preceding all private labels. A suffix word in this family with no stars is a \emph{suffix base}. Put $\beta_j\coloneqq\binom{L}{j}4^{L-j}$. The base count is
\begin{equation}\label{eq:regional-count}
 Z_s\coloneqq\sum_{\substack{0\le j_1,j_2\le M\\j_1+j_2\le2M-t}}
       \beta_{j_1}\beta_{j_2}.
\end{equation}
Each required box lies in this family. Replacing its last star by the private label or an ordinary label preserves the order and both caps. As for a box, the value $\operatorname{val}(\mathbf w)$ of a suffix word $\mathbf w$ is the sum of the products at the words obtained by expanding its stars. Since a star stands for $I$, the sum of the five paired products, these values satisfy the \emph{last-star recurrence}
\begin{equation}\label{eq:last-star}
 \operatorname{val}(\mathbf w)=\operatorname{val}(\mathbf w_{0})+\sum_{a,e\in\{1,2\}}\operatorname{val}(\mathbf w_{ae}),
\end{equation}
where $\mathbf w_{0}$ and $\mathbf w_{ae}$ denote $\mathbf w$ with its last star replaced by the private label $0$ and by the ordinary label $ae$. Replacing every star by the private label gives a base, whose preimages star an initial segment of its private labels. Thus at most $2M+1$ suffix words per base, the \emph{states} of the recurrence~\eqref{eq:last-star}, suffice when we evaluate boxes in increasing star count. The suffix bases are common to all requests on a pair of outer blocks: the table computes their box sums once. A particular request then reads
\begin{equation}\label{eq:high-query-count}
 C_{\rm box}\coloneqq 4^t\binom{2M}t
\end{equation}
values per run, one per choice of the anchors and their ordinary labels.

\begin{lemma}[Low and high parts; after Alman and Vassilevska Williams~{\cite[Lemma~28]{AlmanVassilevskaWilliams2026}} and Majdoddin~{\cite[Lemma~7.1]{Majdoddin}}]\label{lem:low-high}
Fix a requested component entry, its suffix locator with inner levels $\mathcal H$, an integer $1\le t\le2M$ and a run, and let the two arrays paired in that run be arbitrary. The complete contraction at the locator is the sum of two parts. The low part is the contraction with the residual kernel~\eqref{eq:residual-kernel}. The high part sums the values of the $4^t\binom{2M}t$ boxes of the request. For the arrays $\mathcal A_\sigma,\mathcal B_\sigma$ of Section~\ref{sec:walsh}, the low and high parts summed over the $2M+1$ runs give $4^L$ times the requested component entry.
\end{lemma}

\begin{proof}
The complete contraction is the sum of the leaf products over all leaves. The leaves with ordinary labels exactly on a set $U\subseteq\mathcal H$ with $|U|<t$ form the low part, and summing over their ordinary labels gives the left side of~\eqref{eq:residual-kernel}. Every other leaf has at least $t$ ordinary inner labels; its last $t$ ordinary levels and their labels determine the only box whose expansion contains it, so the boxes partition the high leaves. The last claim is Lemma~\ref{lem:aligned-embedding}.
\end{proof}

\subsection{Restricting the operands while preserving the complete sum}\label{sec:walsh-cover}

The shared evaluation of Section~\ref{sec:walsh-sharing} stores, for short intervals of levels, vectors indexed by the sets of levels that use the rare letter ($p_2$, or $w_o$ on the opposite side). Bounding how often this letter appears keeps the vectors short. We partition the original middle summands by translations, so each part uses at most half of the available levels for the letter $p_2$. The first claim of Lemma~\ref{lem:aligned-embedding} then permits the same restriction on $w_o$. Summing the translated parts recovers the component product. We use the operand restrictions of the working note's lemmas \emph{An exact half-weight source cap} and \emph{A separate jointly capped high-box encoder}~\cite{WorkingNote}.

\paragraph{A deterministic list of translations.}
Appendix~\ref{app:translation-list} constructs a list $\delta_1,\ldots,\delta_{J_\delta}\in\F_2^M$, with $J_\delta\coloneqq64M^3$, in $2^M\operatorname{poly}(M)$ bit operations. For every invertible binary $M\times M$ matrix $A$ and every $h\in\F_2^M$, some $j$ satisfies
\[
 \operatorname{wt}\bigl(A(h+\delta_j)\bigr)\le\lfloor M/2\rfloor.
\]
The list may contain repeated translations.

\paragraph{Translated and cropped arrays.}
For a good set $Q$, let $\iota_Q(h)$ be the first translation index $j$ for which
$\operatorname{wt}(\Gamma_Q^{-1}(h+\delta_j))\le\lfloor M/2\rfloor$.
For a pair of translation indices $(j_1,j_2)$ replace a component's inputs, in their original coordinates, by
\[
\begin{aligned}
 X^{j_1j_2}_{Q_L}[(i_L,a_L),(h'_1,h'_2)]
 &\coloneqq X_{Q_L}[(i_L,a_L),(h'_1+\delta_{j_1},h'_2+\delta_{j_2})]\,
       \mathbf1\{\iota_{Q_L}(h'_1+\delta_{j_1})=j_1\},\\
 Y^{j_1j_2}_{Q_R}[(h'_1,h'_2),(a_R,i_R)]
 &\coloneqq Y_{Q_R}[(h'_1+\delta_{j_1},h'_2+\delta_{j_2}),(a_R,i_R)]\,
       \mathbf1\{\iota_{Q_R}(h'_2+\delta_{j_2})=j_2\}.
\end{aligned}
\]
Here $(i_L,a_L)$ and $(a_R,i_R)$ denote the complete component outer addresses, as in Section~\ref{sec:walsh}. Each original middle summand belongs to exactly one translation pair. Apply the Walsh transform after these masks; they are independent of the transformed outer coordinates and commute with that transform.

In the raw arrays of Section~\ref{sec:walsh}, the mask makes every nonzero normal lookup have $p_2$-indicator $\Gamma_Q^{-1}(h')$ of weight at most $\lfloor M/2\rfloor$, where $h'$ is its translated inner coordinate. Retain this normal cap and impose the same cap on occurrences of $w_o$ in opposite words, setting every array entry that violates a cap to zero; we call this \emph{cropping}. The encoders of the low and high parts both use this rule.

\begin{lemma}[Restricted operands]\label{lem:restricted-sum}
For every requested component entry, evaluate the low and high parts of the translated and cropped arrays at the entry's suffix locator. Summing those values over the translation pairs and the $2M+1$ runs gives $4^L$ times the entry. For each translation pair and run, the low part of the cropped arrays is their low numerator~\eqref{eq:cover-low}.
\end{lemma}

\begin{proof}
Fix a translation pair. Its masked blocks are blocks of one component, so Lemma~\ref{lem:aligned-embedding} applies to their arrays before cropping. Every surviving complete monomial has the counts of $w_o$ equal to the normal $p_2$ counts of the other raw word in the same half, by Lemma~\ref{lem:aligned-embedding}, and the mask bounds these by $\lfloor M/2\rfloor$ wherever the other array entry is nonzero. Hence this cropping loses no complete product term, and the complete contraction of the cropped arrays, summed over the runs, is $4^L$ times the entry of the masked blocks. Since each original middle summand belongs to exactly one translation pair, the sum over the translation pairs is $4^L$ times the component entry. By Lemma~\ref{lem:low-high}, which allows arbitrary arrays, the complete contraction in each run is the sum of the low and high parts of the cropped arrays, and the polynomial identity~\eqref{eq:cover-low} applies to them as a tensor identity.
\end{proof}

Because translations and cropping can change the low and high parts individually, the assembly sums both parts over all runs, translations and components before dividing by $4^L$.

\subsection{Shared evaluation of the low part}\label{sec:walsh-sharing}

By Lemma~\ref{lem:restricted-sum}, the low part of a request is the low numerator~\eqref{eq:cover-low} of the translated and cropped arrays, which needs the coefficients of the operand polynomials $l_C(z)$ and $r_C(z)$. The required coefficient at any requested position will be a fixed linear combination of stored operand values, with the same coefficients for all jobs. This is the point at which the common requested positions reduce the cost. We use the working note's \emph{Grouped completion with asymmetric row chunks}~\cite{WorkingNote} and include the sharing across comparison moduli explicitly. We first describe the representation and the cost of evaluating it, and then construct it: we contract the inner levels, and a short window supplies the subset-indexed vectors and selectors.

\paragraph{The representation needed for shared evaluation.}
For each coefficient, we will organize the operand values into groups. A group stores one numerical vector of a common positive integer length $C_*$ for each job and middle component, hence $J'K$ columns. An elementary reading is the product of one selector row of length $C_*$ with the matrix of these columns; summing the appropriate readings will recover the polynomial coefficient. The selector depends on the requested address and the group, and is common to every job and component. Let $P_g$ count the groups and $M_g$ the requested selector rows over all groups. The construction below determines these groups and $C_*$, and Section~\ref{sec:walsh-rates} bounds $P_g$ and $M_g$.

\paragraph{Group products.}
The group products use the algorithm of Section~\ref{sec:ordinary-guarantee} for $C_*^{a_*}\times C_*$ by $C_*\times C_*^{\kappa_*}$ matrices, with the exponents $\alpha_*,\beta_*$ defined there. Check the actual integer inequality $J'K\ge C_*^{\kappa_*}$ with the required ceiling. Divide a group's requested rows into batches of size $\lceil C_*^{a_*}\rceil$ and its columns into batches of size $\lceil C_*^{\kappa_*}\rceil$, padding only ordinary matrix inputs by zero. If there are $\chi$ rows, this uses
$O((1+\chi/C_*^{a_*})J'K/C_*^{\kappa_*})$ products.
For $P_g$ groups and $M_g$ requested rows in total, the cost is at most
\begin{equation}\label{eq:group-products}
 J'K\left(P_g C_*^{1+\alpha_*}+M_g C_*^{\beta_*}\right)
\end{equation}
times a factor polynomial in $L+m+b_{\rm in}+\log(2N)+\log(2J')$ for bit lengths and addresses (Section~\ref{sec:walsh-rates}). This includes dense data formation $J'KP_gC_*$, selector formation $M_gC_*$, every output write $J'KM_g$, and short last batches.
To express the operand polynomials in this form, we use the caps of Section~\ref{sec:walsh-cover}. Each word has a short interval in which few levels use the rare letter. Its selector needs only the subsets of that interval within the cap.

\paragraph{Contracting the inner levels.}
Fix the two good sets and a channel set, with $a_n,a_o$ levels in its two halves. Contract their $2M$ inner levels first. A normal nonchannel level has three raw possibilities $x_1,x_2,p_1$; a channel level uses $p_2$. Opposite inner levels have a single raw possibility. The $2k$ other levels carry ordinary forms, which use only the raw letters $x_1,x_2,p_2$ at a normal level and $s_o,v_o,w_o$ at an opposite level, giving $3^{2k}$ remaining raw entries, and the $M$ normal inner levels give at most $3^M$ lookups per entry; we call the contracted arrays the \emph{raw roots}.
Cropping leaves rare-letter budgets $\lfloor M/2\rfloor-a_n$ and $\lfloor M/2\rfloor-a_o$ outside the inner sets. On the opposite side first apply the integer forward map
$(s_o,v_o,w_o)\mapsto(y_1,y_2,w')\coloneqq(s_o+v_o,s_o-v_o,-w_o)$.
In the original label order its four outputs are $(y_1,y_2,y_1+w',y_2-w')$. Permuting ordinary labels in the order $(21,11,22,12)$ gives the normal pattern $(y_1+w',y_1,y_2-w',y_2)$. We use this fixed permutation on every opposite axis in the \emph{requested keys}, the ordinary labels that the suffix locator of a request prescribes at its outer levels. The forward map preserves rare support and uses no division. Its application to the raw roots, one level at a time, is charged to the root encoder in~\eqref{eq:group-roots} of Section~\ref{sec:walsh-rates}.

\paragraph{Selector rows from short windows.}
Choose the half with more channels, with a fixed tie rule, and write $r_0$ for its remaining rare budget. Among the $k$ cyclic windows of length $h_w\le k$, every word with at most $r_0$ rare letters has one with at most $\lfloor r_0h_w/k\rfloor$ of them. Partition the \emph{raw} arrays by the first such window of each word before applying any transform. There are $k$ window components, one for each of the $k$ cyclic windows, with zeros elsewhere. For each window component transform the other $2k-h_w$ levels to ordinary labels and keep the window raw. Call the transformed part of a word its \emph{outside key} and its raw window part its \emph{window word}. At a window level the four ordinary forms make two pairs, each sharing one nonrare letter, and the rare letter occurs in one form of each pair. A \emph{pair-bit string} chooses one pair at each of the $h_w$ window levels. A \emph{formal coefficient} is the coefficient of one power $z^j$, $0\le j<t$, in the truncated operand polynomials. Within one translation pair, run and formal coefficient, a \emph{group} fixes the two good sets, the channel set, the window component, an outside key and a pair-bit string. Its stored vector is indexed by the sets $J$ of at most $\lfloor r_0h_w/k\rfloor$ window levels: entry $J$ is the value of the window word with the rare letter on $J$ and the chosen nonrare letters elsewhere, multiplied by the signs of the rare letter in the chosen pairs at the levels of $J$. A window word with $j$ rare letters fits $2^j$ pair-bit strings, so it is copied to all of these groups. Each vector has length
\[
 C_w\coloneqq\sum_{j\le\lfloor r_0h_w/k\rfloor}\binom{h_w}{j}.
\]
A requested key has ordinary labels on the window levels. It determines its group and the \emph{activity set} $A$ of window levels whose requested form contains the rare letter. Expanding those forms shows that its value is the sum of the stored entries with $J\subseteq A$, that is, the stored vector times the \emph{selector row} $\mathbf1\{J\subseteq A\}$. This row depends on the requested word, not on an operand's numerical entries. This constructs the shared linear representation required above.

\paragraph{The window length.}\label{sec:walsh-low}
The window length $h_w$ controls both the length of the stored vectors and the number of groups in~\eqref{eq:group-products}. For the binary entropy $H(y)\coloneqq-y\ln y-(1-y)\ln(1-y)$, put
\begin{equation}\label{eq:window-entropy}
 \gamma(a)\coloneqq H((1/2-a)/(c-1))/\ln4 .
\end{equation}
Lemma~\ref{lem:walsh-finite} assumes that its parameters satisfy the two guards
\begin{equation}\label{eq:group-guards}
 \zeta<(c-1)\gamma(2v),\qquad
 \frac{2\ln((1-v)/v)}{\ln4}
       +\frac{\zeta\gamma'(v)}{2\gamma(v)^2}>0 .
\end{equation}
The first lets the windows fit, as shown below; the second is used to count the groups in Section~\ref{sec:walsh-rates}. For channel fractions $a\coloneqq a_n/M$, $a'\coloneqq a_o/M$, choose
$h_w=\zeta M/\gamma(\max(a,a'))+O(1)$.
A finite choice uses the actual $k,r_0$ and a rational upper enclosure $\widehat\gamma$ of $H(r_0/k)/\ln4$ with error $O(M^{-2})$, and takes $h_w\coloneqq\lfloor\zeta M/\widehat\gamma\rfloor$. Then $h_wH(r_0/k)\le\zeta M\ln4$, and since $r_0/k<1/2$ the binomial estimate $\sum_{j\le\lambda h}\binom hj\le e^{hH(\lambda)}$ gives $C_w\le4^{\zeta M}=D_s^\zeta$. Such enclosures require only polynomial time by Appendix~\ref{app:numerical}. The first inequality in~\eqref{eq:group-guards} leaves $h_w<k$ for all sufficiently large $M$; bounded exceptions use the direct algorithm. Pad group lengths to the common integer
$C_*\coloneqq\lceil D_s^\zeta\rceil\le2D_s^\zeta$, which is at least every $C_w$ by the bound above.

\paragraph{Columns of all jobs on the common positions.}
Use the $J'$ synchronized jobs on the same positions $W_p$ from Section~\ref{sec:assembly}, with the common inner dimension $D=KD_s$ and $K=4^{rm}$. In an outer block of width $B\Lambda_s$ on the row axis, a row has a band index $b_L\in[B]$ and a suffix row; similarly a column has $b_R\in[B]$ and a suffix column. The left operand is treated separately for each band index $b_L$, because its root and residual encoders are formed on the operand rows with that band index. Fix $b_L$. The left operand values of a request depend on the outer-block index and the band index $b_L$ of its row and on its two suffix addresses, which fix its locator. They do not depend on its column band $b_R$, which only selects the right inputs $Y_{\nu,b_R}$. For a position of $W_p$ with band index $b_L$, call the outer-block index of its row together with its two suffix addresses its \emph{projected row}; summed over $b_L$, there are at most $|W_p|$ projected rows. Retain a routing list to all original positions. For a fixed $b_L$, the set of operand rows with band index $b_L$ has $O(N/B)$ elements since an outer block fits; summing over the $B$ choices gives $O(N)$ rows. The root and residual encoders are formed on these operand rows, and the projected rows supply the requested rows of the group products in~\eqref{eq:group-products}. The proof of Lemma~\ref{lem:walsh-finite} sums these row sets over the band indices and the $K$ middle components to obtain the rates $A_{\rm root}$ and $A_{\rm res}$, and sums the projected rows in the same way to obtain $q_1$. On the right retain the outer-block index of the right operand, fix $b_R$ and forget $b_L$ instead.

For a fixed group, collect its $J'K$ data columns, one per job and middle component.

Route every completed polynomial back to its original output, job and component. Sum its window components, then use matching operands in~\eqref{eq:cover-low}. All component reads, polynomial coefficients and scatter operations are paid.

\begin{lemma}[Shared low evaluation; across middle components after the working note~\cite{WorkingNote}]\label{lem:shared-low}
Fix a translation pair, a run and a formal coefficient. For every requested position, every channel set $C$ of the inner levels of its suffix locator with $|C|<t$ and both operands, the coefficient of the operand polynomial $l_C(z)$ or $r_C(z)$ is, in every job and every middle component, the sum over the window components of one elementary reading each: the selector row of the request times the stored vector of its group in that job and component. The groups and the selector rows depend only on the requested position and the fixed parameters; they are the same for all $J'$ jobs and all $K$ middle components.
\end{lemma}

\begin{proof}
After the inner levels are contracted, the coefficient is a contraction of the raw roots over the $2k$ outer levels, and the window components partition the raw roots. In one window component, the requested labels outside the window give the outside key and those on the window give the pair-bit string and the activity set $A$; with the two inner sets of the request and the channel set they determine the group. Expanding the requested forms on the window gives the stored vector times the selector row $\mathbf1\{J\subseteq A\}$. None of these choices involves the numerical entries of a job or a component.
\end{proof}

\subsection{High tables and assembly}\label{sec:walsh-high}

The low step shares matrix multiplications across recovery chunks. The high step sums the middle components of each job before applying the last-star recurrence.

\paragraph{The high encoder.}
For each job, component band, component, translation pair and run, construct the high operand dictionary on suffix bases with at most $M$ private levels per half and at most $2M-t$ in total. On the same cropped arrays, use the private factors $H_0(0)$ and $2s_o$, the constant coefficients of the low-part factors $H_0(z)$ and $2s_o$. The number of bases is $Z_s$ from~\eqref{eq:regional-count}. After transforming $j$ levels, retain only prefixes that obey the caps on private levels. Each such prefix has at most $4^{2L-j}$ remaining raw entries and exactly $4^{2L-j}$ completions to bases using ordinary labels. Completion sets for distinct prefixes are disjoint. Each layer thus has at most $Z_s$ entries, and the $2L$ layers of constant fan-in cost $Z_s\operatorname{poly}(L)$ per component band and component. The initial $12^L$ raw entries fit the included $16^L$ bases with only ordinary labels.

\paragraph{The high tables.}
For each job, suffix base, pair of outer blocks, translation pair and run, form the $B\times K$ left and $K\times B$ right matrices of its constant coefficients. Multiply them with the fixed algorithm of shape parameter $\lambda_h$ and loss $\xi_h$ of Section~\ref{sec:ordinary-guarantee}, assuming $K\le B^{\lambda_h}$. Form all $2BK$ input entries and all $B^2$ output entries. The last-star recurrence~\eqref{eq:last-star} is linear in these matched base products. Apply it \emph{after} summing over the $K$ middle components.

\paragraph{Assembly.}
Fix a requested position of one job. For a translation pair $(j_1,j_2)$ and a run $\sigma$, let $\operatorname{low}_\nu(j_1,j_2,\sigma)$ be the low numerator~\eqref{eq:cover-low} of middle component $\nu$, computed as in Section~\ref{sec:walsh-sharing}. Let $\operatorname{high}(j_1,j_2,\sigma)$ be the sum of the box values read from the high tables, which already includes all $K$ components. Sum the complete low and high numerators over runs and translations and divide once by $4^L$:
\begin{equation}\label{eq:final-assembly}
 4^{-L}\sum_{j_1,j_2}\sum_{\sigma=-M}^{M}\Bigl(\sum_{\nu=1}^{K}\operatorname{low}_\nu(j_1,j_2,\sigma)+\operatorname{high}(j_1,j_2,\sigma)\Bigr).
\end{equation}
By Lemma~\ref{lem:restricted-sum}, applied to each middle component, and by~\eqref{eq:blocked-dot}, this is the requested entry of the job. The division is exact because the complete sum equals $4^L$ times the entry; it is applied only to this complete sum.

\subsection{Resource bounds}\label{sec:walsh-rates}

We count the component band width, the suffix bases and the groups of selector rows. These bounds give the operand and request costs in the proof of Lemma~\ref{lem:walsh-finite}.

\paragraph{Band width and suffix bases.}\label{sec:suffix-counts}
The band width gives the fit exponent $T$. The number of suffix bases bounds the costs of the high encoder and the high tables. For $\vartheta>1$ and $0<y<1$, define
\begin{equation}\label{eq:profile-entropy}
\begin{aligned}
 C(\vartheta)&\coloneqq [\vartheta\ln\vartheta-(\vartheta-1)\ln(\vartheta-1)]/\ln4,\quad t_0(\vartheta)\coloneqq\vartheta-1+C(\vartheta),\\
 G(\vartheta,y)&\coloneqq \big[(\vartheta-1+y)\ln(\vartheta-1+y)-(\vartheta-1)\ln(\vartheta-1)\\
 &\hspace{35mm}+(1-y)\ln(1-y)-y\ln4\big]/\ln2 .
\end{aligned}
\end{equation}
Since $\beta_{j-1}/\beta_j=4j/(L-j+1)<4/(c-1)<1$ through $j=M$, the $\beta_j$ increase through $j=M$, and $\beta_M=D^{st_0(c)+O(\log m/m)}$. This gives the band width: by Section~\ref{sec:restricted-supports}, $\Lambda_s=|\calG_s|4^k$ with $81C_s/1024\le|\calG_s|\le C_s$, and $C_s4^k=\binom LM4^{L-M}=\beta_M$; so $81\beta_M/1024\le\Lambda_s\le\beta_M$ and $\Lambda_s=D^{st_0(c)+O(\log m/m)}$. The $\beta_j$ are log-concave. In~\eqref{eq:regional-count} their product is greatest when the two indices sum to $2M-t$ and are the nearest integers to $M-t/2$. There are at most $(M+1)^2$ pairs. Applying the binomial entropy estimate there and at $\beta_M$ gives
\begin{equation}\label{eq:suffix-terminal-estimate}
 \log_D(Z_s/\beta_M^2)=-sG(c,v)+O(\log m/m).
\end{equation}

\paragraph{Number of groups.}
The number of groups and the vector length $C_*$ determine the shared evaluation cost~\eqref{eq:group-products} and hence the exponent $A_{\rm res}$ of the residual encoder. Section~\ref{sec:walsh-low} bounds $C_*$. To count the groups, use the binary entropy $H$ and the function $\gamma$ of~\eqref{eq:window-entropy}, and put
\begin{equation}\label{eq:group-entropy}
 C_K\coloneqq C(c),\qquad T_s\coloneqq t_0(c),\qquad \psi\coloneqq H(v)/\ln2,\qquad
 G_s\coloneqq G(c,v),
\end{equation}
where $C,t_0,G$ are from~\eqref{eq:profile-entropy}. For one component band, choose the two inner sets in at most $\binom LM^2$ ways and the channels in $\binom M{a_n}\binom M{a_o}$ ways. Each choice has $4^{2k-h_w}$ outside keys and $2^{h_w}$ pair-bit strings, with only polynomially many windows, runs, translations and coefficients. Summing over $a_n+a_o<t$ gives the group-count bound, up to polynomial factors,
\begin{equation}\label{eq:group-number}
 D_s^{\,2T_s+
 \max_{a+a'\le2v}\{[H(a)+H(a')]/\ln4
                  -\zeta/[2\gamma(\max(a,a'))]\}} .
\end{equation}
The negative term records the replacement of $4^{h_w}$ requested keys by $2^{h_w}$ pair-bit groups; the $C_*$ stored entries are charged separately. At fixed $a+a'=2\bar a$, concavity of entropy and decrease of $\gamma$ make the maximum occur at $a=a'=\bar a$. Since $\gamma$ is positive and concave, $1/\gamma$ is convex. Thus
$2H(\bar a)/\ln4-\zeta/(2\gamma(\bar a))$ is concave, and its positive derivative at $v$ in~\eqref{eq:group-guards} makes the maximum on $[0,v]$ occur at $v$. The exponent in~\eqref{eq:group-number} is therefore $2T_s+\psi-\zeta/(2\gamma(v))$. Uniform binomial estimates and the polynomially many pairs $(a_n,a_o)$ account for the stated polynomial factors.

\paragraph{Root encoders and dense data.}
The root encoder gives the exponent $A_{\rm root}$. Contracting the inner levels in Section~\ref{sec:walsh-sharing}, over all channels and good sets, charges per component band of width $\Lambda_s$ and data column (one job and middle component)
\begin{equation}\label{eq:group-roots}
 \binom LM^2\,3^{2k+M}\sum_{h<t}\binom{2M}{h}\operatorname{poly}(L).
\end{equation}
The stored vectors are the dense data charged in~\eqref{eq:group-products}. For each outside key, copying every retained window word to its $2^j$ pair-bit groups gives exactly $2^{h_w}C_w$ entries. After $d'$ outside transforms, $0\le d'\le2k-h_w$, the intermediate arrays fit the same final allocation:
\[
 \sum_{j\le\lfloor r_0h_w/k\rfloor}\binom{h_w}j2^{h_w-j}2^j=2^{h_w}C_w,\qquad
 4^{d'}3^{2k-h_w-d'}2^{h_w}C_w\le4^{2k-h_w}2^{h_w}C_w.
\]
The dense data formation charged in~\eqref{eq:group-products} therefore covers all zeros, copies and partial transforms.

\paragraph{The exponents.}
The rates of Lemma~\ref{lem:walsh-finite} are
\begin{equation}\label{eq:walsh-rates}
\begin{aligned}
 E_{\rm root}&\coloneqq2C_K+(2c-1)\log_4 3+\psi-T_s,&
 A_{\rm root}&\coloneqq r+sE_{\rm root},\\
 E_{\rm res}&\coloneqq T_s+\psi-\zeta\!\left(\frac1{2\gamma(v)}-1-\alpha_*\right),&
 A_{\rm res}&\coloneqq r+sE_{\rm res},\\
 E_{\rm box}&\coloneqq T_s-G_s,& A_{\rm box}&\coloneqq r+sE_{\rm box},\\
 T&=u+sT_s,& V&=sG_s-u\xi_h,\\
 q_1&=r+s(\psi+\beta_*\zeta),&q_2&=s(\psi+2v).
\end{aligned}
\end{equation}

\begin{proof}[Proof of Lemma~\ref{lem:walsh-finite}]
\emph{Exactness.} The procedure computes the low numerators of all jobs by Lemma~\ref{lem:shared-low} and the group products~\eqref{eq:group-products}, and the high values from the tables of Section~\ref{sec:walsh-high}. It returns~\eqref{eq:final-assembly}, which is the requested entry of every job by Lemma~\ref{lem:restricted-sum} and~\eqref{eq:blocked-dot}.

\emph{Accounting for operand preparation and low requests.}
The shared maps cost $\binom LM(2^M+2^{L-M})\operatorname{poly}(L)$; translations cost $2^M\operatorname{poly}(M)$. Both fit within the outer width up to polynomial factors. Masked inputs and Walsh transforms cost $J'ND\operatorname{poly}(L)$. By the fit condition and Lemma~\ref{lem:profile-embedding}, the actual number $\lceil N/(B\Lambda_s)\rceil$ of outer blocks is within a constant factor of $N/(B\Lambda_s)$.

Put $\mathsf N_{\rm ch}\coloneqq\sum_{h<t}\binom{2M}h=D_s^{\psi+o(1)}$. The root bound~\eqref{eq:group-roots}, group bound~\eqref{eq:group-number} and suffix-base estimate~\eqref{eq:suffix-terminal-estimate} give, per component band,
\[
\begin{aligned}
 \mathsf N_{\rm root}&\coloneqq D_s^{2C_K+(2c-1)\log_4 3+\psi+o(1)},&
 \mathsf N_{\rm grp}&\coloneqq D_s^{2T_s+\psi-\zeta/(2\gamma(v))+o(1)},\\
 Z_s&=D_s^{2T_s-G_s+o(1)},&
 \Lambda_s&=D_s^{T_s+o(1)} .
\end{aligned}
\]
Here $\mathsf N_{\rm root}$ bounds root work per data column and $\mathsf N_{\rm grp}$ bounds the number of groups. Summing over the band indices gives $B\,O(N/(B\Lambda_s))=O(N/\Lambda_s)$ component bands. Each requested position fixes both good sets. It has $\mathsf N_{\rm ch}$ channel sets, at most $k$ windows, $t$ coefficients and polynomially many runs and translations. Thus $M_g\le |W_p|D_s^{\psi+o(1)}$. After division by $J'$, the three operand charges and the low request charge are, up to polynomial factors,
\[
 \frac{NK \mathsf N_{\rm root}}{\Lambda_s},\qquad
 \frac{NK \mathsf N_{\rm grp} C_*^{1+\alpha_*}}{\Lambda_s},\qquad
 \frac{NK Z_s}{\Lambda_s},\qquad
 wK \mathsf N_{\rm ch} C_*^{\beta_*}.
\]
For each job, summing windows, multiplying the matched polynomials in $O(t^2)$ operations, and routing its $K$ columns to the original occurrences cost $Kw\mathsf N_{\rm ch}\operatorname{poly}(L)$, covered by the low request charge since $C_*^{\beta_*}\ge1$.

Substitute $K=D^r$, $D_s=D^s$ and $C_*=D_s^{\zeta+o(1)}$ to obtain $A_{\rm root},A_{\rm res},A_{\rm box},q_1$ in~\eqref{eq:walsh-rates}. Every middle component is included.

\emph{Accounting for the high tables and reads.}
The outer-block width $B\Lambda_s=D^{u+sT_s+o(1)}$ gives $T$. The high products cost, per job,
\[
 \left(\frac{N}{B\Lambda_s}\right)^2 Z_sB^{2+\xi_h}
 =N^2D^{-sG_s+u\xi_h+o(1)},
\]
which gives $V$. Their input and output formation is included since $K\le B^{\lambda_h}\le B$. The last-star recurrence has at most $(2M+1)Z_s$ states per pair of outer blocks and pair of band indices, covered by the joint product charge. The tables already sum the components. A query reads $4^t\binom{2M}t=D_s^{2v+\psi+o(1)}$ values, which gives $q_2$.

\emph{Integer arithmetic and addresses.}
Appendix~\ref{app:rectangular} implements both rectangular algorithms on the actual integer entries, with fixed coefficients and the chosen positive losses. A linear operand value has at most $b_{\rm in}+O(L)$ bits; a bilinear value, after channel sums and $K$-component accumulation, has at most $2b_{\rm in}+O(L+\log K)$ bits. Translation masks, formal coefficients, signed residual weights, Walsh normalization and all addresses cost a polynomial in $L+m+b_{\rm in}+\log(2N)+\log(2J')$. Index lists and layer allocation are included in the preceding finite bounds. On our schedules all these parameters are $O(\log(2n))$, and the uniform polynomial factors contribute only the displayed vanishing exponents.
\end{proof}

Substituting the rates of~\eqref{eq:walsh-rates} into Section~\ref{sec:envelope} bounds the cost of the comparison counts used by the recovery algorithm.

%% file: exponents.tex
\section{Parameters and the deterministic real-weight bounds}\label{sec:point}

Section~\ref{sec:envelope} gives a sufficient list of savings, and Lemma~\ref{lem:walsh-finite} expresses the ordinary-product costs in terms of its construction parameters. We choose one column of~\eqref{eq:point} for each rectangular multiplication theorem and check two requirements: all finite dimensions eventually fit, and every cost saves more than the target saving $S$. The first requirement makes the direct fallbacks a bounded-size exception; the second absorbs every subpolynomial overhead.

All terminating decimals in the following table are exact rational numbers. In both columns put $s\coloneqq1-r$, $x\coloneqq e/2$, $b\coloneqq x+\eta$, and use candidate-selection shape parameter $\alpha=0.3$ with loss $\varepsilon=0.0001$.
\begin{equation}\label{eq:point}
\begin{array}{c|cc}
 &\text{Uzna\'nski}&\text{Le Gall--Urrutia}\\ \hline
 c&10.728&10.86082855\\
 v&0.06520068&0.06495237\\
 \zeta&0.173&0.17105135\\
 r&0.11535843&0.11452882\\
 u&0.23072&0.3648700\\
 e&0.0912562&0.08906135\\
 \eta&0.025&0.03043\\
 S&0.00702710&0.00693727
\end{array}
\end{equation}
The group products in the first column use $a_*=\kappa_*=2$ and $w_*=4.0000002$. Thus
$\alpha_*=1.0000002$, $\beta_*=0.0000002$.
Indeed a product with long side $C_*^2$ and short side $C_*$ is a product with shape parameter $1/2$, and the loss $10^{-7}$ in its long dimension gives this $w_*$.
For the second column use $a_*=\kappa_*=50/17$ and
\[
 w_*=(2.001118+10^{-7})\,50/17,\qquad
 \alpha_*=\frac{388893}{200000},\qquad
 \beta_*=\frac{11181}{3400000}.
\]
The shape parameter of its group product is $17/50$. Le Gall and Urrutia's Table~3 gives $\omega(1,17/50,1)\le2.001118$~\cite{LeGallUrrutia}; the additional fixed loss is reserved for a finite exact algorithm. The high products use respectively $\lambda_h=1/2$ and $\lambda_h=0.31389$, with $\xi_h=10^{-7}$ in both columns. Appendix~\ref{app:rectangular} provides the rational integer implementation of these two input theorems. The algorithms and their coefficients are fixed before the data arrive; their bounded-integer inputs may depend on comparisons of real values.

\begin{proof}[Proof of Theorem~\ref{thm:main}]
\emph{Valid dimensions.} Choose either column and run Algorithm~\ref{alg:minplus}, which uses the selection, recovery and deterministic prime procedures of Sections~\ref{sec:selection}--\ref{sec:envelope}. By the sufficient cost criterion of Lemma~\ref{lem:criterion}, it suffices that the dimension tests of the entry procedure and the selection tests succeed for all sufficiently large $n$, that $eT<1$, and that every saving exceeds the column's target $S$. Let $m$ range over positive multiples of
\[
 J_m\coloneqq\operatorname{lcm}\bigl(\operatorname{den}(r),\operatorname{den}(s),
 \operatorname{den}(cs),\operatorname{den}(2sv),\operatorname{den}(u)\bigr).
\]
Here $\operatorname{den}$ is the reduced denominator. The two values are respectively $1250000000000000$ and $5000000000000000$. Consequently $\ell=rm$, $M=sm$, $L=cM$, $t=2vM$ and $\log_4 B=um$ are integers, so $m$ lies on the integral grid of Section~\ref{sec:ordinary-guarantee}. The window length $h_w$ is chosen separately by the finite rounding rule in Section~\ref{sec:walsh-low}.

Take the least $D\coloneqq4^m$ on this grid satisfying~\eqref{eq:effective-grid}. Its fixed grid ratio preserves $n^e\le D\le n^{e+o(1)}$, and $D$ accommodates all trial moduli. Before allocating the arrays of the entry procedure, test
\begin{equation}\label{eq:blocked-width-test}
 B\binom LM4^{L-M}\le n,\qquad
 K\le B^{\lambda_h},\qquad
 J'K\ge\lceil C_*^{\kappa_*}\rceil .
\end{equation}
The tests use capped integer powering and comparisons; the rational powers have fixed denominators. First compare $4^{um+L-M}$ to $n$. If it fits, $L\le2(L-M)\le\log_2n$, and an integer recurrence computes the capped binomial. The common group length $C_*$ can likewise be calculated before its arrays are allocated. If any condition of the entry procedure, the prime pool or the selection fails, use the direct product, charging work already done (lines~\ref{alg:mp-guard} and~\ref{alg:mp-fail} of Algorithm~\ref{alg:minplus}).

For all sufficiently large $n$ the tests of the entry procedure succeed uniformly, because Appendix~\ref{app:numerical} verifies
\[
 1-eT>0,\qquad
 \lambda_h u-r>0,\qquad
 \eta+er-\kappa_*es\zeta>0.
\]
The last inequality absorbs the constant factor from rounding $C_*$, using $J'\ge n^\eta$. For a single job at either column, the same strict guard would read $r>\kappa_*s\zeta$, whereas $r\approx0.115$ and $\kappa_*s\zeta\approx0.306,0.445$, respectively; sharing selectors across jobs supplies the factor $J'$. Both guards in~\eqref{eq:group-guards} and $4b<0.3$ also have positive slack. All remaining direct branches are bounded as in Sections~\ref{sec:finish}--\ref{sec:envelope}. Such bounded-size fallbacks do not depend on the magnitudes or the signs of the input reals.

\emph{Exact reconstruction.} Apply Lemma~\ref{lem:walsh-finite} to each synchronized sweep of line~\ref{alg:mp-count} of Algorithm~\ref{alg:minplus}. Every genuine chunk is evaluated with its own modulus and active set. Its exact counts reconstruct its genuine recovery measurements; zero jobs are discarded. Evaluating the recovered improving candidates and the incumbent gives each block minimum, and taking the minimum over blocks gives the desired product.

\emph{Running time.} The three encoders on each side, joint products and two reads have savings
\[
 1-eA_{\rm root},\quad1-eA_{\rm res},\quad1-eA_{\rm box}
 \quad\text{on each side},\qquad
 eV,\quad x-eq_1,\quad x-eq_2 .
\]
Every one of these and every other row of Section~\ref{sec:envelope} exceeds the column's target $S$. Appendix~\ref{app:numerical} gives a minimum slack greater than $8.42\cdot10^{-8}$ in the first column and $2.25\cdot10^{-8}$ in the second. The fixed slack absorbs all uniform subpolynomial factors: synchronized trials, support recovery, rounding in the entry procedure, address and bit costs. This proves deterministic $O(n^{2.9929729})$ and $O(n^{2.99306273})$ product bounds from the two stated rectangular multiplication theorems.

\emph{Return to APSP.} By Section~\ref{sec:apsp}, Algorithm~\ref{alg:apsp} returns all distances exactly with $O(\log(2n))$ calls of Algorithm~\ref{alg:minplus}, plus $O(m_{\rm in})$ further work for a list of $m_{\rm in}$ input edges. The fixed slack also absorbs the $O(\log(2n))$ squarings.
\end{proof}

%% file: conclusion.tex
\section{Conclusion}\label{sec:conclusion}

We compute exact min-plus products and directed all-pairs distances with real weights in deterministic $O(n^{2.9929729})$ time using Uzna\'nski's theorem $\omega(1,\tfrac12,1)=2$. With the rectangular bounds of Le Gall and Urrutia alone, the running time is $O(n^{2.99306273})$.

The algorithm chooses a common candidate set and identifies the improving candidates at each output using certified comparison counts. Those counts reduce to prescribed entries of ordinary integer matrix products. Section~\ref{sec:ordinary} evaluates part of each requested entry jointly across the products, using coefficients that depend on the requested position but not on the numerical entries and are therefore the same even for products formed at different moduli.

All these bounds are asymptotic. In each of the two parameter choices of Section~\ref{sec:point}, the padded inner dimension of the ordinary products is $4^m$, where $m$ must be a positive multiple of a fixed integer of at least $1.25\cdot10^{15}$, so the algorithm uses these products only on astronomically large inputs. The numerical inequalities on which both exponents depend were checked by computer: Appendix~\ref{app:numerical} gives the exact rational interval procedure, and the verification companion~\cite{VerificationCompanion2026} contains the parameters and these calculations.

%% file: numerical.tex
\section{Rational rate and cost checks}\label{app:numerical}

To finish the proof of Theorem~\ref{thm:main}, we need positive slack in both the dimension tests and the running-time inequalities. The first table evaluates the complete rates in~\eqref{eq:walsh-rates} at the two columns of~\eqref{eq:point}; the second checks that the constructions fit; the third checks every charged cost against the target saving. The rational calculation following the tables certifies their endpoints. Each entry in the first table is a lower endpoint $l$ of an interval $[l,l+10^{-12}]$. For costs we use upper endpoints for the three encoders, the fit and both reads, and a lower endpoint for the joint gain. The labels refer to the rectangular multiplication theorem for that whole column.
\begin{center}\begin{tabular}{lrr}
\toprule & Uzna\'nski & Le Gall--Urrutia\\
\midrule
Root encoder $A_{\rm root}$ & $8.279746670440$ & $8.362777249905$\\
Residual encoder $A_{\rm res}$ & $10.875517568105$ & $11.150316713000$\\
High encoder $A_{\rm box}$ & $10.765770160604$ & $10.899976222985$\\
Fit $T$ & $10.958136759419$ & $11.228210852797$\\
Joint gain $V$ & $0.077005005743$ & $0.077893413324$\\
Low read $q_1$ & $0.422994970231$ & $0.422106586652$\\
High read $q_2$ & $0.422994973463$ & $0.422106585980$\\
\bottomrule\end{tabular}\end{center}
The next table bounds the six guards from below. Its rows are respectively $1-eT$, $\eta+er-\kappa_*es\zeta$, $\lambda_hu-r$, $0.3-4b$, $(c-1)\gamma(2v)-\zeta$, and the derivative in~\eqref{eq:group-guards}.
\begin{center}\begin{tabular}{lrr}
\toprule & Uzna\'nski & Le Gall--Urrutia\\
\midrule
Outer-block fit & $0.000002080255$ & $0.000000383365$\\
Grouped data columns & $0.007594928257$ & $0.000955606736$\\
High-product shape parameter & $0.000001570000$ & $0.000000224300$\\
Selection shape parameter & $0.017487600000$ & $0.000157300000$\\
Window length & $0.960382864065$ & $0.967179698602$\\
Endpoint derivative & $2.709506620082$ & $2.716495758179$\\
\bottomrule\end{tabular}\end{center}
The following entries are lower bounds on the savings of Section~\ref{sec:envelope} minus the column's target $S$. Each of the three encoder rows represents two separately charged operands; recovery and score evaluation also remain separate charges. Thus the table accounts for all nineteen costs.
\begin{center}\begin{tabular}{lrr}
\toprule & Uzna\'nski & Le Gall--Urrutia\\
\midrule
Root encoders (two) & $0.237394681892$ & $0.248262498374$\\
Residual encoders (two) & $0.000514493701$ & $0.000000470612$\\
High encoders (two) & $0.010529625069$ & $0.022296132612$\\
Joint tables and ordinary products & $0.000000084205$ & $0.000000022546$\\
Low reads & $0.000000086397$ & $0.000000022548$\\
High reads & $0.000000086102$ & $0.000000022608$\\
Dense inputs and rank lists & $0.901716700000$ & $0.904001380000$\\
Selection and rank products & $0.017872900000$ & $0.023392730000$\\
Recovery construction & $1.922344800000$ & $1.918102055000$\\
Recovery row lists & $1.992972900000$ & $1.993062730000$\\
Measurement assembly and decoding & $0.038601000000$ & $0.037593405000$\\
Recovered scores & $0.038601000000$ & $0.037593405000$\\
Label updates & $0.017972900000$ & $0.023492730000$\\
Block initialization and merging & $0.063601000000$ & $0.068023405000$\\
Cubic selection work & $1.851716700000$ & $1.843141380000$\\
Quadratic selection work & $1.922344800000$ & $1.918102055000$\\
\bottomrule\end{tabular}\end{center}
Using the unrounded rational intervals, the minimum time slack is greater than $8.42\cdot10^{-8}$ for the Uzna\'nski column and $2.25\cdot10^{-8}$ for the Le Gall--Urrutia column. The smallest guard values are $0.00000157$ and $0.0000002243$, respectively. The uniform vanishing errors in the common inner dimension, rates, trial counts, address sizes, recovery and APSP squarings eventually fit within this fixed positive slack.

\paragraph{Exact finite calculations.}
The nonrational quantities in these tests are logarithms of positive rationals. Enclosing each logarithm by rational endpoints and propagating them outward therefore suffices to certify the displayed signs and slack.

For a rational $1\le y_{\log}\le2$ put $z_{\log}\coloneqq(y_{\log}-1)/(y_{\log}+1)$ and set
\[
 L_{96}(z_{\log})\coloneqq
       2\sum_{i=0}^{95}\frac{z_{\log}^{\,2i+1}}{2i+1}.
\]
The series for $\ln((1+z_{\log})/(1-z_{\log}))$ has a positive tail satisfying
\begin{equation}\label{eq:log-enclosure}
 L_{96}(z_{\log})\le\ln y_{\log}
 \le L_{96}(z_{\log})+
       \frac{2z_{\log}^{193}}{193(1-z_{\log}^2)}.
\end{equation}
For any positive rational $w_{\log}$, find the integer $j_{\log}$ such that
$w_{\log}=2^{j_{\log}}y_{\log}$ and $1\le y_{\log}<2$ by rational comparisons.
Apply~\eqref{eq:log-enclosure} to $y_{\log}$ and $2$, rounding outward to multiples of $10^{-65}$.
Add $j_{\log}\ln2$, reversing the endpoints of that term if $j_{\log}<0$, and round outward again. Substitute these intervals in~\eqref{eq:window-entropy}, \eqref{eq:group-entropy} and~\eqref{eq:walsh-rates}, using
\[
 \gamma'(v)=
 -\frac{\ln((c-\tfrac32+v)/(\tfrac12-v))}{(c-1)\ln4}
\]
for the derivative guard. For a product take the minimum and maximum of the four endpoint products; for a positive divisor invert its endpoints in reverse order. Finally round the rates outward to the displayed precision. For the guard and cost tables, substitute the unrounded intervals and round only the lower bound downward.

The exponents in Theorem~\ref{thm:main} depend on these computed inequalities. The displayed rational procedure verifies them. We also carried out these checks by computer in exact rational arithmetic. The verification companion~\cite{VerificationCompanion2026} contains the literal rational parameters and these finite calculations.

%% file: rectangular.tex
\section{Exact integer rectangular products}\label{app:rectangular}

The main algorithm multiplies rectangular matrices with integer entries and pays for each bit operation. To use the cited arithmetic bounds in this model, we account for the sizes of the intermediate integers and for the cost of writing the whole product.

Let $n$ be the long dimension and $H\le n^\alpha$ the short dimension, as in Lemma~\ref{lem:rectangular}, and let
$b_{\rm loc}$ be the input bit length. The entry procedure needs the product of
$n\times H$ and $H\times n$ matrices, including all $n^2$ outputs.
Candidate selection also needs an $n\times n$ by $n\times H$ product. We give
the finite integer implementation of Lemma~\ref{lem:rectangular} for both
orientations.

It suffices to obtain, for each shape parameter and a reserved positive exponent loss, one exact rational bilinear formula of fixed finite dimensions. Such a formula can be chosen before the data arrive, rotated to the second product orientation, and applied recursively using integer arithmetic. The two cited arithmetic bounds supply the required formulas by different arguments. We first show that rational formulas exist, then give the common implementation and bit-cost analysis.

\paragraph{A rational base at shape parameter $1/2$.}
The faster real-weight algorithm uses the following result of Uzna\'nski both for the group products of Section~\ref{sec:walsh-sharing} and for the high products. Uzna\'nski proves $\omega(1,\tfrac12,1)=2$~\cite[Corollary~1.2 and Lemma~6.1]{Uznanski}, working over $\mathbb C$~\cite[p.~1]{Uznanski}. In that exponent convention, for every fixed $\varepsilon>0$ this yields arithmetic cost $O_\varepsilon(n^{2+\varepsilon})$ for short side $H\le n^{1/2}$. In its
tensor formulation, the complex asymptotic rank of
$\mathcal T\coloneqq\langle4,2,4\rangle$ is at most $16$. Here
$\langle a,b,c\rangle$ denotes the trilinear form
$\sum_{ijk}X_{ij}Y_{jk}Z_{ki}$ of these dimensions.
This bound follows from $\omega(1,\tfrac12,1)=2$ and the definition of asymptotic rank: rectangular exponents are the same for arithmetic cost and for tensor rank, so $\mathcal T^{\otimes j}=\langle4^j,2^j,4^j\rangle$ has rank $4^{(2+o(1))j}$. Directly, in the notation of~\cite[(10)--(11)]{Uznanski}, where a tensor character~\cite[(7)]{Uznanski} is a point of Strassen's asymptotic spectrum, the proof of~\cite[Theorem~1.1]{Uznanski} for the exponents $(1,1,\tfrac12)$ shows that the dot-product parameters of every tensor character satisfy $p_X+p_Y+\tfrac12p_Z\le2$. By~\cite[(11)]{Uznanski}, every character therefore takes the value $4^{p_Y}2^{p_Z}4^{p_X}\le16$ on $\mathcal T$, and Strassen's spectral theorem~\cite[Theorem~3.2]{Uznanski} bounds the asymptotic rank of $\mathcal T$ by the largest character value. Equivalently, take $\mathbf h=(1,1,\tfrac12)$ and $N=4$ in~\cite[(26)]{Uznanski}; since $\lceil N^{h_i}\rceil=N^{h_i}$, the factor $8$ in that bound is not needed.

Uzna\'nski states, citing Strassen, that his exponent bounds hold over every field of characteristic zero~\cite[p.~1]{Uznanski}. We give an explicit finite-basis argument for the rational implementation. Fix an exact complex rank-$R_{\mathbb C}$ decomposition
of $\mathcal T^{\otimes j}$. Its coefficient equations are polynomial equations over
$\mathbb Q$. The ideal they generate is proper because there is a complex
solution. A maximal ideal containing it gives, by Zariski's lemma, a
residue field $F$ finite over $\mathbb Q$. It supplies an exact solution
in $F$. Choose a basis of $F$ over $\mathbb Q$ starting with
$1$. For every $v$, tensor the same solution $v$ times over this one
field. Each of the $R_{\mathbb C}^v$ products of two $F$-linear forms costs at most
$[F:\mathbb Q]^2$ products of rational coordinate forms. Recombination through the
multiplication table of the basis and projection to the coefficient of
$1$ recover the rational tensor $\mathcal T^{\otimes jv}$ on rational inputs.
Thus its rational rank is at most $[F:\mathbb Q]^2R_{\mathbb C}^v$. Letting $v$ grow with $j$
fixed, then taking the infimum over $j$, bounds the rational asymptotic
rank of $\mathcal T$ by its complex asymptotic rank.

\paragraph{Rational bases at the Le Gall--Urrutia shape parameters.}
Use Le Gall and Urrutia over $\mathbb Q$~\cite[Theorem 1.1, Table 3 and Section 2.2]{LeGallUrrutia}; their exponent $\omega(1,k,1)$ is defined through the border rank of the exact shapes $\langle n,\lfloor n^k\rfloor,n\rangle$. Their lower bound on the dual exponent of matrix multiplication exceeds $0.31389$, giving $\omega_*=2$ at the two smaller shape parameters in Lemma~\ref{lem:rectangular}. At $\alpha=17/50$, Table 3 gives the upper bound $\omega_*=2.001118$. For each rational shape parameter $\alpha\eqqcolon p_0/q_0$ use exact shapes $\langle2^{q_0j},2^{p_0j},2^{q_0j}\rangle$. More explicitly, for a fixed
positive rational $\tau$ one may take a sufficiently large such shape
with a rational polynomial degeneration having fewer than
$(2^{q_0j})^{\omega_*+\tau/2}$ terms. If the
degeneration has leading term $z_{\rm deg}^{d_{\rm deg}} \mathcal T_*$, its $v$th power has leading
coefficient $\mathcal T_*^{\otimes v}$ at degree $d_{\rm deg}v$. Following the interpolation argument of Bini~\cite{Bini1980}, evaluation at $O(v)$
distinct rationals extracts that coefficient; the implicit constant
depends on the fixed polynomial degrees. The resulting exact rank
has only this linear overhead over the $v$th power of that number of terms.
Choose $v$ large enough to absorb it in the reserved exponent.
The powered dimensions still have exactly the same shape parameter.

\paragraph{Choosing the fixed formula and its orientation.}
For each shape parameter and desired fixed loss $\varepsilon>0$, choose a
rational $0<\sigma<\varepsilon/4$. The preceding arguments give
integers $B_0,H_0$ with $H_0=B_0^\alpha$ and an exact rational
rank-$R_0$ formula with $R_0<B_0^{\omega_*+\sigma}$. For shape parameter $1/2$ take $(B_0,H_0)\coloneqq(4^j,2^j)$, and for a Le Gall--Urrutia shape parameter $p_0/q_0$ take $(2^{q_0j},2^{p_0j})$. Dovetail over $j$, possible ranks and all rational
coefficient lists of increasing height. For each list, expand the finite
bilinear identity and compare every coefficient. The strict rank test
uses integer powers since $\sigma$ is rational. Some finite list passes,
so this constructor terminates. Its cost, coefficient sizes and
storage are constants for the chosen shape parameter and loss.

The identity for $\langle B_0,H_0,B_0\rangle$ gives the orientation used by the entry procedure.
Rotate the trilinear variable groups from $(X,Y,Z)$ to $(Z,X,Y)$
to give $\langle B_0,B_0,H_0\rangle$. Thus the appropriate coefficient
family becomes the decoder, with its local row/column indices permuted.
The rank and rational coefficients carry over to this cyclic orientation used by candidate selection. For each orientation let $D_0$ clear all denominators in
the two input maps and the decoder. Multiplying each map by $D_0$
scales a one-level product by $D_0^3$. At recursion depth $t$ the
integer result is $D_0^{3t}XY$, by bilinearity and induction. Decode
and divide exactly before accumulating entries for a different run,
pair of outer blocks or orientation.

\paragraph{Integer recursion and bit cost.}
Use the least $t\ge0$ for which $B_0^t\ge n$, found by integer
comparison. Then $H_0^t\ge H$ and, for $n>1$, $B_0^t<B_0n$.
Every padded dense array has at most $B_0^{2t}$ entries. Flattening
the tensor along its $B_0^2$-dimensional leg gives rank $B_0^2$:
each coordinate is separated by matrix units on the two other legs.
Consequently $R_0\ge B_0^2$. At recursion level $j$ there are at most $R_0^j$ nodes, each with remaining depth $t-j$. The fixed linear combinations, copies and output updates at one node cost $O_{B_0}(B_0^{2(t-j)})$ scalar operations. Summing over $0\le j\le t$ gives $O_{B_0}((t+1)R_0^t)$ in total.
Processing one child at a time uses $O_{B_0}((t+1)B_0^{2t})$ integer
slots. The absolute row sums of the fixed encoding and decoding maps
bound the growth per level by a fixed factor. A leaf scalar product
doubles the encoded bit length, so every intermediate and partial sum
has $O_{B_0}(b_{\rm loc}+t+1)$ bits. The common denominator power has
$O_{B_0}(t+1)$ bits. Schoolbook arithmetic covers the signed additions,
multiplications and final exact divisions with a polynomial bit factor.

For candidate selection, $b_{\rm loc}=O(\log(2n))$ and all addresses have
$O(\log(2n))$ bits. Padding, output writes and these bit costs fit
in $O_\varepsilon(n^{\omega_*+\varepsilon})$. Inside a job the product costs at most
$O_\varepsilon(n^{\omega_*+\varepsilon})
\operatorname{poly}(L+b_{\rm in}+\log(2+N))$, including its global
addresses, where $N$ and $b_{\rm in}$ are the long dimension and the bit length of the input entries of the surrounding
ordinary product and $L$ is the number of levels in each half of a word of Section~\ref{sec:ordinary}. The local
matrix size need not control every global index or input bit length.
These bounds charge all $nH$ inputs on each side and all $n^2$
outputs of this product, including their permutation into
the requested outer-label order.

This supplies both orientations of Lemma~\ref{lem:rectangular} and the longer-entry variants used in Lemma~\ref{lem:walsh-finite}.

%% file: recovery-proof.tex
\section{Construction and decoding of the measurement matrix}\label{app:recovery-proof}

Lemma~\ref{lem:recovery} supplies the binary measurements used in Section~\ref{sec:finish}. Recall that one can deterministically construct a binary matrix $A_{d,\kappa}$ with at most $\kappa f(d)$ rows and at most $df(d)$ nonzero entries, together with its row and column lists, such that every $z\in\{0,1\}^d$ of support size at most $\kappa$ is recovered from $A_{d,\kappa}z$; the construction and decoding bit costs are the bounds $O(df(d))$ and $O(\kappa f(d))$ stated in Lemma~\ref{lem:recovery}. The matrix must isolate an element of every nonempty support of size at most $\kappa$, while bounding both the number of rows and the number of rows containing any index. We establish those bounds and the exact construction and peeling costs using the explicit disperser of Ta-Shma, Umans and Zuckerman~\cite[Definition 1.3, Theorem 1.4 and Section 6]{TaShmaUmansZuckerman2007}.

We use maps that meet the following requirements. For small $K$ a map with one right vertex meets them, and for larger $K$ the cited construction supplies the map; the next paragraph gives the threshold. For every $N=2^\ell$ with $\ell\ge1$ and power of two $1\le K\le N$, there is a uniform explicit map
\[
g_K:[N]\times[\Delta]\longrightarrow[M],
\qquad \Delta\le C\ell^a,
\qquad \frac{K\Delta}{C\ell^c}\le M\le2K\Delta,
\]
with fixed positive integers $C,a,c$, such that every set of at least $K$ left vertices has at least $M/2$ distinct neighbors. Given $\ell$ and $\log_2 K$, one computes $\Delta,M$. Given also the $\ell$-bit encoding of $u-1$ for a left vertex $u\in[N]$ and a seed $i\in[\Delta]$, one computes the right label $g_K(u,i)\in[M]$. Both operations take $\operatorname{poly}(\ell)$ bit operations: the seed uses $O(\log(2\ell))$ bits and the right label uses $O(\ell)$ bits. The constants may cover the bounded small values of $\ell$.

This is a constant-error disperser, with degree polynomial in the input bit length; the displayed lower bound only needs a fixed polynomial loss. A $K$-element set has at least $M/2$ and at most $K\Delta$ distinct neighbors, giving the stated upper bound on $M$. We use the map with $\Delta=M=1$ for $K\le C\ell^c$, where it meets all these requirements, and otherwise the dimensions returned by the uniform construction in the cited Section 6.

\begin{proof}[Proof of Lemma~\ref{lem:recovery}]
For $d=1$ use the identity matrix. Otherwise put $\ell\coloneqq\lceil\log_2d\rceil$ and embed $[d]$ in $[2^\ell]$. The subsets and measurements below use the one-based labels $u\in[d]$.

\emph{Sets that isolate an index.} We first construct a family $\mathcal P$ of subsets of $[d]$ such that every nonempty set $S$ of size at most $\kappa$ meets some member in exactly one element. In addition, $|\mathcal P|=\kappa\operatorname{poly}(\ell)$ and each element occurs in $\operatorname{poly}(\ell)$ members.

Include $[d]$ itself, which handles $|S|=1$. For every dyadic $K$ with $2\le K\le\kappa$, use the displayed disperser and put
\[
Q\coloneqq\max\left(2,\left\lceil\frac{4K\Delta}{M}\right\rceil\right).
\]
This is polynomial in $\ell$. If $K\le|S|<2K$, the set $S$ has fewer than $2K\Delta$ incident edges and at least $M/2$ active right vertices. Choose an active right vertex $v$ with at most $Q$ incident edges from $S$, and let $i\in[\Delta]$ be the seed of any one of these edges. The fiber $\{u\in S:g_K(u,i)=v\}$ therefore contains between one and $Q$ distinct elements.

Let $\mathcal Q$ consist of the first $(Q-1)(\ell-1)+1$ primes. For each seed $i$, right label $v\in[M]$, prime $q\in\mathcal Q$ and residue $0\le b<q$, include the set
\[
\{u\in[d]:g_K(u,i)=v,\ u\bmod q=b\}.
\]
Every nonzero difference of two indices has absolute value less than $2^\ell$ and thus at most $\ell-1$ distinct prime divisors. Some $q\in\mathcal Q$ therefore separates a chosen element of the fiber from its other at most $Q-1$ elements. The corresponding set meets $S$ once. Retaining all right labels and seeds handles parallel disperser edges.

Since $Q=\operatorname{poly}(\ell)$, the prime number theorem bounds the largest prime by $\operatorname{poly}(\ell)$, and sieving costs $\operatorname{poly}(\ell)$ bit operations. This scale has $\Delta M\sum_{q\in\mathcal Q}q=K\operatorname{poly}(\ell)$ sets, and each element occurs in $\Delta|\mathcal Q|=\operatorname{poly}(\ell)$ of them. The dyadic $K$ values sum to less than $2\kappa$ and there are at most $\ell+1$ scales. Enumerating $(u,i,q)$ fills the row and column lists in $d\operatorname{poly}(\ell)$ bit operations, including modular reduction, neighbor evaluation and all row headers. The indices and addresses have $O(\log d)$ bits.

\emph{Identifying an isolated index.} Let $L\coloneqq\lceil\log_2(d+1)\rceil$. For each $B\in\mathcal P$, put into $A_{d,\kappa}$ its indicator row and the $L$ indicator rows
\[
B_j\coloneqq\{u\in B:\text{bit }j\text{ of }u\text{ is }1\},
\qquad 0\le j<L.
\]
All are binary; repeated or empty rows are allowed. From their measurements form
\[
c_B\coloneqq\sum_{u\in B}z_u,
\qquad t_B\coloneqq\sum_{j=0}^{L-1}2^j\sum_{u\in B_j}z_u
=\sum_{u\in B}u z_u.
\]
The second identity is an integer sum with carries. Counts are at most $d$ and index sums at most $d(d+1)/2$, so these numbers have $O(\log d)$ bits.

\emph{Removing the recovered indices.} Initialize a queue with all original rows $B$ having $c_B=1$. On taking an entry from the queue, test its current count; discard it if the count is no longer one. Otherwise save $u\coloneqq t_B$ before changing any row, output $u$, and visit the sets $B\in\mathcal P$ containing $u$ via the column list of $u$, subtracting one from $c_B$ and $u$ from $t_B$. The rows $B_j$ supplied the initial index sums and require no further updates. Enqueue a row when its count first becomes one. Counts decrease, so each row is enqueued at most once. A stale queue entry has count zero. The maintained values are exactly the counts and index sums of the residual support. Whenever that support is nonempty, the defining property of $\mathcal P$ supplies a count-one row. Therefore the process ends exactly when the support is empty, and every output index is emitted once.

\emph{Cost.} There are at most $\kappa$ emitted indices, each with $\operatorname{poly}(\ell)$ column-list occurrences. Reading all measurements, initializing row records, testing stale entries and updating sums take $\kappa\operatorname{poly}(\ell)$ bit operations. This includes the empty support and requires no fresh $d$-entry visited array for an individual vector. Multiplication of the row and occurrence bounds by $L+1$ for the rows $B_j$ is absorbed in $f$. The same lists and matrix serve every vector.
\end{proof}

The resulting matrix and lists therefore meet every construction and decoding bound in Lemma~\ref{lem:recovery}.

%% file: original-embedding.tex
\section{The original squared construction}\label{app:original-embedding}

The original five-term forms give the same complete grid of block products as the aligned embedding of Section~\ref{sec:walsh}, with coefficient one. We give their arrays, runs and inverse locator to make explicit the version of Majdoddin's squared construction with restricted inner sets. The common maps and their construction cost are supplied by Lemma~\ref{lem:profile-embedding}; here $L,M,k,\calF,\calG,K',K_g,D_s,N_0$ and $\Lambda$ have the meanings in that lemma and its setup. The cancellation of extra opposite letters is the argument also used for the aligned forms, whose complete contraction is scaled by $4^L$. The proof of Theorem~\ref{thm:main} does not use this appendix.

\paragraph{Encoding the operand blocks.}
In Majdoddin's squared construction a word has two halves of $L$ levels. A left raw word $u'u''$ has first-half letters in the normal alphabet $\{x_1,x_2,p_1,p_2\}$ and second-half letters in the opposite alphabet $\{y_1,y_2,q_{\rm loc}\}$. A right raw word $v'v''$ uses these alphabets in the reverse order. Thus each operand array has $4^L3^L=12^L$ slots.

In a normal level the inherited forms are
\[
\begin{aligned}
f_{11}&\coloneqq x_1+p_1,&f_{12}&\coloneqq x_1-p_2,&f_{21}&\coloneqq x_2,\\
f_{22}&\coloneqq x_2+p_2,&f_0&\coloneqq-x_1-x_2.
\end{aligned}
\]
The opposite forms are $g_{11}\coloneqq y_1$, $g_{12}\coloneqq y_2$, $g_{21}\coloneqq y_1-q_{\rm loc}$, $g_{22}\coloneqq y_2+q_{\rm loc}$ and $g_0\coloneqq y_1+y_2$. Their paired sum is
\[
 I\coloneqq p_1y_1+p_2q_{\rm loc}=\Pi+\sum_{a,e\in\{1,2\}}f_{ae}g_{ae},
 \qquad \Pi\coloneqq f_0g_0 .
\]
An ordinary label $ae$ specifies the two factors $f_{ae},g_{ae}$, and the private label specifies $f_0,g_0$, in the orientation of that half.

For a normal word let $Q$ be its set of $p_1,p_2$ levels, $T$ its set of $p_2$ levels, and $\iota$ the sequence of $x$ subscripts in level order. For an opposite word let $T$ be its set of $q_{\rm loc}$ levels and $S$ its set of $y_2$ levels. For a binary matrix $U$, write $U[J]$ for the sum over $\F_2$ of its columns indexed by $J$. Recall from Lemma~\ref{lem:profile-embedding} that for a good set $Q$, $\Gamma$ maps subsets of $Q$ to inner-index bits and $\Xi$ maps subsets of its complement to outer-address bits.

For each $Q\in\calG$, denote its two operand blocks by $X_Q$ of size $N_0\times D_s$ and $Y_Q$ of size $D_s\times N_0$. Index rows by $\{1,2\}^k\times\F_2^k$, inner indices by $\F_2^M\times\F_2^M$, and columns by $\F_2^k\times\{1,2\}^k$. In each run, $a_\sigma$ is paired only with $b_\sigma$. For $\sigma\in\{-M,\ldots,M\}$ the two arrays are
\[
\begin{aligned}
 a_\sigma[u'u'']&\coloneqq X_{Q(u')}
 \big[(\iota(u'),\Xi[S(u'')]),(\Gamma[T(u')],\Gamma[T(u'')])\big],\\
 b_\sigma[v'v'']&\coloneqq Y_{Q(v'')}
 \big[(\Gamma[T(v')],\Gamma[T(v'')]),(\Xi[S(v')],\iota(v''))\big].
\end{aligned}
\]
The first lookup requires $Q(u')\in\calG$ and $|T(u'')|-|T(u')|=\sigma$.
The second requires $Q(v'')\in\calG$ and $|T(v'')|-|T(v')|=\sigma$. Set a slot to zero whenever its test fails. Computing the letter sets, binary matrix products and good-set dictionary index takes $\operatorname{poly}(L)$ local bit operations per slot, plus the charged input access.

\paragraph{Reconstructing the entry.}
An output word $w'w''$ uses the five letters $z_0,z_{11},z_{12},z_{21},z_{22}$ in each half. For an output half $w$, let $Q(w)$ be the levels of $z_0$, $\iota(w)$ the sequence of first subscripts outside $Q(w)$, and $S(w)$ the levels with second subscript $2$. Put $Q_1\coloneqq Q(w')$, $Q_2\coloneqq Q(w'')$, both in $\calG$. At an outer letter $z_{ab}$ the bilinear kernel is $f_{ab}g_{ab}$ with the forms defined above. At an inner letter $z_0$ it is $p_1y_1+p_2q_{\rm loc}$, with operand roles exchanged in the second half. A run value contracts the product of these level kernels against $a_\sigma,b_\sigma$: sum over raw word pairs, weighting the two array entries by the corresponding kernel coefficient. 
\begin{lemma}[Original squared construction; after Majdoddin~{\cite[Theorem~5.1]{Majdoddin}}]\label{lem:original-embedding}
The original squared construction on the list $\calG$ computes the complete cross grid of $K_g$ left and $K_g$ right blocks, each of outer width $N_0$. Its band width is $\Lambda=K_g4^k$. For arbitrary operand blocks, summing its $2M+1$ \emph{runs} $\sigma\in\{-M,\ldots,M\}$, each a contraction of one left and one right operand array as defined above, gives every block product with coefficient one. The construction cost in Lemma~\ref{lem:profile-embedding} covers the maps and lists of inner sets. For outer dimension $N$, the condition $N\ge K'4^k$ is sufficient for fitting a band and implies $L\le\log_2 N$.
\end{lemma}

\begin{proof}
The coefficient argument in the proof of Majdoddin's Theorem 5.1~\cite{Majdoddin} gives the following scalar entry on our restricted list after summing over the runs:
\[
 (X_{Q_1}Y_{Q_2})
 \big[(\iota(w'),\Xi[S(w'')]),(\Xi[S(w')],\iota(w''))\big].
\]
To see why the restriction to $\calG$ preserves this identity, note that each inner level forces a normal $p$ letter. The $M$ inner levels in each half therefore determine that normal inner set. For normal $p_2$ subsets $A\subseteq Q_1,B\subseteq Q_2$, the run tests give
\[
 |B|+e_L-|A|=\sigma=|B|-|A|-e_R,
\]
where $e_L,e_R$ count extra $q_{\rm loc}$ letters outside the inner levels of the opposite halves. Both vanish. At an outer letter $z_{ab}$ the remaining normal and opposite forms pick $x_a$ and $y_b$ respectively, each with coefficient one. Thus the outside subscripts fix the $\iota$ sequences and the two sets of $y_2$ levels used in the displayed row and column. Each surviving monomial has coefficient one and occurs in the unique run $\sigma=|B|-|A|$. The restricted $\Gamma$ maps send $(A,B)$ bijectively to all $4^M$ inner indices. The argument uses just the two queried good sets. Each requested entry determines its bands and blocks by division with remainder by $\Lambda$ and $N_0$. The two stored inverses of the complementary $\Xi$ restrictions recover the second subscripts from its outer address, and the $\iota$ sequences give the first subscripts. This constructs its unique output word in polynomial time in $L+\log(2N)$.

A left run array depends only on its row band, and a right run array only on its column band. They can be reused over all opposite bands. If $N\ge\Lambda$, padding to $b_N\coloneqq\lceil N/\Lambda\rceil$ bands gives $b_N\Lambda<2N$. Under the sufficient fit $N\ge K'4^k$, setup is $O(N\operatorname{poly}(L))$, because $k\ge M$. Also $L=M+k\le2k\le\log_2N$. The actual width $\Lambda$ and sufficient threshold $K'4^k$ differ by at most the fixed factor $1024/81$.
\end{proof}

\section{Constructing the translation list}\label{app:translation-list}

Following the proof of the working note's lemma \emph{An exact half-weight source cap}~\cite{WorkingNote}, we construct the list of $J_\delta=64M^3$ translations used in Section~\ref{sec:walsh-cover}: for every invertible binary $M\times M$ matrix $A$ and every $h\in\F_2^M$, some $j$ satisfies $\operatorname{wt}(A(h+\delta_j))\le\lfloor M/2\rfloor$. Set $\varepsilon_0\coloneqq1/(2M)$ and $q_0\coloneqq1+1/(4M)$. For each nonzero character $a$ let $S_a\coloneqq\sum_j(-1)^{a\cdot\delta_j}$, and minimize
\[
 \sum_{a\ne0}(q_0^{S_a}+q_0^{-S_a})
\]
by revealing the $J_\delta M$ bits by conditional expectations. Its initial mean is
$2(2^M-1)((q_0+q_0^{-1})/2)^{J_\delta}<q_0^{\varepsilon_0J_\delta}$.
Indeed $\ln q_0\ge7/(32M)$ and
$\ln((q_0+q_0^{-1})/2)\le1/(32M^2)$, so the logarithmic gap is at least $5M>\ln(2(2^M-1))$.
Every $|S_a|<\varepsilon_0J_\delta$. For an invertible binary $M\times M$ matrix $A$ and $h\in\F_2^M$, each output bit of $A(h+\delta_j)$ is determined by a nonzero character, with a fixed sign from $h$. The average Hamming weight of $A(h+\delta_j)$ over the list is therefore less than $M/2+1/4$. Integrality gives some $j$ with $\operatorname{wt}(A(h+\delta_j))\le\lfloor M/2\rfloor$. Exact conditional expectations over the $2^M-1$ characters take $2^M\operatorname{poly}(M)$ bit operations, with rational operands of polynomially many bits.